\documentclass[a4paper,11pt]{article}
\pdfoutput=1 
\RequirePackage[2019/11/18]{latexrelease}
\usepackage{jheppub} %
\usepackage[T1]{fontenc}
\usepackage[utf8]{inputenc}
\usepackage{lmodern}
\usepackage{amsmath}
\usepackage{cancel}
\usepackage{bm}
\usepackage{xcolor}
\usepackage{graphicx}
\usepackage{subcaption}

\def\ca{C_{\rm A}}
\def\cf{C_{\rm F}}

\allowdisplaybreaks
\newcommand{\p}{{\bm{p}}}
\newcommand{\ke}{{\bm{k}}}
\newcommand{\q}{{\bm{q}}}

\newcommand{\norm}[1]{\frac{d^3 #1}{(2\pi)^3}}

\title{Medium-induced fragmentation and equilibration of highly energetic partons.}
\abstract{
We investigate the energy loss and equilibration of highly energetic particles/jets inside a QCD medium. Based on an effective kinetic description of QCD, including $2\leftrightarrow 2$ elastic processes, radiative $1\leftrightarrow 2$ processes, as well as the back-reaction of jet constituents onto the thermal medium, we describe the in-medium evolution of jets from the energy scale of the jet $\sim E$ all the way to the medium scale $\sim T $. While elastic processes and back-reaction are important to describe the equilibration of soft fragments of the jet, we find that the energy loss is dominated by an inverse turbulent cascade due to successive radiative branchings, which has interesting implications for the energy spectra and chemistry of jet fragments.
}
\author[a]{Soeren Schlichting}
\emailAdd{sschlichting@physik.uni-bielefeld.de}
\author[a]{Ismail Soudi}
\emailAdd{isma@physik.uni-bielefeld.de}

\affiliation[a]{Fakult\"at f\"ur Physik, Universit\"at Bielefeld, D-33615 Bielefeld, Germany}

\begin{document}

\maketitle
\section{Introduction}
Experimental studies of heavy-ion collisions at RHIC  \cite{Adcox:2001jp,Adler:2002xw} and LHC \cite{Aad:2010bu,Chatrchyan:2012nia,Aad:2014bxa,Abelev:2013kqa} have established that a deconfined QCD medium is formed at the early stages of the collision. Besides soft observables which display collective behavior~\cite{Schenke:2010nt,Baier:2007ix}, another crucial signatures of the Quark-Gluon-Plasma (QGP) in heavy-ion collisions is the energy loss and suppression of highly energetic particles or jets. Due to the interaction with the medium, highly energetic particles or jets can lose energy, and perhaps even thermalize inside the medium, 
thereby providing an important tool to characterize the QGP medium and a unique opportunity to gain insights into the non-equilibrium dynamics of hot and dense QCD matter. Since the same processes underlying jet quenching are also believed to be responsible for the equilibration of the QGP at early times \cite{Baier:2000sb,Schlichting:2019abc,Berges:2020fwq}, establishing a conclusive picture of the in-medium jet evolution all the way from the energy scale of the jet $\sim~E$ to the scale of thermal medium $\sim T$ is an important task, whose experimental and theoretical study provide a unique opportunity to explore the out-of-equilibrium dynamics of the QGP.


Starting with the original calculations of the rates for medium-induced radiation  \cite{Baier:1996kr,Baier:1996sk,Zakharov:1996fv,Zakharov:1997uu,Wiedemann:2000za} and more recent refinements \cite{Arnold:2008iy,Arnold:2008zu,CaronHuot:2010bp,Sievert:2018imd,Kang:2016mcy,Barata:2020sav,Andres:2020vxs}, there have been several theoretical studies tackling the description of the jet evolution in QCD medium, based on (semi-)analytic approaches \cite{Baier:2001yt,Mehtar-Tani:2014yea,Arleo:2017ntr}, effective kinetic descriptions \cite{Jeon:2003gi,Fochler:2011en,Blaizot:2013hx,Blaizot:2015jea,Mehtar-Tani:2018zba,Adhya:2019qse} as well as sophisticated Monte Carlo simulations \cite{Zapp:2011ya,Zapp:2012ak,Casalderrey-Solana:2014bpa,Casalderrey-Solana:2016jvj,Caucal:2018ofz,Chen:2017zte,Putschke:2019yrg,Schenke:2009gb} some of which even include a fully coupled jet-medium evolution. 
Although, all of these approaches provide a solid description of experimental measurements in their respective range of applicability, a general challenge in the description of in-medium jet evolution is to devise a consistent simultaneous description of the near-thermal soft and hard constituents of the jet. Evidently, a complete description of jet evolution in high-energy heavy-Ion collisions is complicated as it requires the combination of different physical mechanisms starting from the initial jet production, via the medium modified vacuum shower and medium-induced splittings all the way to hadronization. Since each stage demands different physics, and experimental observables which are measured at the end can in principle be sensitive to all stages of the evolution, we will not attempt to make direct comparisons with experimental data. Instead our study will solely concentrate on the in-medium evolution of highly energetic partons, in order to establish a detailed microscopic understanding of the in-medium modification of hard fragments, the equilibration of soft fragments with the surrounding medium and the connections of the process of jet quenching with that of the thermalization of near equilibrium excitations of the QGP.

While previous works have shown that energy loss in the medium is governed by successive radiative branchings driving an inverse turbulent energy cascade \cite{Blaizot:2013hx,Blaizot:2015jea,Mehtar-Tani:2018zba}, the focus of these earlier studies was on the effects on the hard constituents and did not properly take into account the equilibration of soft fragments and energy balance with the medium. We improve on this analysis, by using the full medium-induced radiation kernel (computed in the infinite medium), and also include elastic energy loss and medium recoils in the small angle approximation, allowing us to follow the jet evolution from high energies ($\sim E$) all the way the soft medium sector ($\sim T$). Based on this effective kinetic description, which parallels earlier studies in the context of jet quenching~\cite{Blaizot:2013hx,Blaizot:2015jea,Mehtar-Tani:2018zba,Adhya:2019qse} and thermalization of the QGP~\cite{Schlichting:2019abc,Kurkela:2014tea,Baier:2000sb,Kurkela:2018oqw,Kurkela:2018wud}, we carefully examine the evolution of the in-medium jet shower starting from early collisional and radiative energy loss all the way to complete equilibration of jets inside the medium. We establish three different regimes corresponding to initial elastic and radiative energy loss, turbulent energy loss via multiple successive branchings and equilibration, and provide detailed analytic discussions of the underlying physics mechanisms in each regime. By comparing the results for the in-medium evolution of highly energetic partons to that of the low lying excitations of a (weakly coupled) thermal QGP, we finally determine to what extent the physics of near-equilibrium excitations is relevant to the problem of jet quenching.

This paper is organized as follows: We first introduce the effective kinetic description in Section \ref{sec:Kinetic-Equation}, followed by a short discussion of the parametrical behavior of the various processes. We present our main results in Section \ref{sec:Equilibration}, where we discuss different phases of the in-medium jet evolution, with an emphasis on how a turbulent cascade dominates the energy loss and affects the chemical composition of medium modified jets throughout the evolution. We conclude in Section \ref{sec:Conclusion}, with a summary of our main results and their possible consequences, and provide additional details of our study in Appendices \ref{ap:NumericalImplementation} and \ref{ap:SmallAngle}.

\section{Effective kinetic description of in-medium jet evolution }\label{sec:Kinetic-Equation}

\begin{figure}
    \centering
    \includegraphics[width=\textwidth]{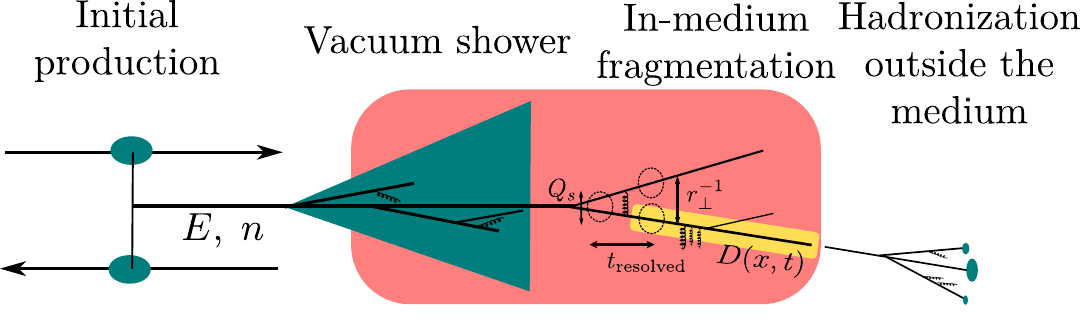}
    \caption{Visualization of jet evolution in heavy-ion collisions. We focus on the in-medium evolution of the partons once they are resolved by the medium. }
    \label{fig:JetDiagram}
\end{figure}

\subsection{Jet evolution in heavy-ion collisions}
Jets are defined, via a suitable clustering algorithm, as collimated sprays of hadrons that are produced from a hard collision process. Starting from a highly virtual partons produced in an initial hard scattering, in the vacuum the jet constituents evolve toward mass-shell following a collinear parton shower (akin to the time-like DGLAP equation) \cite{Altarelli:1977zs,Gribov:1972ri}.  In the presence of a QCD medium, the evolution involves additional processes and different approaches have been used to study the modification of the jet shower by the medium (see e.g. \cite{Cao:2020wlm,Blaizot:2015lma,Mehtar-Tani:2013pia} for a review). One approach consists of a modification of the DGLAP evolution to include medium modifications as done in some MonteCarlo event generators \cite{Majumder:2013re,Armesto:2009fj,Renk:2009nz,Cao:2020wlm}; a different approach is to follow the kinematics of hard partons inside the medium and study how the jet constituents interact with the medium, as e.g. done in \cite{Blaizot:2013hx,Blaizot:2015jea,Mehtar-Tani:2018zba,Adhya:2019qse}. Even though our work is not intended to provide a comprehensive description of in-medium jet evolution, we will generally follow the latter approach, by studying the energy loss of hard partons inside a thermal medium.

Besides the scales  relevant for vacuum dynamics, the dynamics of jets inside the medium are sensitive to additional scales emerging from the medium (as discussed in \cite{CasalderreySolana:2012ef,Mehtar-Tani:2013pia}), including the typical transverse momentum acquired during scatterings $Q_s = \sqrt{\hat{q} L}$ where $L$ is the medium size and ${\hat q}$ is the momentum broadening parameter; and the typical inverse size of the jet in the medium $r_\perp^{-1} =(\Theta L)^{-1}$, where $\Theta$ is the opening angle of the jet \cite{CasalderreySolana:2012ef,Mehtar-Tani:2013pia}.  Vacuum-like emissions can be factored out \cite{Mehtar-Tani:2017web,Caucal:2018dla} and degrade the virtuality of the partonic constituents as in the vacuum; however individual jet constituents embedded in the medium are only resolved by the medium when the separation length is larger than the medium resolution scale $Q_s^{-1} \ll r_\perp$, and it is thus useful to define a decoherence time  $t_{\rm resolved} \sim \left(\frac{1}{\hat{q}\theta_{12}^2}\right)^{1/3}$ \cite{Mehtar-Tani:2013pia} where constituents of the jet become resolved by the medium.
Subsequently the resolved constituents interact with the medium as uncorrelated colored partons, which undergo elastic and in-elastic interactions with the medium leading to energy loss of the highly energetic partons and energy transfer to the thermal medium. Throughout the in-medium evolution, vacuum-like emissions at smaller and smaller angles continue to form, effectively providing a source of partons as they become resolved by the medium. Eventually, after a time $\sim L$ the hard constituents of the jet leave the medium and the vacuum-like evolution continues outside the medium \cite{Caucal:2018dla}, until it reaches non-perturbative scales where the constituents confine into hadrons.

Based on this discussion, it is clear that a complete picture of jet evolution in heavy-ion collisions will have to start from the jet creation all the way to hadronization, as is schematically illustrated in Fig.~\ref{fig:JetDiagram}. Since such a description necessarily involves a variety of different processes at different scales, it is essential to develop a robust theoretical understanding of each stage, in order to devise suitable observables to probe e.g. the properties of medium induced emissions or study connections between jet quenching and equilibration. We will therefore not attempt to develop a complete description of in-medium jet evolution, but rather focus on the particular aspect of the energy loss and equilibrium of highly energetic partons due to interactions with the surrounding medium. While our discussion thus ignores effects of vacuum like emissions and the (lack of) color coherence of the vacuum shower, which are known to be important to describe some experimental observables such as e.g. fragmentation functions~\cite{Mehtar-Tani:2014yea,Caucal:2018dla}, 
we anticipate that in a more complete theoretical description of jets vacuum-like emissions can be absorbed into the initial conditions or factored out into a source term depending on their decoherence time. By investigating how uncorrelated colored partons lose their energy and eventually become part of the medium, our results can therefore be seen as a Green's function propagating a single medium-resolved high energetic parton through the medium\footnote{Note that we consider initial conditions as a very narrow gaussian which effectively correspond to a single parton with energy fraction $x=1$}. Since the evolution is linear, more realistic initial conditions including an early vacuum-like cascade can be implemented by a simple convolution with the corresponding Green's function.
Vacuum-like splittings could in principle appear at any spacial position along the path-length of the medium, producing new sources not included in our approach. However, such vacuum-like emissions inside the medium are sub-leading in the double log approximation \cite{Caucal:2018dla}.
Even though the complicated interplay of initial conditions, source terms and in-medium evolution will inevitably modify observables such as momentum spectra, the qualitative conclusions of this paper regarding energy loss and equilibration mechanism are thus not expected to be modified by such a more refined implementation. 

\subsection{Effective kinetic description of highly energetic partons}
We study the evolution and equilibration of high energetic partons inside a thermal Quark-Gluon plasma (QGP) based on an effective kinetic description, where both the hard particles and the thermal QGP medium are described by a phase-space distribution $f_{a}(p,x,t)$ of on-shell partons, with $a=g,q,\bar{q}$ denoting the parton species.
Starting from the effective kinetic theory of QCD , the time evolution of the phase space distribution functions $f_{a}(p,x,t)$ is governed by  the Boltzmann equation
\begin{eqnarray} 
\label{eq:QCDBoltzmann}
\left(\partial_t + \frac{\mathbf{p}}{|\mathbf{p}|} \mathbf{\nabla} \right) f_a(\p,\mathbf{x},t)=C_a^{2\leftrightarrow 2}[\{f_i\}]+C_a^{1\leftrightarrow 2}[\{f_i\}]\;,
\end{eqnarray}
where at leading order of the coupling constant $g$ one needs to include number conserving $2\leftrightarrow 2 $ processes and effective collinear  $1 \leftrightarrow 2 $ processes. While in principle Eq.~\ref{eq:QCDBoltzmann} can provide a detailed account of the space-time dynamics of the coupled hard particles-medium system, it is also notoriously difficult to solve and extract all the features of the rich underlying dynamics. In order to gain insights into the evolution and equilibration of the hard partons inside the QGP, we will therefore consider the evolution of hard particles inside a static homogenous thermal medium. Since the hard partons are dilute compared to the soft thermal particles, we can describe the phase-space distribution of the hard partons $\delta f_a(\p,\mathbf{x},t)$ as a linearized perturbation on top of the static medium and write
\begin{eqnarray}
f_a(\p,\mathbf{x},t) &=& n_a(p) +\delta f_a(\p,\mathbf{x},t)\;,
\label{eq:Linearization}
\end{eqnarray}
where $n_a(p)$ is the thermal distribution, i.e. depending on the particle species $n_g(p)=\frac{1}{e^{p/T}-1}$ is the Bose-Einstein distribution or $n_{q/\bar{q}}(p)=\frac{1}{e^{p/T}+1}$ is the Fermi-Dirac distribution\footnote{We consider vanishing chemical potentials for all quark flavors.}. By linearizing Eq.~(\ref{eq:QCDBoltzmann}) around thermal equilibrium and integrating over the position $\mathbf{x}$, we then obtain a closed set of evolution equations for the momentum distributions $\delta \bar{f}_a(\p,t)= \int d^3\mathbf{x}~\delta f_a(\p,\mathbf{x},t)$ of hard partons
\begin{eqnarray}
\label{eq:QCDLinBoltzmann}
\partial_t  \bar{f}_a(\p,t)=\delta C_a^{2\leftrightarrow 2}[\{f_i\},\{\delta\bar{f}_{i}\}]+\delta C_a^{1\leftrightarrow 2}[\{f_i\},\{\delta\bar{f}_{i}\}]\;,
\end{eqnarray}
which we will solve numerically to analyze the in-medium evolution of the parton shower. When investigating the in-medium evolution of the parton shower, it is often more convenient to study the re-distribution of energy, which can be quantified in terms of 
\begin{eqnarray}
\label{eq:FragmentationFct}
D_a(x,t)\equiv x \frac{dN_a}{dx}=\nu_a\int \norm{p}~\frac{|\p|}{E}~\delta\Big(\frac{|\p|}{E}-x\Big) \delta \bar{f}_a(\p,t)\;,
\end{eqnarray}
where $E$ is the total energy of the energy distribution and $x \equiv \frac{|\p|}{E}$ is the energy fraction carried by each parton in the energy distribution. The number of degrees of freedom are $\nu_g = 2(Nc^2-1)$ and $\nu_q = 2 N_c$. We note that the distribution $D_a(x,t)$ to some extent analogous to a fragmentation function in the vacuum  \cite{Mehtar-Tani:2018zba}; in particular the distributions satisfy the following sum rules, related to energy $E$ and charge $(Q_f)$ conservation 
\begin{eqnarray}
\sum_{a} \int dx~ D_{a}(x,t) = 1\;, \qquad \int \frac{dx}{x}~ \Big(D_{q_f}(x,t) - D_{\bar{q}_{f}}(x,t) \Big) = Q_{f}\;.
\end{eqnarray}

Based on Eq.~(\ref{eq:QCDLinBoltzmann}), the evolution of the momentum/energy distributions of partons $D_a(x,t)$ is entirely driven by interactions with the medium constituents, 

\begin{eqnarray}
\partial_t D_a(x,t) &=& C_a^{2\leftrightarrow 2}[\{D_i\}]+ C_a^{1\leftrightarrow 2}[\{D_i\}]\;,
\end{eqnarray}
where as in Eq.~(\ref{eq:FragmentationFct}) we have defined $ C_a[\{D_i\}]\equiv \nu_a\int \norm{p}~\frac{|\p|}{E}~\delta\Big(\frac{|\p|}{E}-x\Big) C_a[\{f_i\},\{\delta\bar{f}_{i}\}]$. We will include all contributions from $1 \leftrightarrow 2 $ inelastic processes and (small-angle) $2\leftrightarrow 2 $ elastic processes involving quark and gluon degrees of freedom. Below we provide a short summary of the contributions from the individual processes, along with some further details of the concrete implementation in our study.

\subsection{Small-angle scatterings}
Contributions to the collision integrals for elastic $2\leftrightarrow 2$ scattering processes can be further separated into large-angle scatterings and small-angle scatterings
\begin{eqnarray}
C_a^{2\leftrightarrow 2}[\{f_i\}] = C_a^{\rm large}[\{f_i\}]+C_a^{\rm small}[\{f_i\}]\;,
\end{eqnarray} 
by introducing a cut-off $\mu$ on the energy-momentum transfer in the $t$ and $u$ channels \cite{Blaizot:2014jna,Ghiglieri:2015ala}. When the infrared cutoff for the large-angle scattering is matched with ultraviolet cutoff for the small-angle scattering, it can be shown that the cut-off dependence cancels, and one recovers the full in-medium matrix elements at leading and next-to-leading order \cite{Ghiglieri:2015ala}. Since large-angle elastic scatterings exhibit the same parametric dependencies as small angle processes \cite{Arnold:2002zm,Schlichting:2019abc,Kurkela:2011ti}, we will only consider small-angle scatterings in the following and leave the conceptually straightforward inclusions of large-angle scatterings as a future task.

By considering the limit of small momentum transfer, the collision integral for small angle $2\leftrightarrow2$ scatterings reduces to a Fokker-Planck equation \cite{Blaizot:2014jna,Ghiglieri:2015ala}
\begin{eqnarray}
C_a^{\rm small}[\{f_i\}]=- \nabla_p \mathcal{J}_a+S_a\;, \label{eq:Fokker}
\end{eqnarray}
which features two distinct contributions, associated with drag and momentum diffusion ($- \nabla_p \mathcal{J}_a$) and conversion between quark and gluon degrees of freedom $(S_a)$. Drag and momentum diffusion arise from soft $u,t$-channel gluon exchanges, and can be characterized by the momentum currents 
\begin{eqnarray}
\label{eq:CurrentJG}
\mathcal{J}_g &=& -\frac{C_A}{4}\Big[\hat{\bar q} \nabla_\p f_g(\p) + \bar\eta_D \frac{\p}{|\p|} f_g(\p)(1+f_g(\p))\Big]\;,\\
\label{eq:CurrentJQ}
\mathcal{J}_{q_f} &=& -\frac{C_F}{4}\Big[\hat{\bar q} \nabla_\p f_{q_f}(\p) + \bar\eta_D \frac{\p}{|\p|} f_{q_f}(\p)(1-f_{q_f}(\p))\Big]\;,\\
\label{eq:CurrentJB}
\mathcal{J}_{\bar q_f} &=& -\frac{C_F}{4}\Big[\hat{\bar q} \nabla_\p f_{\bar {q}_f}(\p) + \bar\eta_D \frac{\p}{|\p|}f_{\bar{q}_f}(\p)(1-f_{\bar{q}_f} (\p))\Big]\;,
\end{eqnarray}
where $\hat{\bar q } $ and $\bar\eta_D$ are the momentum diffusion constant and the drag coefficient stripped of their respective color factor. One finds that at leading order
\begin{eqnarray}
\label{eq:qHatLO}
&\hat{\bar q} \equiv \frac{g^4}{\pi} \mathcal{L}\int \frac{d^3k}{(2\pi)^3}\left\{ \ca f_g(\ke)(1+f_g(\ke)) +\frac{1}{2}\sum_f \Big[f_{q_f}(\ke)(1-f_{q_f}(\ke))+f_{\bar{q}_f}(\ke)(1-f_{\bar{q}_f}(\ke)) \Big] \right\}\;,& \nonumber\\ 
\end{eqnarray}
\begin{eqnarray}
\label{eq:EtaDLO}
\bar\eta_D &\equiv&  \frac{g^4}{\pi} \mathcal{L}\int \frac{d^3k}{(2\pi)^3} \frac{2}{|\ke|}\left\{ \ca f_g(\ke)+\frac{1}{2}\sum_f  \Big[f_{q_f}(\ke)+f_{\bar{q}_f}(\ke) \Big] \right\},
\end{eqnarray}
where $\mathcal{L} = \int_{m_D}^{\mu} \frac{dq}{q}$ denotes the logarithmic phase-space for small angle scatterings, which we will take to be of order unity setting $\mathcal{L}=1$ in our analysis.

Similarly, conversion terms in Eq.~(\ref{eq:Fokker}) stems from soft $u$-channel quark exchanges in the $g q \leftrightarrow gq$, $g \bar{q} \leftrightarrow g\bar{q}$ and  $g g \leftrightarrow q\bar{q}$ processes, which effectively convert between particle flavors without significantly affecting their momenta. By following \cite{Ghiglieri:2015ala,Blaizot:2014jna}, we find that the corresponding terms in the Fokker-Planck equation can be written as 
\begin{eqnarray}
S_g&=&\frac{1}{8|\p|} \sum_{f} \left(\bigg[f_{q_f}(\p)(1+f_g(\p))-f_g(\p)(1-f_{\bar{q}_f}(\p)) \bigg]  \mathcal{I}_{\bar{q}_f} \right. \\
&& \qquad \qquad ~ +  \left. \bigg[f_{\bar{q}_f}(\p)(1+f_g(\p))-f_g(\p)(1-f_{q_f}(\p)) \bigg] \mathcal{I}_{q_f}\right), \nonumber \\ \label{eq:SourceGluon}
S_{q_f} &=&\frac{\nu_{g}}{\nu_{q}}\frac{1}{8|\p|}\Bigg[f_g(\p)(1-f_{q_f}(\p))\mathcal{I}_{q_f}-f_{q_f}(\p)(1+f_g(\p))\mathcal{I}_{\bar{q}_f}\Bigg]\;, \label{eq:SourceQuark}\\
S_{\bar{q}_f} &=&\frac{\nu_{g}}{\nu_{q}}\frac{1}{8|\p|}\Bigg[f_g(\p)(1-f_{\bar{q}_f}(\p))\mathcal{I}_{\bar{q}_f}-f_{\bar{q}_f}(\p)(1+f_g(\p))\mathcal{I}_{q_f}\Bigg]\;, 
\end{eqnarray}
where $\mathcal{I}_{q_f}$ and $\mathcal{I}_{\bar{q}_f}$ are given by the following moments of the phase-space distribution
\begin{eqnarray}
\label{eq:IqDef}
\mathcal{I}_{q_f}&=& \frac{g^4C_{F} \mathcal{L}}{\pi}\int \frac{d^3k}{(2\pi)^3}\frac{1}{|\ke|}\Big[f_{q_f}(\ke)(1+f_g(\ke))+f_g(\ke)(1-f_{\bar{q}_f}(\ke))\Big]\;,\\
\label{eq:IqBDef}
\mathcal{I}_{\bar{q}_f}&=&\frac{g^4C_{F} \mathcal{L}}{\pi}\int \frac{d^3k}{(2\pi)^3}\frac{1}{|\ke|}\Big[f_{\bar {q}_f}(\ke)(1+f_g(\ke))+f_g(\ke)(1-f_{q_f}(\ke))\Big]\;.
\end{eqnarray}
We note that while the conversion terms in Eq.~(\ref{eq:Fokker}), affect the chemistry of the QGP, they do not directly contribute to the redistribution of energy as the relevant linear combination
\begin{eqnarray}
\nu_{g} S_{g} + \nu_{q} \sum_{f} (S_{q_f} +S_{\bar{q}_f}) =0\;,
\end{eqnarray}
vanishes identically.

Based on the collision integrals for small angle scattering processes in Eq.~(\ref{eq:Fokker}), we then proceed with the linearization around the static homogenous equilibrium background.\footnote{Evidently the collision integral vanishes for the equilibrium background due to detailed balance} When linearizing the momentum current $(\mathcal{J}_a)$ around the equilibrium distribution, one obtains two distinct types of contributions, which can be associated with changes of the phase-space density in Eqns.~(\ref{eq:CurrentJG},\ref{eq:CurrentJQ},\ref{eq:CurrentJB}) or respectively with the changes of the momentum diffusion constant $\hat{\bar q}$ and the drag coefficient $\bar\eta_D$ in Eq.~(\ref{eq:qHatLO},\ref{eq:EtaDLO}). Physically, the first part $\mathcal J_a[\{D_i\}]$ acts primarily on the hard sector, diffusing the particle momentum and dragging it to the infrared. Conversely, the second part $\delta \mathcal J_a[\{D_i\}]$ associated with the changes of $\hat{\bar q}$ and $\bar\eta_D$, corresponds to the recoil response of the medium, and describes how the energy lost from the hard sector is deposited into the softer medium particles. 

Expressing the result in terms of the energy distribution $D_{a}(x,t)$, the hard particles currents $- \nabla_p\mathcal J_a[\{D_i\}]$ are given by
\begin{eqnarray}
\label{eq:JHardG}
- \nabla_p\mathcal J_g[\{D_i\}] &=&\frac{\ca\hat{\bar q}_{\rm eq}}{4 T^2}x\partial_x\Bigg[\frac{T^2}{E^2}x^2\partial_x  + \frac{T}{E}x^2  (1+ 2n_B(xE)) \Bigg]\frac{D_g(x)}{x^3}\;,\\
\label{eq:JHardQ}
- \nabla_p\mathcal J_{q_f}[\{D_i\}] &=&\frac{\cf\hat{\bar q}_{\rm eq}}{4 T^2}x\partial_x\Bigg[\frac{T^2}{E^2}x^2\partial_x  + \frac{T}{E}x^2  (1-2n_F(xE)) \Bigg]\frac{D_{q_f}(x)}{x^3}\;,\\
\label{eq:JHardB}
- \nabla_p\mathcal J_{\bar{q}_f}[\{D_i\}] &=&\frac{\cf\hat{\bar q}_{\rm eq}}{4 T^2}x\partial_x\Bigg[\frac{T^2}{E^2}x^2\partial_x  + \frac{T}{E}x^2  (1-2n_F(xE)) \Bigg]\frac{D_{\bar{q}_f}(x)}{x^3}\;,
\end{eqnarray}
where $\hat{\bar q}_{\rm eq}$ is the equilibrium momentum diffusion constant
\begin{eqnarray}
\hat{\bar q}_{\rm eq}&=&\frac{g^4}{\pi} \int \norm{p}\Bigg[\ca n_B(p)(1+n_B(p)) + N_f n_F(p)(1-n_F(p))\Bigg]
= \frac{g^4}{\pi} \frac{T^3}{2}\left(\frac{N_c}{3}+\frac{N_f}{6}\right)\;, \nonumber \\
\end{eqnarray}
and we have made use of the Einstein relation $\bar{\eta}_D=\hat{\bar q}_{\rm eq}/T$ to eliminate the drag coefficient from Eqns.~(\ref{eq:JHardG},\ref{eq:JHardQ},\ref{eq:JHardB}).
Similarly, the recoil terms $\delta \mathcal J_a[\{D_i\}]$ are written as 
\begin{eqnarray}
-\nabla_p\delta  \mathcal J_g[\{D_i\}]
&=& \frac{\ca \hat{\bar q}_{\rm eq}}{4 T^2}  \frac{ T\delta\bar\eta_D-\delta \hat{\bar q}}{\hat{\bar q}_{\rm eq}} \frac{\nu_g}{2\pi^2}  \frac{T}{E} x\partial_x x^2 n_B(xE)(1+ n_B(xE)),\nonumber\\ \label{eq:recoil1}\\
-\nabla_p\delta \mathcal J_{{q_f}/\bar{q}_f}[\{D_i\}]&=&  \frac{C_{F}\hat{\bar q}_{\rm eq}}{4 T^2}\frac{ T\delta\bar\eta_D-\delta \hat{\bar q}}{\hat{\bar q}_{\rm eq}}  \frac{\nu_q}{2\pi^2}  \frac{T}{E} x\partial_x x^2 n_F(xE)(1- n_F(xE))\;,\nonumber\\ \label{eq:recoil2}
\end{eqnarray}
where the recoil coefficients are given by

\begin{eqnarray}
&\delta \hat{\bar q}= \frac{g^4}{\pi} E^3 \int  dx~ \frac{1}{x}\Bigg[ \ca  \nu_{g}^{-1} D_g(x)(1+2n_B(xE)) + \frac{1}{2} \sum_{f} \nu_{q}^{-1} (D_{q_f}(x)+D_{\bar{q}_f}(x))(1-2n_F(xE)) \Bigg]\;,&\nonumber\\
\end{eqnarray}
\begin{eqnarray}
\delta\bar\eta_D=  \frac{g^4}{\pi} E^2 \int dx~  \frac{2}{x^2}\Big[ \ca  \nu_{g}^{-1}  D_g(x) +\frac{1}{2} \sum_{f} \nu_{q}^{-1} (D_{q_f}(x)+D_{\bar{q}_f}(x)) \Big]\;.\label{eq:DeltaEta}
\end{eqnarray}
Since $\delta \hat{\bar q}$ and $\delta\bar\eta_D$ are determined by the non-equilibrium contributions from the energy distribution, they do not satisfy an Einstein relation, i.e. $\delta\bar\eta_D \neq \delta \hat{\bar q}/T$, giving rise to a finite recoil contribution in Eqns.~(\ref{eq:recoil1},\ref{eq:recoil2}).

Similarly, when linearizing the conversion terms around equilibrium one finds that the contributions can be separated into conversions of hard particles $S_a[\{D_i\}]$ and (recoil) conversions of thermal constituents $\delta S_{a} [\{D_i\}]$ in an analogous fashion. Evaluating the action of conversions on the hard particles, one finds
\begin{eqnarray}
    S_g[\{D_i\}] = 
    \qquad\qquad\qquad\qquad\qquad\qquad\qquad\qquad\qquad\qquad\qquad\qquad\qquad\qquad\qquad\qquad
    \nonumber\\
    \nu_{g} \frac{\mathcal{I}_{q_f}^{\rm eq}}{8T^2} \frac{T}{xE}  \sum_{f} \left\{  \nu_{q}^{-1}\Big[D_{q_f}(x)+D_{\bar{q}_f}(x)\Big](1+2n_B(xE)) -  2\nu_{g}^{-1}D_g(x) (1-2n_F(xE))\right\}\;, \nonumber\\ \label{eq:ConversionGluon}
\end{eqnarray}
\begin{eqnarray}
    &S_{q_f,\bar{q}_f} [\{D_i\}] =\nu_{g} \frac{\mathcal{I}_{q_f}^{\rm eq}}{8T^2} \frac{T}{xE} \left\{  \nu_{g}^{-1}D_g(x) (1-2n_F(xE))-\nu_{q}^{-1} D_{q_f,\bar{q}_f}(x)(1+2n_B(xE)) \right\}\;,&\nonumber\\
    \label{eq:ConversionQuark}
\end{eqnarray}
where in accordance with Eqns.~(\ref{eq:IqDef},\ref{eq:IqBDef}), we denote 
\begin{eqnarray}
\mathcal{I}_{q_f}^{\rm eq}=\mathcal{I}_{\bar{q}_f}^{\rm eq}=\frac{g^4 C_F \mathcal{L} T^2}{8\pi}\;,
\end{eqnarray}
for a charge neutral plasma. Due to the identity $n_F(\p)(1+n_B(\p)) = n_B(\p) (1+n_F(\p))$ one finds that the recoil contribution to the source term $S_{g}$ in Eq. \eqref{eq:SourceGluon} vanishes identically,
\begin{eqnarray}
\delta S_{g} [\{D_i\}] &=&0\;,
\end{eqnarray}
and only the quark and antiquark channels acquire a recoil contribution given by
\begin{eqnarray}
    \delta S_{q_f} [\{D_i\}] &=& \frac{\nu_{g}}{2\pi^2} \frac{x^2}{8E}\left(\delta\mathcal{I}_{q_f}-\delta\mathcal{I}_{\bar{q}_f}\right) ~n_B(xE)(1-n_F(xE)) \;,\\
    \delta S_{\bar{q}_f} [\{D_i\}] &=&\frac{\nu_{g}}{2\pi^2} \frac{x^2}{8E} \left(\delta\mathcal{I}_{\bar{q}_f}-\delta\mathcal{I}_{q_f}\right)~n_B(xE)(1-n_F(xE)) \;,
\end{eqnarray}
where $\delta\mathcal{I}_{q_f}$ and $\delta\mathcal{I}_{\bar{q}_f}$ are the linearization of the integrals in Eqns.~(\ref{eq:IqDef},\ref{eq:IqBDef}), whose difference is given by
\begin{eqnarray}
    \left(\delta\mathcal{I}_{\bar{q}_f}-\delta\mathcal{I}_{q_f}\right) = \frac{g^4 C_{F} \mathcal{L}}{\pi} E^2\int~dx~\frac{1}{ x^2}  \left(1 + 2n_B(xE)\right) \nu_q^{-1}\left(D_{q_f}(x) -D_{\bar{q}_f}(x) \right) \;.
\end{eqnarray}
Since  $\sum_{a} \delta S_{a} [\{D_i\}]=0$, the conversions of thermal constituents do not affect the energy distribution of parton fragments $\sum_{a} D_{a}(x,t)$. However, for charged partons, they do affect the distribution of valence charge, as described by the charge distributions $\frac{1}{x}(D_{q_f}(x,t)-D_{\bar{q}_f}(x,t))$.
\subsection{Collinear radiation}\label{sec:CollinearRadiation}

Elastic interactions of jet particles with the constituents of the thermal QGP, give rise to medium induced radiation which provides an important contribution to their energy loss \cite{Jeon:2003gi,Baier:1996kr,Gyulassy:2000fs}. Generally, the interplay between vacuum like emissions which are tied to the production vertex and medium induced emissions which can occur anywhere inside the medium can be rather complicated \cite{Arnold:2008iy}, and results in an explicit path length $L$ or time $t$ dependence of the medium induced radiation rates \cite{Arnold:2008iy,Andres:2020vxs,CaronHuot:2010bp,Mehtar-Tani:2017web}. Since we are particularly interested in hard partons which lose a large fraction of their energy to the medium, there is however a clear separation of time scales between the initial vacuum like shower and the subsequent energy loss of the partons inside the medium. We will therefore not include the effects of vacuum like emissions, anticipating that they can be absorbed into initial conditions or source terms for the in-medium evolution. Since we are particularly interested in the evolution on large time scales, we will also not consider the explicit path length $L$ dependence of the medium induced radiation rates, and instead employ the large $L$ limit of the medium induced radiation rates, following the approach of Arnold, Moore and Yaffe (AMY) \cite{Arnold:2002zm}, where medium induced radiation is described by collinear $1\leftrightarrow 2$ splittings/mergings with an effective rate $ \frac{d\Gamma^a_{bc}(p,z)}{dz} = \int d^2\ke \frac{d\Gamma^a_{bc}(p,z)}{d^2\ke dz}$ obtained from integrating the fully differential rate over the transverse momentum $\ke$ of the splitting. Based on the approach of \cite{Arnold:2002zm,Kurkela:2011ti,Schlichting:2019abc}, the collinear in-medium radiation rates $ \frac{d\Gamma^a_{bc}(p,z)}{dz}$\footnote{We follow the notation of P. Arnold~\cite{Arnold:2008iy}, and refer to~\cite{Arnold:2008iy} for comparison to other notations.} are then obtained from 
\begin{eqnarray}
\label{eq:EffectiveRate}
\frac{d\Gamma^a_{bc}(p,z)}{dz} =  \frac{\alpha_s P_{bc}(z)}{[2pz(1-z)]^2} \int \frac{d^2 \p_b}{(2\pi)^2} ~\text{Re}\left[ 2 \mathbf{p}_b \cdot \mathbf{g}_{(z,P)}( \mathbf{p}_b) \right]\;,
\end{eqnarray}
where $\mathbf{g}_{(z,P)}$ is a solution to the integral equation
\begin{eqnarray}
    2\mathbf{p}_b &=& i \delta E(z,P,\mathbf{p}_b) \mathbf{g}_{(z,P)}(\mathbf{p}_b) + \int \frac{d^2q}{(2\pi)^2}~\frac{d\bar{\Gamma}^{\rm el}}{d^2q}~\left\{ C_{1} \left[ \mathbf{g}_{(z,P)}(\mathbf{p}_b) - \mathbf{g}_{(z,P)}(\mathbf{p}_b -\mathbf{q}) \right] +  \right. \nonumber \\
    && \left. C_{z} \left[ \mathbf{g}_{(z,P)}(\mathbf{p}_b) - \mathbf{g}_{(z,P)}(\mathbf{p}_b -z\mathbf{q}) \right]  + C_{1-z} \left[ \mathbf{g}_{(z,P)}(\mathbf{p}_b) - \mathbf{g}_{(z,P)}(\mathbf{p}_b -(1-z)\mathbf{q}) \right] \right\}\;. \nonumber\\
\end{eqnarray}
The energy $\delta E(z,P,\mathbf{p}_b)$ is defined by
\begin{eqnarray}
    \delta E(z,P,\mathbf{p}_b) &=& \frac{\mathbf{p}_b^2}{2Pz(1-z)}+\frac{m^2_{\infty.(z)}}{2zP}+\frac{m^2_{\infty.(1-z)}}{2(1-z)P}-\frac{m^2_{\infty.(1)}}{2P}\;.
\end{eqnarray}
We define the factors $C_1 = \tfrac{1}{2} \left( C_{z}^{R} + C_{1-z}^{R} - C_{1}^{R} \right)$,
$C_z  =  \tfrac{1}{2} \left( C_{1-z}^{R} + C_{1}^{R} - C_{z}^{R} \right)$ and 
$C_{1-z} = \tfrac{1}{2} \left( C_{1}^{R} + C_{z}^{R} - C_{1-z}^{R} \right) $, using color factor of the species with momentum fraction $1$,$z$ and $1-z$ respectively.
For the elastic broadening kernel $\frac{d\bar{\Gamma}^{\rm el}}{d^2q}$, we use the leading order expression 
\begin{eqnarray}
    \frac{d\bar{\Gamma}^{\rm el}}{d^2q} = \frac{m_D^2}{q^2(q^2+m_D^2)}\;.
\end{eqnarray}
We solve Eq.~(\ref{eq:EffectiveRate}) self-consistently, obtaining a resumation of multiple scatterings to all orders, encompassing the Bethe-Heitler (BH) regime at low energy $z(1-z)p\lesssim\omega_{BH}\sim T$ as well as the Landau-Pomeranchuk-Migdal (LPM) regime at high energy $z(1-z)p\gg\omega_{BH}\sim T$. Based on this formalism, the effect of medium induced radiation is then described by the $1\leftrightarrow2$ collision integral
\begin{eqnarray}
&C_a^{1\leftrightarrow 2}[\{f_i\}]=\nonumber
\qquad \qquad \qquad \qquad \qquad \qquad \qquad \qquad \qquad \qquad \qquad \qquad \qquad \qquad \qquad 
&\\
&\sum_{bc}\Bigg\{-\frac{1}{2}\int_0^1 dz \frac{d\Gamma^a_{bc}(\p,z)}{dz}\Big[ f_a(\p)(1\pm f_b(z\p))(1\pm f_c(\bar z \p)) - f_b(z\p)f_c(\bar z p)(1\pm f_a(\p)) \Big] \nonumber &\\
&+ \frac{\nu_b}{\nu_a}\int_0^1 \frac{dz}{z^3}\frac{d\Gamma^b_{ac}(\frac{\p}{z}\;,z)}{dz}
\Big[ f_b\left(\frac{\p}{z}\right)(1\pm f_a(\p))\left(1\pm f_c\left(\frac{\bar z}{z}\p\right)\right) -f_a(\p)f_c\left(\frac{\bar z}{z}\p\right)\left(1\pm f_b\left(\frac{\p}{z}\right)\right) \Big] \Bigg\}\;, \nonumber &\\
\end{eqnarray}
where $ \frac{d\Gamma^a_{bc}(p,z)}{dz} $ is the effective rate for particle $a$ to split into $ b $ and $ c $ with energy $ zp $ and $ \bar z p$ respectively, and we will denote $ \bar z=1-z $ in the following. Details of the re-construction of the in-medium rates $ \frac{d\Gamma^a_{bc}(p,z)}{dz} $ are provided in Appendix \ref{ap:NumericalImplementation}. By linearizing the collision integrals, and expressing the contributions to the evolution equation for the energy distribution, the contributions to the evolution of $D_{g}(x,t)$ are given by the sum of $g \to gg$, $q \to qg$, $\bar{q} \to \bar{q}g$ and $g \to q \bar{q}$ processes
\begin{eqnarray}
C_g^{g\leftrightarrow gg}[\{D_i\}]&= & \int_0^1 dz  \frac{d\Gamma^g_{gg}(\left(\frac{xE}{z}\right),z)}{dz}\Bigg[ D_g\left(\frac{x}{z}\right) \left(1+ n_B(xE)+ n_B\left(\frac{\bar zxE}{z}\right)\right)
\nonumber\\
&&+\frac{D_g(x)}{ z^3}\left(n_B\left(\frac{xE}{z}\right) - n_B\left(\frac{\bar zxE}{z}\right)\right) + \frac{D_g\left(\frac{\bar zxE}{z}\right)}{ \bar z^3}\left(n_B\left(\frac{xE}{z}\right) -  n_B(xE)\right) \Bigg]\nonumber\\ 
&&-\frac{1}{2} \int_0^1 dz  \frac{d\Gamma^g_{gg}(xE,z)}{dz}\Bigg[ D_g(x)(1+ n_B(zxE)+n_B(\bar z xE))
\nonumber\\
&&+\frac{D_g(zx)}{ z^3}(n_B(xE)-n_B(\bar zxE))+\frac{D_g(\bar zx)}{\bar z^3}(n_B(xE)-n_B(zxE)) \Bigg]\;, \nonumber\\  \label{eq:Inelastic1st}
\end{eqnarray}
\begin{eqnarray}
    C_g^{q\leftrightarrow qg}[\{D_i\}]= \sum_{f} \int_0^1 dz & \frac{d\Gamma^{q}_{gq}\left(\frac{xE}{z}\;,z\right)}{dz}\Bigg[ D_{q_f}\left(\frac{x}{z}\right) \left(1+ n_B(xE)- n_F\left(\frac{\bar zxE}{z}\right)\right) \qquad \qquad \qquad \qquad&
    \nonumber\\
    + \frac{\nu_q }{\nu_g }\frac{D_g(x)}{z^3}& \left( n_F\left(\frac{xE}{z}\right) - n_F\left(\frac{\bar zxE}{z}\right)\right)-\frac{D_{q_f}\left(\frac{\bar zx}{z}\right)}{ \bar z^3}\left( n_F\left(\frac{xE}{z}\right) + n_B(xE)\right) \Bigg]\;, \qquad & \nonumber\\
\end{eqnarray}
\begin{eqnarray}
    C_g^{\bar{q}\leftrightarrow \bar qg}[\{D_i\}] =  \sum_{f} \int_0^1 dz & \frac{d\Gamma^{q}_{gq}\left(\frac{xE}{z}\;,z\right)}{dz}\Bigg[ D_{\bar {q}_f}\left(\frac{x}{z}\right) \left(1+ n_B(xE)- n_F\left(\frac{\bar zxE}{z}\right)\right) \qquad \qquad \qquad \qquad&
    \nonumber\\
    + \frac{\nu_q }{\nu_g }\frac{D_g(x)}{z^3}& \left( n_F\left(\frac{xE}{z}\right) - n_F\left(\frac{\bar zxE}{z}\right)\right)-\frac{D_{\bar {q}_f}\left(\frac{\bar zx}{z}\right)}{ \bar z^3}\left( n_F\left(\frac{xE}{z}\right) + n_B(xE)\right) \Bigg]\;, \qquad & \nonumber\\
\end{eqnarray}
\begin{eqnarray}
    C_g^{g\leftrightarrow q\bar q}[\{D_i\}] = - \sum_{f}\int_0^1 dz &  \frac{d\Gamma^g_{q\bar q}(xE,z)}{dz}\Bigg[  D_g(x)(1- n_F(zxE)- n_F(\bar z xE)) \qquad \qquad \qquad \qquad&
    \nonumber\\
    - \frac{\nu_g}{\nu_q}\frac{D_{q_f}(zx)}{ z^3} &( n_B(xE) + n_F(\bar z xE))-\frac{\nu_g}{\nu_q}\frac{D_{\bar{q}_f}(\bar z x)}{ \bar z^3}( n_B(xE) + n_F(zxE)) \Bigg]\;. \qquad & \nonumber\\
\end{eqnarray}
where both $1 \to 2$ and inverse $2 \to 1$ processes are included along with the appropriate final state Bose enhancement and Fermi suppression, such that the above also include the (linearized) back-reaction of the high energetic particles onto the medium and automatically satisfy energy-momentum conservation. Similarly, the contributions to the evolution of the energy distribution of quarks and anti-quarks $D_{q}(x,t)$ and $D_{\bar q}(x,t)$ are given by the sum of $q \to qg$ or respectively $\bar{q}\to \bar{q}g$, and $g \to q \bar{q}$ processes, which take the form
\begin{eqnarray}
C_{q_f}^{q\leftrightarrow qg}[\{D_i\}]&=&
- \int_0^1 dz  \frac{d\Gamma^q_{gq}(xE, z)}{dz}\Bigg[ D_{q_f}(x)(1+ n_B(zxE)- n_F(\bar z xE))
\nonumber\\
&&+ \frac{\nu_q}{\nu_g }\frac{D_g(zx)}{z^3} ( n_F(xE) - n_F(\bar zxE))-\frac{D_{q_f}(\bar zx)}{ \bar z^3}( n_F(xE) + n_B(xE)) \Bigg]\nonumber\\
&&+ \int_0^1 dz  \frac{d\Gamma^q_{gq}\left(\frac{xE}{z}\;,\bar z\right)}{dz}\Bigg[ D_{q_f}\left(\frac{x}{z}\right) \left(1+ n_B\left(\frac{\bar zxE}{z}\right) -n_F(xE)\right)
\nonumber\\
&&+ \frac{\nu_q}{\nu_g }\frac{D_g\left(\frac{\bar zx}{z}\right)}{z^3} \left( n_F\left(\frac{x}{z}\right) - n_F\left(\frac{xE}{z}\right)\right)-\frac{D_{q_f}(x)}{ \bar z^3}\left( n_F\left(\frac{xE}{z}\right) + n_B\left(\frac{\bar zxE}{z}\right)\right) \Bigg]\;,\nonumber\\
\end{eqnarray}

\begin{eqnarray}
C_{q_f}^{g\leftrightarrow q\bar q}[\{D_i\}]=\int_0^1 dz & \frac{d\Gamma^g_{q\bar q}\left(\frac{xE}{z}\;,z\right)}{dz}\Bigg[ D_g\left(\frac{x}{z}\right)\left(1- n_F(xE)- n_F\left(\frac{\bar z xE}{z}\right)\right) \qquad \qquad \qquad \qquad & 
\nonumber\\
- \frac{\nu_g}{\nu_q}\frac{D_{q_f}(x)}{ z^3}&\left( n_B\left(\frac{xE}{z}\right) + n_F\left(\frac{\bar z xE}{z}\right)\right)-\frac{\nu_g}{\nu_q}\frac{D_{\bar{q}_f}\left(\frac{\bar z x}{z}\right)}{ \bar z^3}\left( n_B\left(\frac{xE}{z}\right) + n_F(xE)\right) \Bigg]\;. & \nonumber\\ \label{eq:Inelasticlast}
\end{eqnarray}
and similarly for anti-quarks, with $q_f$ replaced by $\bar{q}_f$ in the above expressions.
\subsection{Conservation laws and scaling}
Before we proceed to our analysis of the evolution for the energy distribution, we briefly note that by explicitly taking into account the backreaction of the high energetic particles on the thermal QGP constituents, the above evolution equations satisfy the following sum rules
\begin{eqnarray}
\partial_{t} \sum_{a} \int_{0}^{\infty} dx D_a(x,t)=0\;, \qquad \partial_{t} \int_{0}^{\infty} dx \frac{1}{x} \left(D_{q_{f}}(x,t)-D_{\bar{q}_f}(x,t) \right)=0\;,
\end{eqnarray}
associated with the energy ($E$) and net charge ($Q_f$) conservation. While for typical excitations with energies $\omega=xE \sim T$ of the order of the temperature of the QGP, all contributions of elastic and inelastic processes to the collision integrals are parametrically of the same order $ \sim g^4 T$ \cite{Arnold:2002zm,Kurkela:2011ti}, the situation is markedly different for high-momentum particles with $\omega=xE \gg T$, where the various contributions to the collision integrals behave parametrically as
\begin{eqnarray}
C_{a}^{\rm inelastic} &\sim& g^4T \sqrt{\frac{T}{xE}} D_{a}(x,t)\;,\label{eq:ParametricEstimateFst} \\
C^{\rm Drag}_{a} &\sim& g^4 T \left(\frac{T}{xE}\right) x\partial_{x} D_{a}(x,t)\;, \\
 C_{a}^{\rm Conversion} &\sim& g^4 T \frac{T}{xE} D_{a}(x,t)\;, \\
C^{\rm Diffusion}_{a} &\sim& g^4 T \left(\frac{T}{xE}\right)^2 (x\partial_{x})^2 D_{a}(x,t)\;, \label{eq:ParametricEstimateLst}
\end{eqnarray}
indicating that the evolution of high-momentum particles $x \gg T/E$ will be dominated by inelastic processes, with power suppressed contributions from elastic processes. Nevertheless, including the effects of all leading order processes is important to study the evolution of softer fragments of the energy distribution all the way to the temperature scale $x \sim T/E$, where elastic and inelastic contributions eventually become comparable in magnitude~\cite{Arnold:2002zm}.

\section{Energy loss and equilibration}\label{sec:Equilibration}

We now proceed to the study of energy loss and equilibrium of hard partons inside a thermal QGP, starting from an initial condition, where the initial energy distribution of partons $D_{a}(x,t)$ is given by a narrow Gaussian of width $\sigma/E=10^{-3}/\sqrt{2}$ centered around the energy $E$, which is normalized to $ \int dx \sum_{a} D_{a}(x,0)=1$. Since the evolution equations are linear, the evolution of a general solution can be decomposed into a basis set of excitations. We choose the initial condition as a Gaussian profile which is transparent to the physics. With regard to the chemical composition, we will consider two types of initial conditions, corresponding to high energetic gluon and quark where initially all the energy is stored either in the gluon or quark distribution respectively, Although we will refer to these two initial condition as gluon/quark jet, we would like to emphasize that they do not correspond to the usual jet definitions as this requires definitions of cones sizes which are aspects beyond the scope of this study. For a gluon jet
\begin{align}
\label{eq:gluon-jet}
D_{g}^{g-jet}(x,0)=\frac{2e^{-\frac{(xE-E)^2}{2\sigma^2 }}}{\sqrt{ 2\pi}  \sigma/E \left(\text{erf}\left(\frac{E}{ \sqrt{2}\sigma}\right)+1\right)}\;, \qquad D_{q}^{g-jet}(x,0)=0\;, \qquad D_{\bar{q}}^{g-jet}(x,0)=0\;,
\end{align}
whereas for a quark jet
\begin{align}
\label{eq:quark-jet}
D_{g}^{q-jet}(x,0)=0\;, \qquad D_{q}^{q-jet}(x,0)=\frac{2e^{-\frac{(xE-E)^2}{2\sigma^2 }}}{\sqrt{ 2\pi}  \sigma/E \left(\text{erf}\left(\frac{E}{ \sqrt{2}\sigma}\right)+1\right)}\;, \qquad D_{\bar{q}}^{q-jet}(x,0)=0\;,
\end{align}
where the error function is given by $\text{erf}(x) = \frac{2}{\sqrt{\pi}} \int_0^x dt~ e^{-t^2}$.

If not stated otherwise, we will present results for the evolution of jets with energy $E=1000T$, and express all time scales in terms of the dimensionless time variable
\begin{eqnarray}
    \tau = \frac{t}{t_{\rm split}(E)}= g^4 T \sqrt{\frac{T}{E}}t\;.
\end{eqnarray}
Since $t_{\rm split}(E)=\frac{1}{g^4T} \sqrt{\frac{E}{T}}$ corresponds to the typical timescale for an initial hard parton to undergo a quasi-democratic ($z\sim1/2$) splitting,\footnote{Note that due LPM suppression, the timescale $t_{\rm split}(E)$ is enhanced by a factor $\sqrt{\frac{E}{T}}$ relative to the typical mean free path or relaxation time $\sim \frac{1}{g^4T}$ of near-thermal excitations.} which will ultimately dictate the energy loss \cite{Baier:2000sb,Kurkela:2011ti,Schlichting:2019abc,Blaizot:2013hx}, we can expect that this normalization takes into account the leading dependence on the jet energy. We will further address the jet energy dependence in Sec.~\ref{sec:Turbulence}, where we compare results for different values of $E/T=10,30,100,1000$.

With regard to the quark distributions, it is convenient to decompose the energy distributions $D_{q_{f}}(x)$ and $D_{\bar q_{f}}(x)$ into flavor singlet (S) and valence (V) distributions, which are obtained by the following linear combinations
\begin{eqnarray}
    D_S(x)= \sum_f D_{q_{f}}(x)+D_{\bar q_{f}}(x), \qquad
    D_{V_{f}}(x)= D_{q_{f}}(x)-D_{\bar q_{f}}(x),
\end{eqnarray}
such that the singlet distribution $D_{S}(x)$ characterizes the energy distribution of quarks inside the jet, whereas the valence distribution describes the distribution of valence charge inside the jet. By careful inspection of the evolution equations, one finds that at the linearized level, the evolution of $D_{V_{f}}(x)$ decouples from the evolution of $D_{S}(x)$ and $D_{g}(x)$, indicating that different mechanisms will ultimately be responsible for the equilibration of energy and valence charge of the jet.

\begin{figure}[ht]
    \centering
    \includegraphics[width=0.9\textwidth]{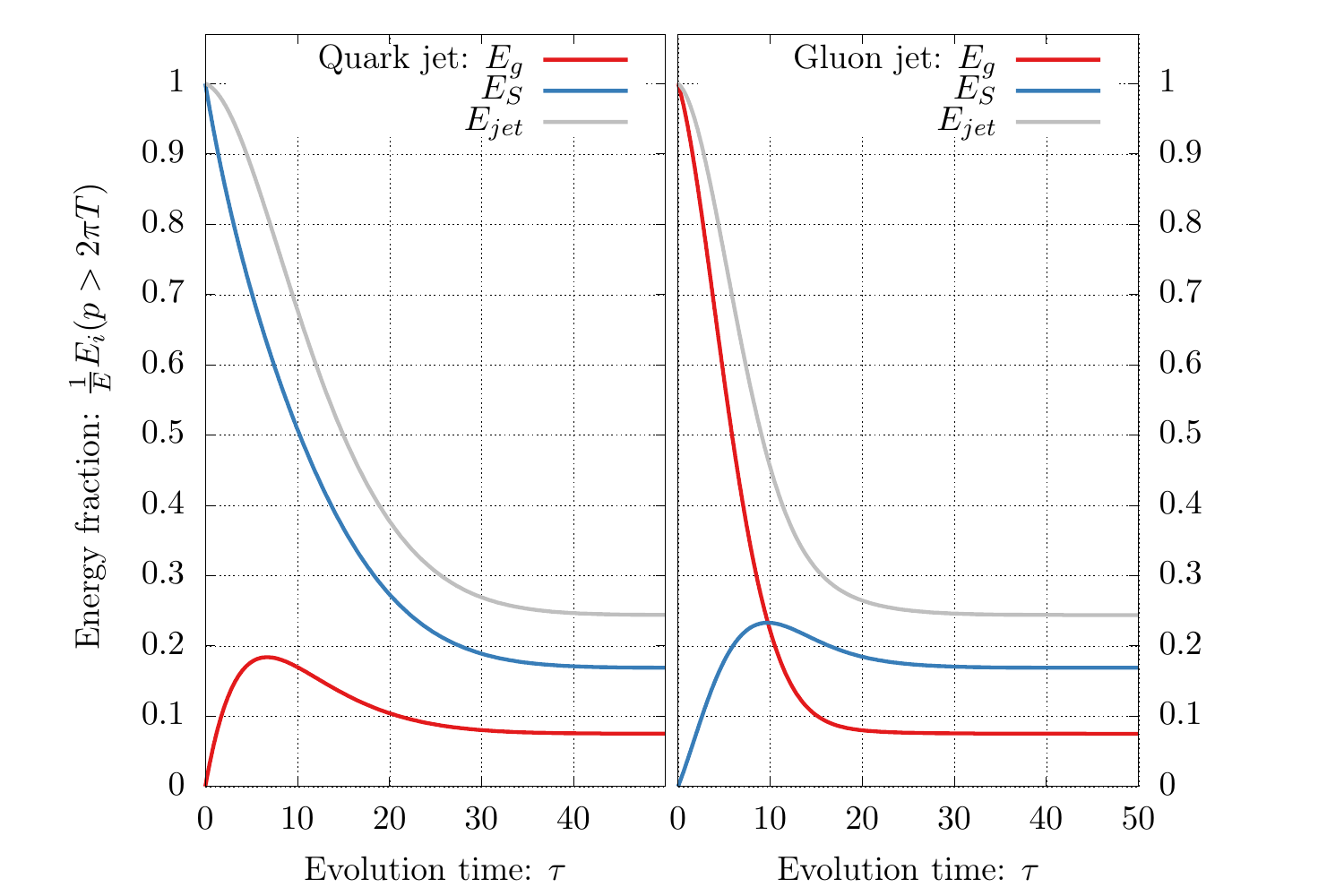}
    \includegraphics[width=0.9\textwidth]{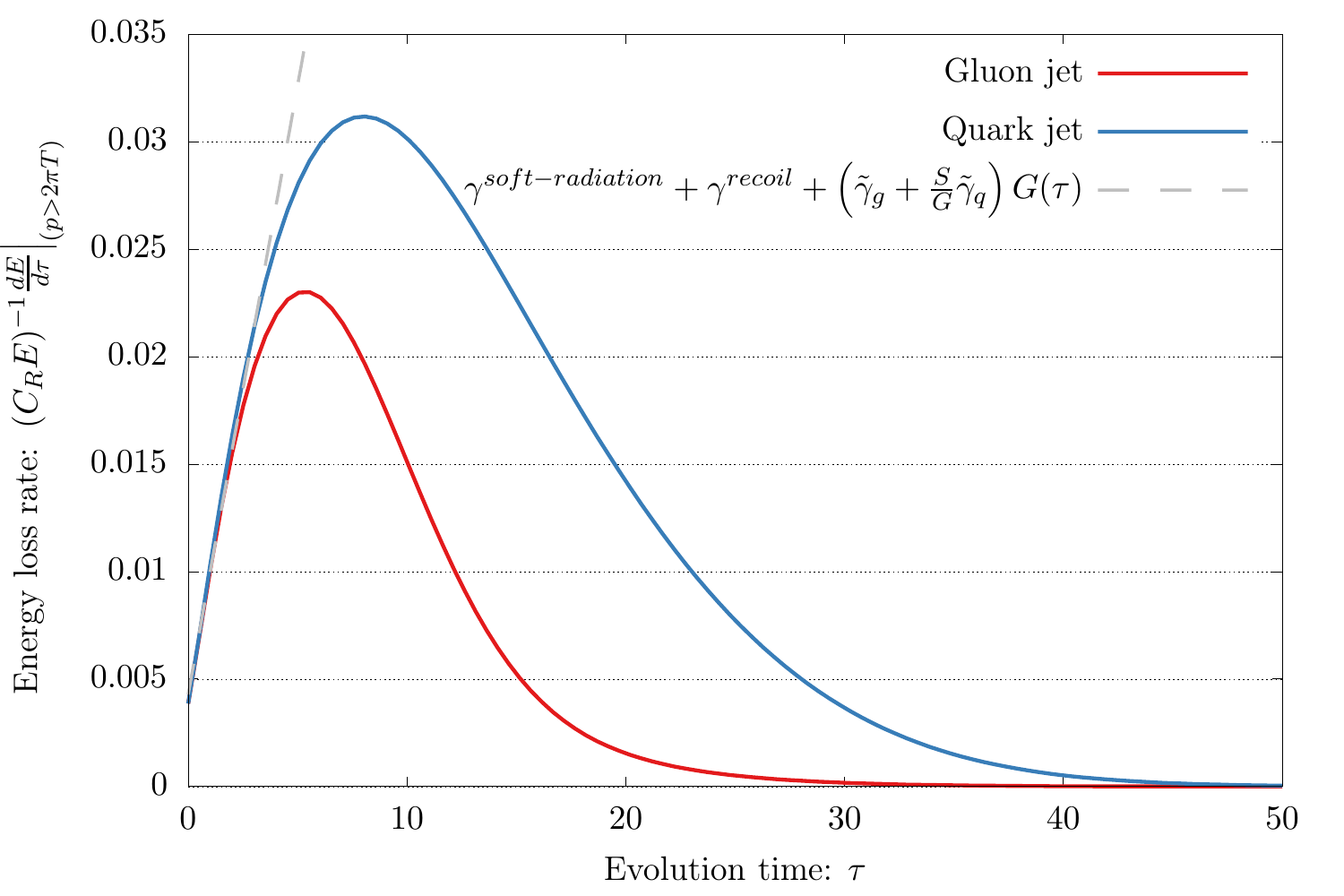}
    \caption{(top) Evolution of the energy carried by particles with momentum $p>2\pi T$ for quark (left) and gluon (right) jets with $E=1000T$.  Different curves labeled $E_{g,S,jet}$ represent the energy fraction of gluons (g), quarks plus anti-quarks (S)  and the sum of all species (jet). (bottom) Differential energy loss rate $dE_{\rm jet}/d\tau$ divided by the corresponding Casimir factor ($C_{R}=C_{A}=N_c$ for gluons jets and $C_{R}=C_{F}=\frac{N_c^2-1}{2N_c}$ for quark jets). }
    \label{fig:JetEnergyLoss}
\end{figure}

Since our effective kinetic description explicitly takes into account the medium response, the total energy $E$ as well as the set of all valence charges $Q_{f}$ are explicitly conserved.
Nevertheless, over the course of the evolution jet energy and valence charge are re-distributed from high-energy $(\omega \sim E)$ to low energy $(\omega \sim T)$, where the soft constituents of the jet will eventually thermalize with the surrounding medium. Hence, in order to analyze jet energy loss, we define a cut-off scale $\mu=2\pi T$, such that the hard constituents with $\omega > \mu$ are to be considered as part of the jet, whereas the soft constituents with $\omega<\mu$ are considered as part of the equilibrated medium.\footnote{ We note that in thermal equilibrium, around $\sim 75 \%$ of the total energy are contained in the energy range $[0,2\pi T]$.} Based on this procedure, the individual contributions of each species to the jet energy and valence charge is then evaluated as
\begin{eqnarray}
    E_i = \int_{\mu/E}^\infty dx~  D_i(x), 
    \qquad
    Q_{f} = \int_{\mu/E}^\infty \frac{dx}{x}~ D_{V_{f}}(x).
\end{eqnarray}

We present our results for jet energy loss in Fig. \ref{fig:JetEnergyLoss}, where the two upper panels show the evolution of the different contributions to the energy for quark and gluon jets.  Different curves $E_{S}$, $E_{g}$ in each panel show the individual contributions of hard quarks and gluons, as well as the total energy of hard constituents $E_{\rm tot}$. While initially quarks(gluons) dominate the energy budget of quark (gluon) jets, strong changes in the chemical composition of the jet take place over the course of the evolution. Eventually, by the time $\tau \gtrsim 20$ the chemical composition of quark and gluon jets becomes nearly identical; however, at this point the jet has already lost a significant fraction of its energy to the thermal medium.

By taking the time derivatives of the total energy of hard constituents $E_{\rm tot}$, we can further compare the differential energy loss rate $dE/d\tau$ for quarks and gluon jets, which are presented in the lower panel of Fig. \ref{fig:JetEnergyLoss}. Starting from a small but non-zero energy loss rate at very early times $\tau\simeq 0$, the energy loss rate $dE/d\tau$ exhibits an approximately linear increase with evolution time $\tau$, which follows the expected Casimir scaling such that $\frac{1}{C_F}dE/d\tau|_{q-jet} \approx \frac{1}{C_A}dE/d\tau|_{g-jet}$ at early times $\tau \lesssim 3$. Subsequently, as the hard constituents of the jet start to be significantly affected by the presence of the medium, the energy loss rate experiences a broad maximum and the Casimir scaling of the energy loss breaks down. Eventually, the energy loss rate $dE/d\tau$ decays exponentially at very late times, as the few remaining constituents equilibrate with the thermal medium.

Based on the behavior observed in Fig.~\ref{fig:JetEnergyLoss}, we find that the in-medium evolution of the jet can be roughly divided into three distinct stages, characterized by direct energy loss, inverse turbulent cascade, and the eventual approach to equilibrium, which we will now discuss in more detail.

\begin{figure}[t!]
    \includegraphics[width=0.5\textwidth]{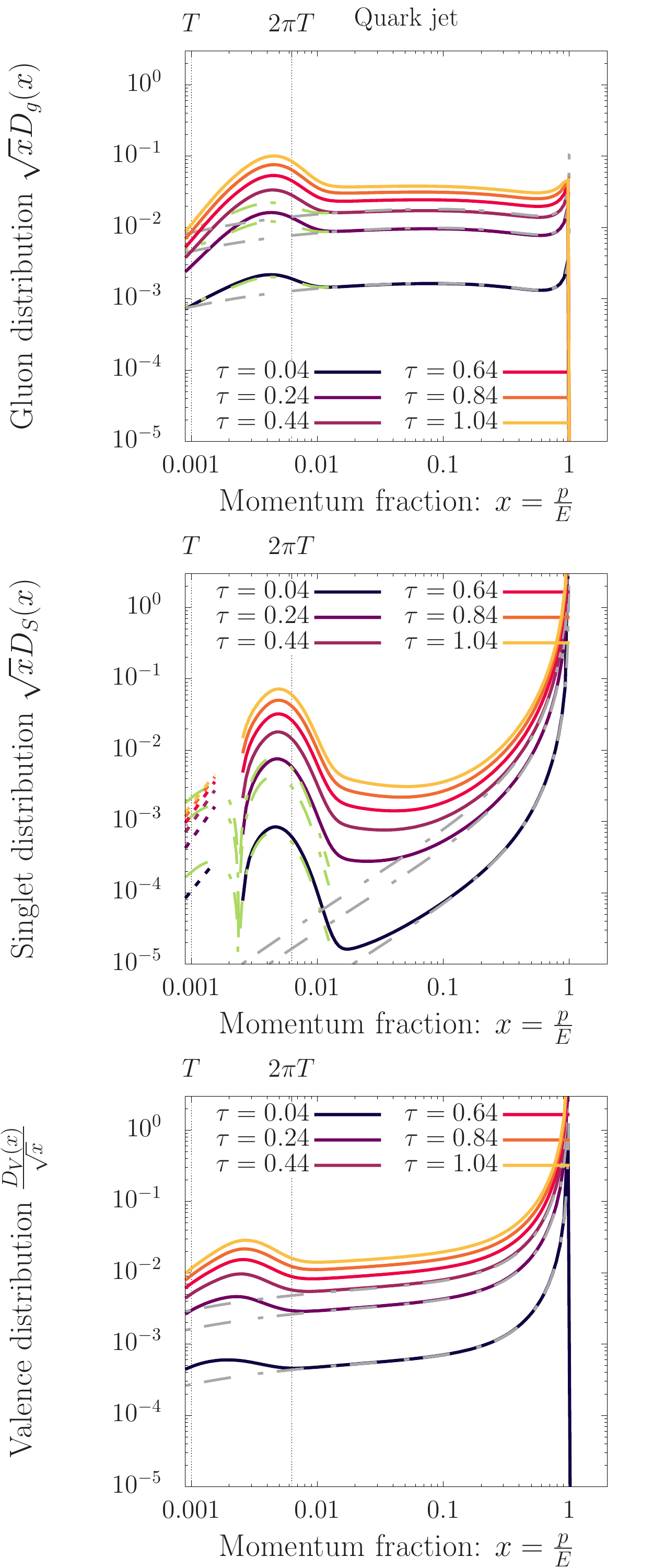}   \includegraphics[width=0.5\textwidth]{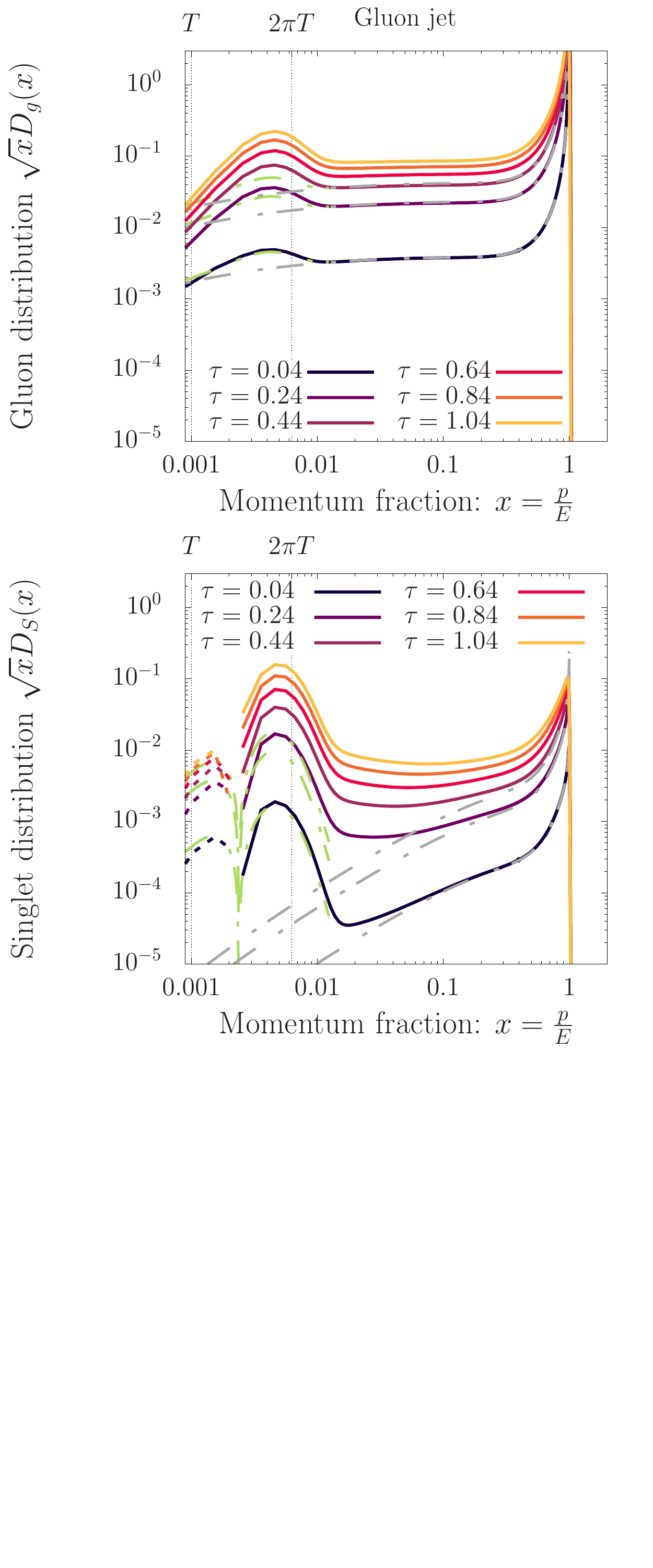}    
    \caption{Early time behavior of the energy distribution for a quark jet (left) and a gluon jet (right). Gray dashed lines represent single splitting as written in Eqns. (\ref{eq:SingleSplittingGl1}-\ref{eq:SingleSplittingGl2}) and (\ref{eq:SingleSplittingQu1}-\ref{eq:SingleSplittingQu2}), while the green dashed lines represent the same splitting plus the elastic recoil terms from Eqns. (\ref{eq:RecoilEarlyTime1}-\ref{eq:RecoilEarlyTime2}). }
    \label{fig:EarlyTime}
\end{figure}

\subsection{Early stages of the evolution}
During the early stages of in-medium jet evolution, elastic and inelastic processes give rise to (longitudinal) momentum broadening of the hard components of the jet, as can be seen from the widening of the distribution peak around $x\sim1$ in Fig. \ref{fig:EarlyTime}, where we present the evolution of the energy distributions $D_{g}(x),D_{S}(x)$ and $D_{V}(x)$ at early times. Even though these processes initially have a small effect on the hard ($x\sim 1$) components of the jet, they can still lead to a sizeable deposition of energy into soft ($x\sim T/E$) modes due to the emission of soft radiation and recoil of the thermal medium. In order to further quantify the energy loss at early times, we can compute the energy deposition below the scale $\mu$ perturbatively, i.e. assuming that at early times $\tau \ll 1$ the distributions $D_{i}(x)$ of hard fragments are unmodified. By inserting the initial conditions in Eqns.~(\ref{eq:gluon-jet}, \ref{eq:quark-jet}) into the evolution equations (\ref{eq:recoil1}-\ref{eq:DeltaEta}, \ref{eq:Inelastic1st}-\ref{eq:Inelasticlast}) for the energy distribution and integrating over momentum fractions $x$ up to the cut-off scale $\mu/E \ll 1$, one finds an approximately constant energy loss at early times
\begin{eqnarray}
\left.\frac{dE}{d\tau}\right|_{\tau \ll 1}=\gamma^{\rm soft-radiation} + \gamma^{\rm recoil}\;,\label{eq:ConstantEstimate}
\end{eqnarray}
where $\gamma^{\rm soft-radiation}$ is the contribution from the emission of soft radiation and $\gamma^{\rm recoil}$ describes the contribution from elastic recoils. 

Evaluating the inelastic contributions in the limit $x \ll 1$, one finds that Bose-enhancement and Fermi-suppresion factors cancel between gain and loss terms, such that for $x \ll 1$ the radiative contributions to the energy distributions are approximately given by
\begin{eqnarray}
  \text{gluon jet:}  \qquad D_g^{\rm soft-radiation}(x,t) &=&xt \left.\frac{d\Gamma^g_{gg}(E,z)}{dz}\right|_{z=x}\;,\label{eq:SingleSplittingGl1}\\
    D_S^{\rm soft-radiation}(x,t) &=&xt N_f \left.\frac{d\Gamma^g_{q\bar q}(E,z)}{dz}\right|_{z=x}\label{eq:SingleSplittingGl2} \;, \\
    \text{quark jet:} \qquad  D_g^{\rm soft-radiation}(x,t) &=&xt \left.\frac{d\Gamma^q_{gq}(E,z)}{dz}\right|_{z=x}\;,\label{eq:SingleSplittingQu1}\\
    D_S^{\rm soft-radiation}(x,t) &=& D_V^{\rm soft-radiation}(x,t)= xt \left.\frac{d\Gamma^q_{gq}(E,z)}{dz}\right|_{z=1-x} \label{eq:SingleSplittingQu2}\;,
\end{eqnarray}
which is indicated in Fig.~\ref{fig:EarlyTime} by a gray dashed line for the earliest three times. Based on the above expressions, the resulting contributions to energy loss evaluate to
\begin{eqnarray}
    \gamma_{g-jet}^{\rm soft-radiation} &=& \frac{1}{t_{\rm split}(E)}\int_0^\mu dx~ x\left[  \left.\frac{d\Gamma^g_{gg}(E,z)}{dz}\right|_{z=x}+N_f \left.\frac{d\Gamma^g_{q\bar q}(E,z)}{dz}\right|_{z=x} \right]\;,\\
    &=& ( \underbrace{0.0072}_{g\rightarrow gg}+ \underbrace{1.16~10^{-6}}_{g\rightarrow q\bar q}N_f )\;,\nonumber 
\end{eqnarray}
\begin{eqnarray}
    \gamma^{\rm soft-radiation}_{q-jet} &=&  \frac{1}{t_{\rm split}(E)} \int_0^\mu dx~ x  \left[\left.\frac{d\Gamma^q_{gq}(E,z)}{dz}\right|_{z=x}+\left.\frac{d\Gamma^q_{gq}(E,z)}{dz}\right|_{z=1-x}\right]\;, \\
        &=&\underbrace{0.0038}_{q\leftrightarrow gq}\;,\nonumber 
\end{eqnarray}
where the quoted values correspond to numerical evaluations for $E=1000T$ and $\mu=2\pi T$ as usual. 

Similarly, from Eq.\eqref{eq:recoil1} and \eqref{eq:recoil2} we can estimate the effect of the elastic recoil at early times as 
\begin{eqnarray}
    \gamma^{\rm recoil} &=&\frac{1}{t_{\rm split}(E)} \sum_i\int_0^{\mu} dx~ \delta \mathcal{J}_i[\{D_i\}] \simeq
    \frac{1}{t_{\rm split}(E)} 
    \frac{2d_A}{4TE^4} \frac{\hat{\bar{q}}}{(g^4/\pi)} \left[ \delta \hat{\bar q}-T\delta\bar\eta_D  \right]\;,
\end{eqnarray}

where in the last step we have approximated $\int_0^{2\pi} dx~ x^2 n_a(xE)(1\pm n_a(xE)) \simeq \int_0^{\infty} dx ~x^2 n_a(xE)(1\pm n_a(xE))$. Evaluating the contributions to $ \delta \hat{\bar q}$ and $\delta\bar\eta_D$ based on the initial conditions for gluon and quark jets in Eq. \eqref{eq:gluon-jet} and \eqref{eq:quark-jet}, one finds that
\begin{eqnarray}
    \text{gluon jet:}&&\qquad \delta \hat{\bar q} = \frac{g^4}{\pi} \nu_{g}^{-1} C_A E^3 \;, \quad \delta\bar\eta_D =  \frac{g^4}{\pi} 2\nu_{g}^{-1} C_A E^2\;,\\
    \text{quark jet:}&&\qquad \delta \hat{\bar q} = \frac{g^4}{\pi} \frac{\nu_{q}^{-1} E^3}{2} \;, \qquad \delta\bar\eta_D =  \frac{g^4}{\pi} \frac{2\nu_{q}^{-1} E^2}{2}\;,\\
\end{eqnarray}
such that
\begin{eqnarray}
    \gamma^{\rm recoil} &=&\frac{1}{t_{\rm split}(E)} \frac{\hat{\bar q}_{\rm eq} C_R}{4E} \left[ \frac{1}{T}-\frac{2}{E} \right] \simeq \frac{\hat{\bar q}_{\rm eq} C_R}{4TE} \;,\nonumber\\
\end{eqnarray}
where $C_R$ corresponds to the particle carrying all the energy in the initial condition.
We also provide the behavior of the recoil contribution to the energy distribution at the early times
\begin{eqnarray}
    D_g^{\rm recoil}(x,t) &\simeq& \frac{\hat{\bar q}_{\rm eq} C_A}{4TE}x^2t~ n_B(xE)(1+ n_B(xE))\;,\label{eq:RecoilEarlyTime1}\\
    D_S^{\rm recoil}(x,t) &\simeq& 2\frac{\hat{\bar q}_{\rm eq} C_F}{4TE}x^2t~ n_F(xE)(1- n_F(xE))\;.\label{eq:RecoilEarlyTime2}
\end{eqnarray}
which is indicated in Fig. \ref{fig:EarlyTime} by a green dashed line for the earliest three times.
While the sum soft radiation and recoil contributions provides an excellent description of the evolution of the energy distributions in Fig.~\ref{fig:EarlyTime} and the initial energy loss rate in Fig.~\ref{fig:JetEnergyLoss} at very early times, clear deviations of the spectrum at small $x\lesssim T/E$ and intermediate scales $ T/E \ll x \ll 1$ start to develop rather quickly, especially in the flavor singlet quark channel $(D_S)$. Similarly, the early-time estimate in Eq.~(\ref{eq:ConstantEstimate}), also fails to explain the linear rise of the energy loss rate seen in Fig.~\ref{fig:EarlyTime}, which as we will discuss now can be attributed to multiple successive splittings, which ultimately provide a more efficient energy loss mechanism \cite{Kurkela:2014tla,Blaizot:2013hx,Baier:2000sb,Schlichting:2019abc}. 

\subsection{Successive splittings \& evolution at intermediate scales}\label{sec:Turbulence}
Besides contributing to the energy loss, radiative emissions from the original hard partons in Eqns.~(\ref{eq:SingleSplittingGl1},\ref{eq:SingleSplittingGl2},\ref{eq:SingleSplittingQu1},\ref{eq:SingleSplittingQu2}) also establish a spectrum of intermediate energy particles, as is clearly seen from Fig.~\ref{fig:EarlyTime}, where all intermediate scales are populated starting at early times.
Such radiated quanta at intermediate energy scales $T/E \ll x \ll 1$ typically have a higher interaction rate, and they can therefore undergo subsequent interactions with the thermal medium to lose their energy. 

Based on the parametric estimates in Eqns.~(\ref{eq:ParametricEstimateFst}-\ref{eq:ParametricEstimateLst}), one expects the evolution at scales $T/E \ll x \ll 1$, to be dominated by inelastic scatterings and one can therefore approximate the collision integrals as follows 
\begin{eqnarray}
\label{eq:InelasticApproximateEqsFst}
    C_g[\{D_i\}] &=&  \int_0^1 dz~  \frac{d\Gamma^g_{gg}\left(\frac{xE}{z},z\right)}{dz} D_g\left(\frac{x}{z}\right)-\frac{1}{2} \frac{d\Gamma^g_{gg}(xE,z)}{dz} D_g(x)\nonumber \\
    &&+\int_0^1 dz~ \frac{d\Gamma^q_{gq}\left(\frac{xE}{z}\;, z\right)}{dz} D_S\left(\frac{x}{z}\right)- N_f\int_0^1 dz~ \frac{d\Gamma^g_{q\bar q}(xE,z)}{dz} D_g(x)\;, \nonumber\\\\
    C_S[\{D_i\}]&=&\int_0^1 dz~ \frac{d\Gamma^q_{gq}\left(\frac{xE}{z}\;, \bar z\right)}{dz} D_S\left(\frac{x}{z}\right)
    -\frac{d\Gamma^q_{gq}(xE, z)}{dz} D_S(x)\nonumber\\
    &&+2N_f\int_0^1 dz~ \frac{d\Gamma^g_{q\bar q}\left(\frac{xE}{z}\;,  z\right)}{dz} D_g\left(\frac{x}{z}\right)\;,\\
    C_V[\{D_i\}]&=& 
    \int_0^1 dz~ \frac{d\Gamma^q_{gq}\left(\frac{xE}{z}\;, \bar z\right)}{dz} D_V\left(\frac{x}{z}\right)
    -\frac{d\Gamma^q_{gq}(xE, z)}{dz} D_V(x)\;,
\label{eq:InelasticApproximateEqsLst}
\end{eqnarray}
where we neglected the contributions from Bose enhancement and Fermi suppression, which are exponentially suppressed for energies $xE \gg T$. Since at sufficiently high jet energies the relevant splitting rates $\Gamma^{a}_{bc}$ are in the deep LPM regime~\cite{Arnold:2008zu,Arnold:2008iy}, they can further be approximated by the leading-log solutions \cite{Mehtar-Tani:2018zba,Arnold:2008iy}\footnote{Note that this approximation is also commonly referred to as harmonic oscillator approximation, and that the functions $\mathcal{K}_{ij}$ agree with the definitions in \cite{Mehtar-Tani:2018zba}.}
\begin{eqnarray}
\label{eq:SplittingRatesHOApproxFst}
    \frac{d\Gamma^g_{gg}(xE,z)}{dz} &\simeq& \frac{1}{\sqrt x} \mathcal{K}_{gg}(z) 
    = \frac{\alpha_s}{2\pi}P_{gg}(z) \sqrt{\frac{\hat{\bar q}(xE)}{xE}} \sqrt{\frac{(1-z)C_A+z^2 C_A}{z(1-z)}}\;, \nonumber\\ \\
    \frac{d\Gamma^q_{gq}(xE,z)}{dz} &\simeq& \frac{1}{\sqrt x} \mathcal{K}_{gq}(z) 
    =   \frac{\alpha_s}{2\pi}P_{qg}(z)\sqrt{\frac{\hat{\bar q}(xE)}{xE}} \sqrt{\frac{(1-z)C_A+z^2 C_F}{z(1-z)}}\;, \nonumber \\ \\
    \frac{d\Gamma^g_{q\bar q}(xE,z)}{dz} &\simeq& \frac{1}{\sqrt x}\mathcal{K}_{qg}(z) 
    = \frac{\alpha_s}{2\pi} P_{gq}(z)\sqrt{\frac{\hat{\bar q}(xE)}{xE}}\sqrt{\frac{C_F-z(1-z)C_A}{z(1-z)}}\;, \nonumber \\
\label{eq:SplittingRatesHOApproxLst}
\end{eqnarray} 
where in the above expressions $\hat{\bar q}(xE)$ should be fixed to match the full splitting kernel at the relevant energy scale (see Appendix \ref{ap:NumericalImplementation} for a comparison). Based on the above expressions for the splitting rates, and the initial conditions in Eqns.~\eqref{eq:gluon-jet} and \eqref{eq:quark-jet} the single emission spectrum then takes the approximate form 
\begin{eqnarray}
D_g(x,t) \simeq \frac{G(t)}{\sqrt{x}}\;,  \label{eq:GluonKolmogorovSpectrum}\\
D_S(x,t) \simeq S(t) \sqrt{x}\;,
\end{eqnarray}
featuring the characteristic $1/\sqrt{x}$ and $\sqrt{x}$ power laws in the gluon and singlet quark channels with linearly rising amplitudes $G(t)$ and $S(t)$ given by 
\begin{eqnarray}
\label{eq:SingleEmissionHOApprox}
    \text{gluon jet:}&& 
    G(t)=C_A^{3/2} \frac{\alpha_s}{2\pi}\sqrt{\frac{\hat{\bar q}(E)}{E}}t\;,  \quad
    S(t)=2C_F^{1/2} N_fT_R \frac{\alpha_s}{2\pi} \sqrt{\frac{\hat{\bar q}(E)}{E}} t\;, \\
    \\
\label{eq:SingleEmissionHOApprox2}
\text{quark jet:}&& 
    G(t)=C_F C_A^{1/2} \frac{\alpha_s}{2\pi}\sqrt{\frac{\hat{\bar q}(E)}{E}}t\;,  \quad
    S(t)=C_F^{3/2} \frac{\alpha_s}{2\pi} \sqrt{\frac{\hat{\bar q}(E)}{E}} t\;. \\
\end{eqnarray}

Beyond early times, the perturbative description in Eq.~(\ref{eq:SingleEmissionHOApprox}) breaks down, as the radiated quanta undergo successive splittings; the spectrum at intermediate scales $T/E \ll x \ll 1$ no longer follows the single emission spectra from hard ($x\sim1$) primaries, but is instead determined by the dynamics of multiple successive branchings of semi-hard ($T/E \ll x \ll 1$) fragments with a continuous influx of energy and valence charge due to continued emissions from the hard ($x\sim1$) primaries.

In this context, it is important to point out that the set of evolution equations for multiple successive branchings of semi-hard ($T/E \ll x \ll 1$) fragments in Eqns.~(\ref{eq:InelasticApproximateEqsFst}-\ref{eq:InelasticApproximateEqsLst}) features a stationary solution of the form
\begin{eqnarray}
\label{eq:Kolmogorov}
    D_g(x) = \frac{G}{\sqrt{x}}\;, \quad D_S = \frac{S}{\sqrt{x}}\;, \quad D_V = V\sqrt{x}\;,
\end{eqnarray}
which following earlier works \cite{Mehtar-Tani:2018zba,Blaizot:2013hx} corresponds to the Kolmogorov-Zhakarov (KZ) spectrum of weak-wave turbulence, and is associated with the stationary transport of energy and valence charge towards lower energies, i.e. an inverse energy and respectively particle cascade. Despite the fact that the spectral shape $\propto 1/\sqrt{x}$ of the stationary gluon spectrum in Eq.~(\ref{eq:Kolmogorov}) agrees with that of single gluon emission spectra in Eq.~(\ref{eq:SingleEmissionHOApprox}), this agreement is to some extent accidental, as the spectral shape of the KZ spectrum is determined by the characteristic energy $x$-dependence of the splitting rates $\Gamma(xE,z)\sim\sqrt{\frac{\hat{\bar{q}}}{xE}}$ in Eq.~(\ref{eq:SplittingRatesHOApproxFst}) rather than the specific $z$-dependence of the splitting functions \cite{Mehtar-Tani:2018zba,Blaizot:2015jea,nazarenko_2011,zakharov2012kolmogorov}. Similarly, the stationary Kolmogorov spectrum for the singlet quark distribution, also features the same  $\propto 1/\sqrt{x}$ behavior as the gluon distribution, with the ratio quark and gluon distributions $\frac{D_{S}(x)}{D_{g}(x)}=\frac{S}{G}$ determined by the (local) balance of $g\to q\bar{q}$ and $q\to q q$ processes\cite{Mehtar-Tani:2018zba}
\begin{eqnarray}
\frac{S}{G}=\frac{2N_{f} \int dz~z~\mathcal{K}_{qg}(z)}{\int dz~z~\mathcal{K}_{gq}(z)} \approx 0.07 \times 2N_f \label{eq:KolmogorovRatio}
\end{eqnarray}
which is in sharp contrast to single emission spectra in Eq.(\ref{eq:SingleEmissionHOApprox}), where quark emission is power suppressed compared to gluon emission at small $x$. 

\begin{figure}[t!]
    \centering
    \includegraphics[width=0.49\textwidth]{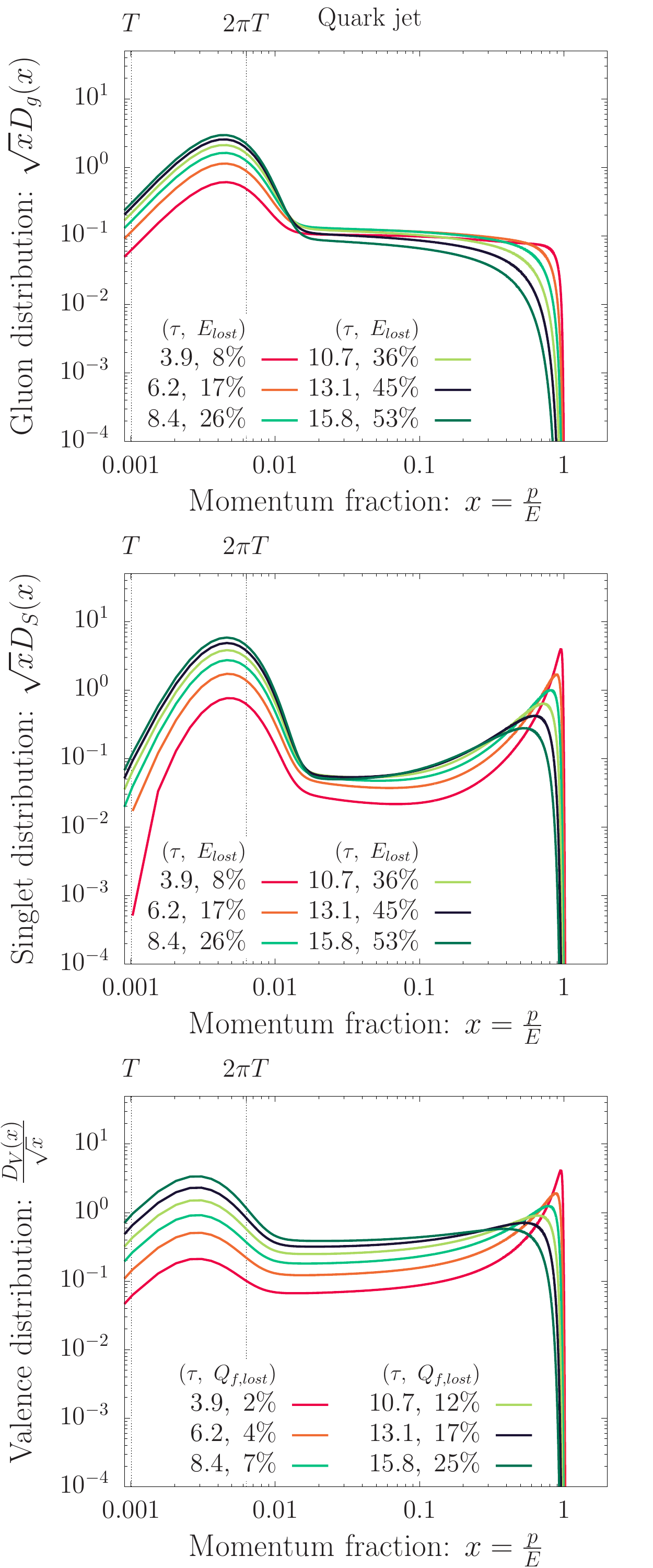} \includegraphics[width=0.49\textwidth]{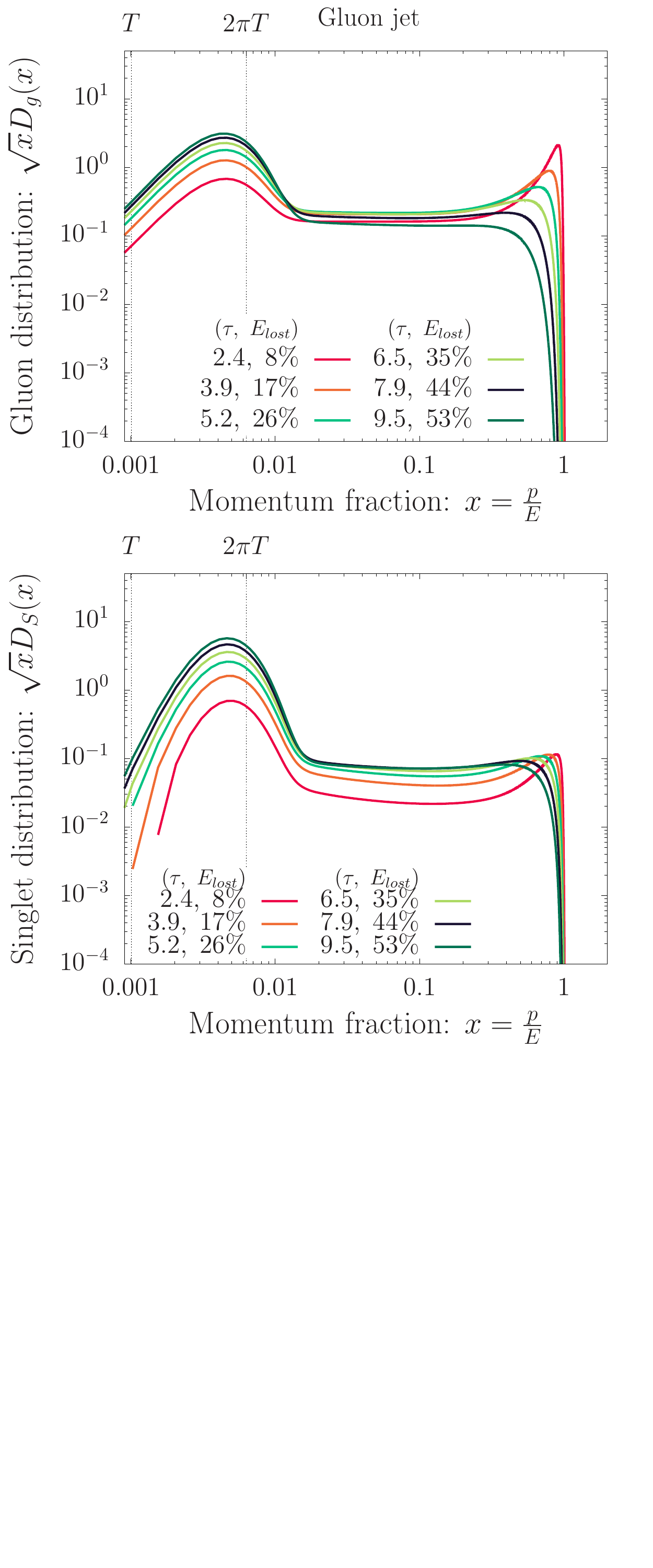}
    \caption{Evolution of the energy distribution at intermediate times for a quark jet (left) and a gluon jet (right). One clearly observes the Kolmogorov spectra in Eq.~(\ref{eq:Kolmogorov}) at intermediate energies $T/E \ll x \ll 1$. }
    \label{fig:Turbulence}
\end{figure}

Numerical results for the evolution of the in-medium energy distributions at intermediate times $\tau$ are presented in Fig.~\ref{fig:Turbulence}, where the different panels show the distributions $D_{g}(x),D_{S}(x)$ and $D_{V}(x)$ for quarks jets (left column) and gluon jets (right column). Despite the fact that the numerical results include both elastic and radiative processes with full in-medium splitting rates, the turbulent spectra in Eq.~(\ref{eq:Kolmogorov}) are clearly visible at intermediate energy scales and persist over the entire range of evolution times shown in Fig.~\ref{fig:Turbulence}. Especially in the subdominant channels i.e. for the singlet quark distribution inside a gluon jet, or the gluon distribution inside a quark jet, the turbulent spectrum persist over a large range of energy fractions $0.02 \lesssim x \lesssim 0.5$ while for the dominant channels, it is not as prominent due to the additional contributions  from the jet peak around $x\sim1$. Strong deviations from the turbulent spectrum also emerge at small $x\sim T/E$, where the effective description in Eqns.~(\ref{eq:Kolmogorov}) breaks down, as other contributions from elastic and inelastic processes become equally important and ultimately lead to the thermalization of the soft fragments. 

\begin{figure}[t!]
    \centering
    \includegraphics[trim={0 0 0 13cm}, clip, width=0.5\textwidth]{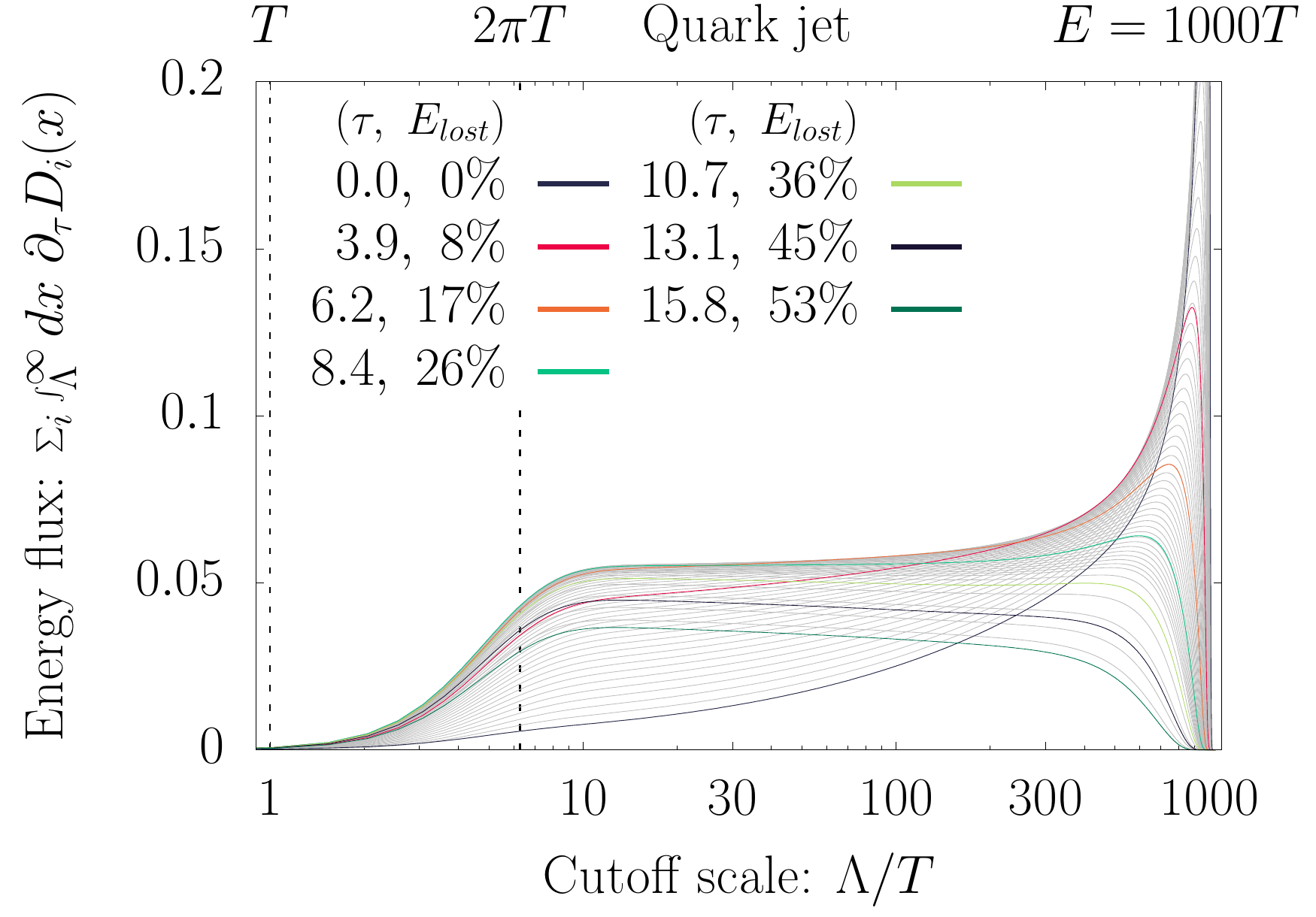}\includegraphics[trim={0 12.5cm 0 0}, clip, width=0.5\textwidth]{plots/EnergyFlux/EnergyFlux-1000.pdf}
    \includegraphics[trim={0 0 0 13cm}, clip, width=0.5\textwidth]{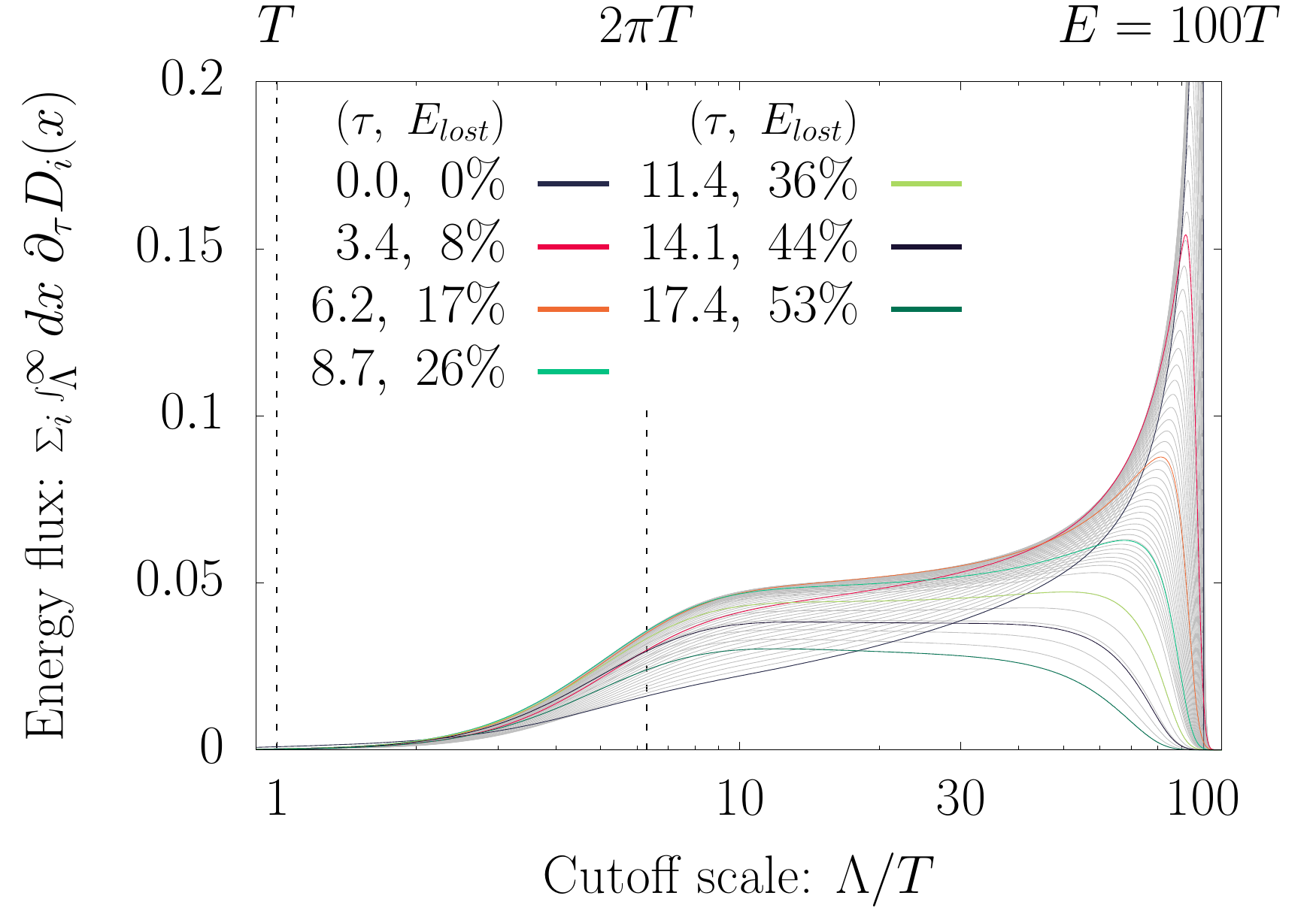}\includegraphics[trim={0 12.5cm 0 0}, clip, width=0.5\textwidth]{plots/EnergyFlux/EnergyFlux-100.pdf}
    \includegraphics[trim={0 0 0 13cm}, clip, width=0.5\textwidth]{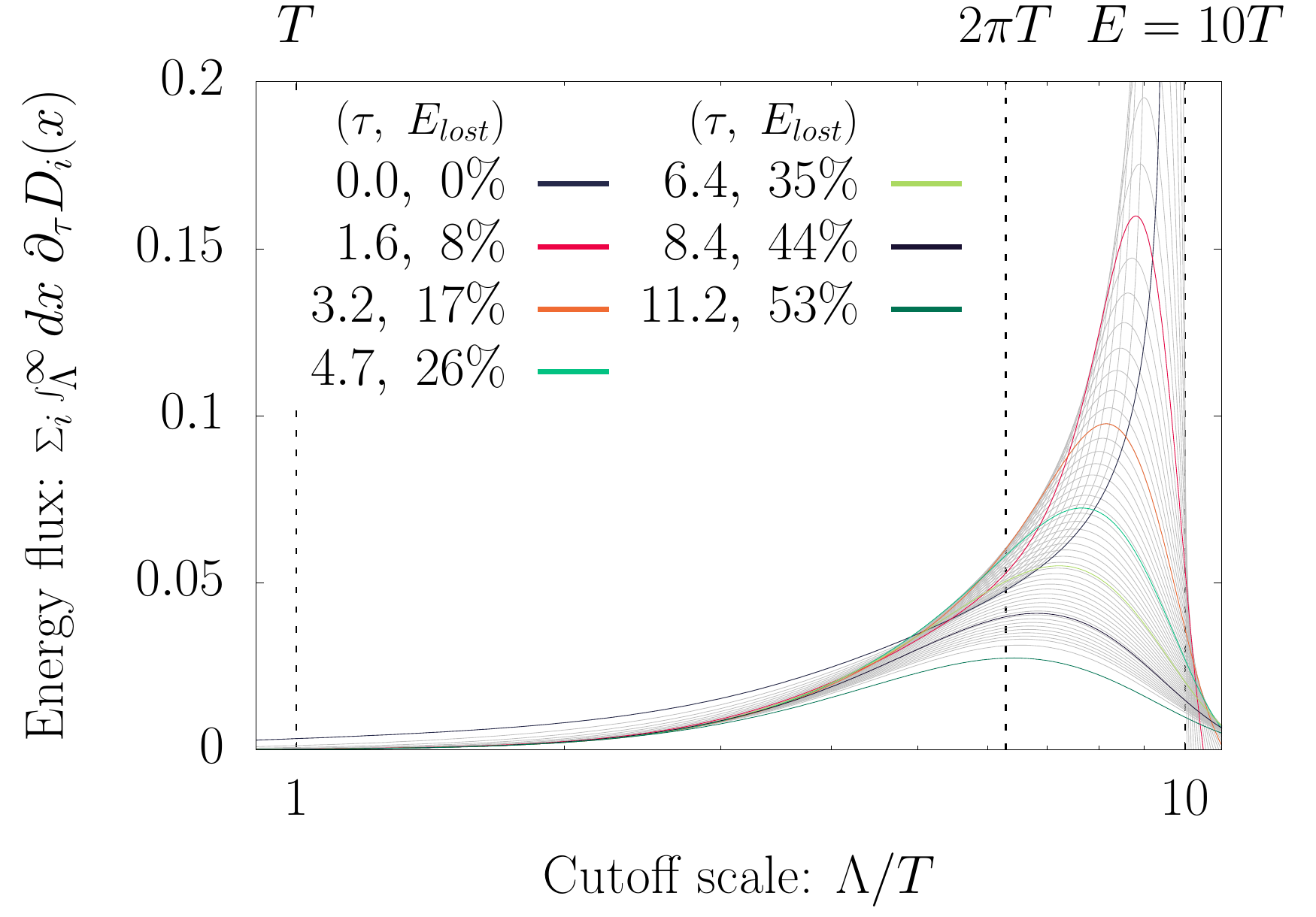}\includegraphics[trim={0 12.5cm 0 0}, clip, width=0.5\textwidth]{plots/EnergyFlux/EnergyFlux-10.pdf}
    \caption{Evolution of the energy flux in Eq.~\ref{eq:EnergyFlux} for quark (left) and a gluon (right) jets with different initial energies $E= 1000,100,10T$ from top to bottom. Different curves in each panel show the energy flux at different times with gray lines corresponding to intermediate times.}
    \label{fig:EnergyFlux}
\end{figure}

Clearly, the onset of turbulence has important consequences for the jet energy loss \cite{Kurkela:2014tla,Blaizot:2013hx,Baier:2000sb,Schlichting:2019abc,Mehtar-Tani:2018zba}. Since semi-hard fragments with $T/E \ll x \ll 1$ can efficiently lose energy to the thermal bath via multiple successive quasi-democratic ($z\sim1/2$) splittings, the energy that is injected into the cascade due to semi-hard ($T/E \ll x \ll 1$) primary emissions is efficiently transferred all the way to the scale of the thermal medium $x\sim T/E$, thus providing a highly efficient energy loss mechanism. One characteristic feature of this turbulent transport, is the fact that it can be described by an energy flux 
\begin{eqnarray}
    \frac{dE}{d\tau}(\Lambda) &=& \sum_i \int_{\Lambda/E}^\infty dx~ \partial_{\tau} D_i(x)\;.\label{eq:EnergyFlux}
\end{eqnarray} 
from high-momenta ($x\sim 1$) to low momenta ($x\sim T/E$), which is independent of the momentum scale $\Lambda$ where the energy flux is evaluated. Numerical results for the energy flux $\frac{dE}{d\tau}(\Lambda)$ are presented in Fig.~\ref{fig:EnergyFlux}, where we show the dependence of $\frac{dE}{d\tau}(\Lambda)$ on the momentum scale $\Lambda$ for three different jet energies $E=10,100,1000 T$ at various different stages of the evolution. When the separation of scales between the jet energy $E$ and the medium temperature $T$ is large, we clearly see a plateau in the energy flux, which is virtually constant within an inertial range of momenta between the jet energy and the medium temperature. Such scale invariance of the energy flux ensures the energy injected into the cascade is transported from high-energy ($x\sim 1$) to low-energy ($x\sim T/E$) fragments, without an accumulation of energy at any intermediate scale. Conversely, the variations of the energy flux with the scale $\Lambda$ indicate the regions where energy is dissipated from the hard components of the jet $(x\sim 1)$ and accumulated at the scale of the medium temperature $x\sim T/E$. By comparing the behavior for different jet energies in Fig.~\ref{fig:EnergyFlux}, we find that even for jets with moderately high energies, $E=100T$, there is still a sizeable momentum range where an approximately scale invariant energy flux is formed at intermediate times, during which the jet loses most of its energy to the thermal medium. However, for very low energy jets, $E=10T$, the energy flux strongly varies with the momentum scale $\Lambda$, indicating that without a significant separation of scales the energy of the jet is directly transferred to the medium without resorting to a turbulent energy cascade.

Based on the approximate form of the kinetic equations for $T/E \ll x \ll E$,  we can estimate the energy loss $\frac{dE}{d\tau}$ in the turbulent regime as the scale invariant energy flux, which can be computed as
\begin{eqnarray}
    \frac{dE}{d\tau} &=& \int^1_{\mu/E} dx~\int_x^1 dz~  \left[\mathcal{K}_{gg}(z) +2N_f\mathcal{K}_{qg}(z) \right] \sqrt{\frac{z}{x}} D_g\left(\frac{x}{z}\right) \nonumber \\
    &&- \int^1_{\mu/E} dx~\int_0^1 dz~ \left[\mathcal{K}_{gg}(z) +2N_f\mathcal{K}_{qg}(z) \right] \frac{z}{\sqrt{x}}D_g(x)\;, \nonumber \\
    &&+ \int^1_{\mu/E} dx~\int_x^1 dz~  \left[\mathcal{K}_{gq}(z)+\mathcal{K}_{gq}(1-z)\right] \sqrt{\frac{z}{x}} D_S\left(\frac{x}{z}\right) \nonumber \\
    &&- \int^1_{\mu/E} dx~\int_0^1 dz~ \mathcal{K}_{gq}(z) \frac{1}{\sqrt{x}}D_S(x)\;,
\end{eqnarray}
By changing the order of integration and performing a change of variable $x\rightarrow x/z$ to combine the gain and loss terms, the energy flux  can be re-expressed as \cite{Mehtar-Tani:2018zba} 
\begin{eqnarray}
    \frac{dE}{d\tau} &=& -\int_{\mu/E}^1 dz ~ z [\mathcal{K}_{gg}(z)+2N_f\mathcal{K}_{qg}(z)] \int_{\mu/E}^{\mu/zE} dx~ \frac{D_g(x)}{\sqrt{x}}\nonumber\\
    &&-\int_{0}^{\mu/E} dz ~ z [\mathcal{K}_{gg}(z)+2N_f\mathcal{K}_{qg}(z)] \int_{\mu/E}^1 dx~  \frac{D_g(x)}{\sqrt{x}} \nonumber\\
    && -\int_{\mu/E}^1 dz~  2 z\left[\mathcal{K}_{gq}(z)+\mathcal{K}_{gq}(1-z)\right] \int_{\mu/E}^{\mu/zE} dx~ \frac{D_S(x)}{\sqrt{x}} \nonumber\\
    &&-\int_0^{\mu/E} dz~2z[\mathcal{K}_{gq}(z)+\mathcal{K}_{gq}(1-z)] \int^1_{\mu/E} dx~  \frac{D_S(x)}{\sqrt{x}}\;.
\end{eqnarray}
Such that upon making use of the explicit form of the Kolmogorov-Zhakarov spectrum in Eq.~(\ref{eq:Kolmogorov}), one obtains the scale invariant energy flux in the limit $\mu/E \ll 1$ as  
\begin{eqnarray}
\frac{dE}{d\tau} =  \tilde\gamma_g  G +  \tilde\gamma_q S\;,\label{eq:EnergyFluxInelasticEarlyTime}
\end{eqnarray}
with the flux constants
\begin{eqnarray}
    \label{eq:TildGammag}
     \tilde\gamma_g &=& \int_{0}^1 dz ~ z [\mathcal{K}_{gg}(z)+2N_f\mathcal{K}_{qg}(z)]~\log(z)
    = \frac{\alpha_s}{2\pi}\sqrt{\frac{\hat{\bar q}(\sqrt{TE})}{E}}(25.78 +2N_f0.177)\;, \\
    \label{eq:TildGammaq}
    \tilde\gamma_q &=& \int_{0}^1 dz ~ 2z [K_{gq}(z)+K_{qq}(z)] \log(z) 
    ~~~~=  \frac{\alpha_s}{2\pi}\sqrt{\frac{\hat{\bar q}(\sqrt{TE})}{E}}(11.595)\;, 
\end{eqnarray}
where we chose to evaluate $\hat{\bar q}(\sqrt{TE})$ at an intermediate scale between the jet energy $E$ and the medium temperature $T$. While the splitting functions in Eq.~(\ref{eq:SplittingRatesHOApproxFst}) exhibit a singular behavior for soft emissions ($z,1-z \to 0$), it is important to point out that the energy flux in Eqns.~(\ref{eq:EnergyFluxInelasticEarlyTime}) is in fact dominated by quasi-democratic $(z\sim1/2)$ splittings, and we refer the interested reader to \cite{Mehtar-Tani:2018zba} for further discussion and additional details of the above calculation.

By making the amplitude $G(\tau)$ time dependent, in order to account for the injection of energy into the cascade due to radiation from the hard $(x\sim1)$ primaries as in Eq.~(\ref{eq:GluonKolmogorovSpectrum}), 
and adding the contributions from soft-radiation and recoil, the energy loss in the turbulent regime can then be estimated as
\begin{eqnarray}
\frac{dE}{d\tau} = \gamma^{\rm soft-radiation} + \gamma^{\rm recoil} + \left(\tilde\gamma_g  +\frac{S}{G} \tilde\gamma_q\right) G(\tau)\;,  
\end{eqnarray}
which is shown in Fig.~\ref{fig:JetEnergyLoss} as a gray dashed line and provides an excellent description of the numerical results up to times $\tau \lesssim 5$ where jets have deposited about $30\%$ of their energy to the thermal medium.

\begin{figure}[t!]
    \centering
    \includegraphics[width=\textwidth]{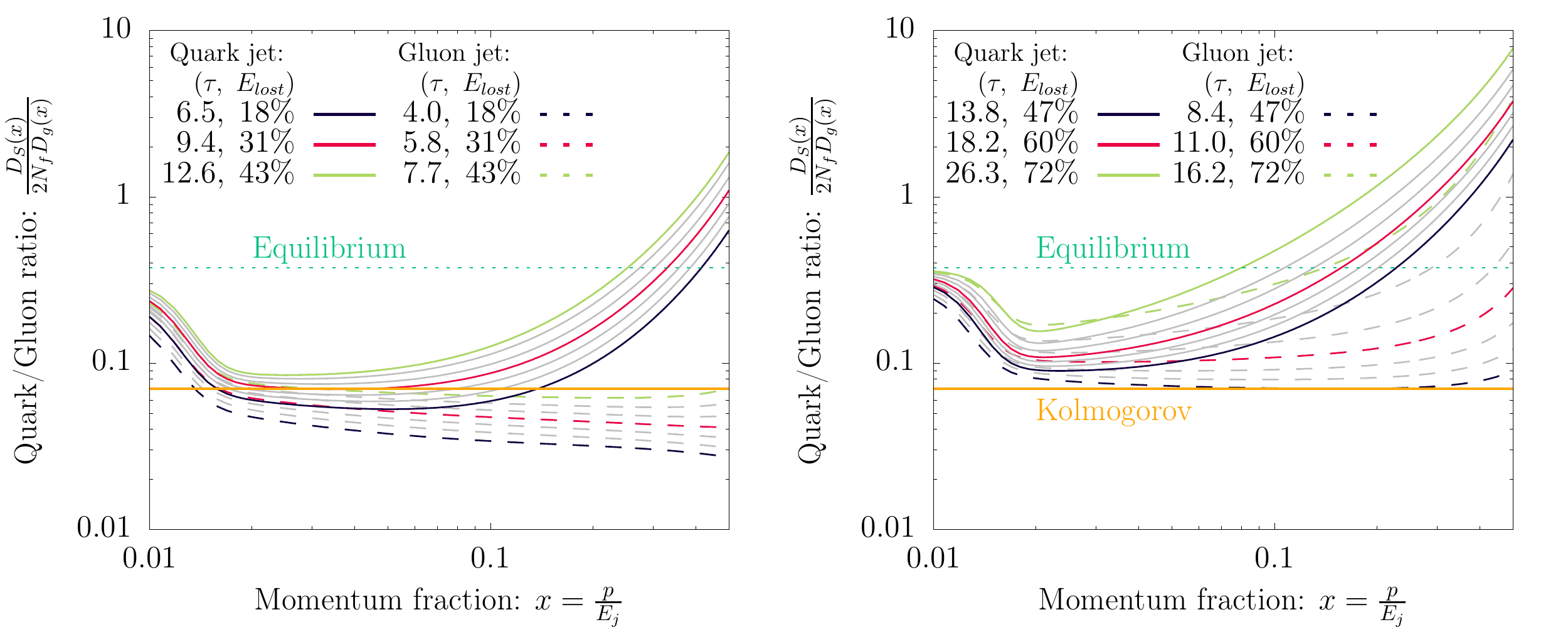}
    \caption{Quark to gluon ratio $D_{S}(x)/2N_f D_{g}(x)$ at different times as a function of the momentum fraction $x$. Different curves in each panel correspond to a quark jet (solid lines) and a gluon jet (dashed lines), at evolution times indicated by the amount of energy that the jet has lost. Horizontal lines correspond to the equilibrium ratio $D_{S}(x)/2N_f D_{g}(x)=\nu_q/\nu_g$ which is approached at small $x$, and the universal Kolmogorov ratio in  Eq.~(\ref{eq:KolmogorovRatio}) which is approached at intermediate values of $T/E \ll x \ll 1$ for a transient period of time.}
    \label{fig:QtoGRatio}
\end{figure}

One striking prediction of the turbulent energy loss mechanism, is the universal ratio of quark and gluon energy distributions $D_{S}(x)/D_{g}(x)$ in Eq.~(\ref{eq:KolmogorovRatio}) within an inertial range of energy $T/E \ll x \ll 1$. Now in order to verify to what extent this behavior can be realized over the course of the jet medium evolution, we present our numerical results for the quark to gluon ratio in Fig.~\ref{fig:QtoGRatio}, which includes all effects due to elastic and in-elastic interactions as described in Sec.~\ref{sec:Kinetic-Equation}. Different curves in Fig.~\ref{fig:QtoGRatio}, show the results for $D_{S}(x)/2N_f D_{g}(x)$ for quark (solid lines) and gluon jets (dashed lines), at various stages of the evolution. Indeed, one finds that starting around the times when the jet has lost about $20\%$ of its total energy, the quark to gluon ratio at intermediate values of $0.02\lesssim x \lesssim 0.1$ is rather well described by the universal Kolmogorov ratio in Eq.~(\ref{eq:KolmogorovRatio}) indicated by a solid orange line in Fig.\ref{fig:QtoGRatio}.  Vice versa, for small momentum fractions $x \sim T/E$ on the order of the medium temperature, the quark to gluon ratio approaches its equilibrium value of $D_{S}(x)/2N_f D_{g}(x)=\frac{\nu_{q}}{\nu_{g}}$ indicating that the soft fragments of the jet have had sufficient time to undergo chemical equilibration. While at early times the large $x$ components of the jet are dominated by the primary jet peak, and the jet chemistry is dominated by the primary species, i.e. by gluons for gluon jets and by quarks for quark jets, the situation is different at late times when the jet has lost a significant amount of its energy. Due to the fact that hard gluons lose their energy more efficiently as compared to hard quarks, one finds that the medium effectively acts as a chemical filter, such that even for gluon jets, the hardest constituents of strongly quenched jets are more likely to be quarks, as can be inferred from the steep rise of the quark to gluon ratio in the right panel of Fig.~\ref{fig:QtoGRatio}.

\begin{figure}[h]
    \centering
    \includegraphics[width=0.49\textwidth]{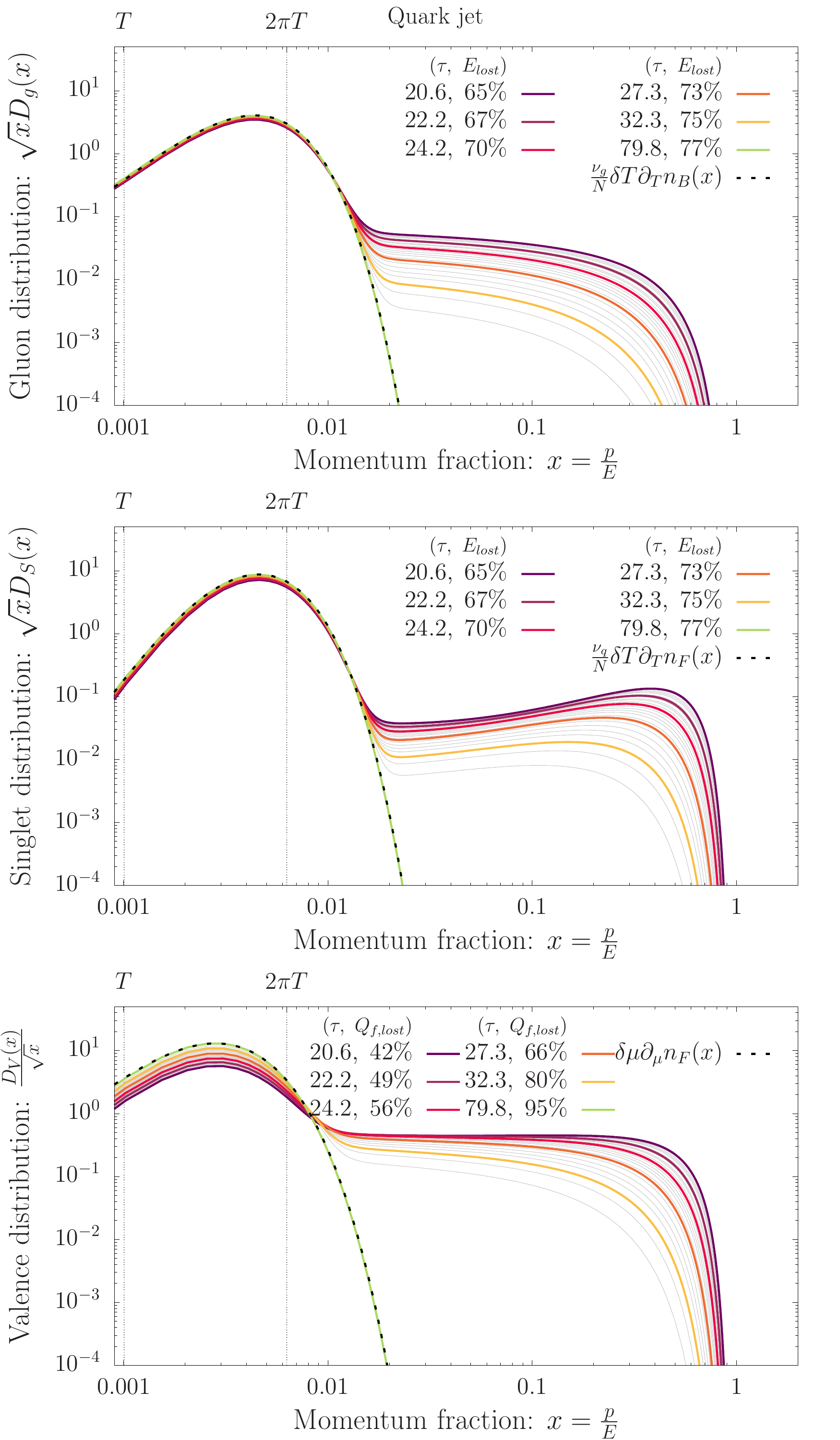} \includegraphics[width=0.49\textwidth]{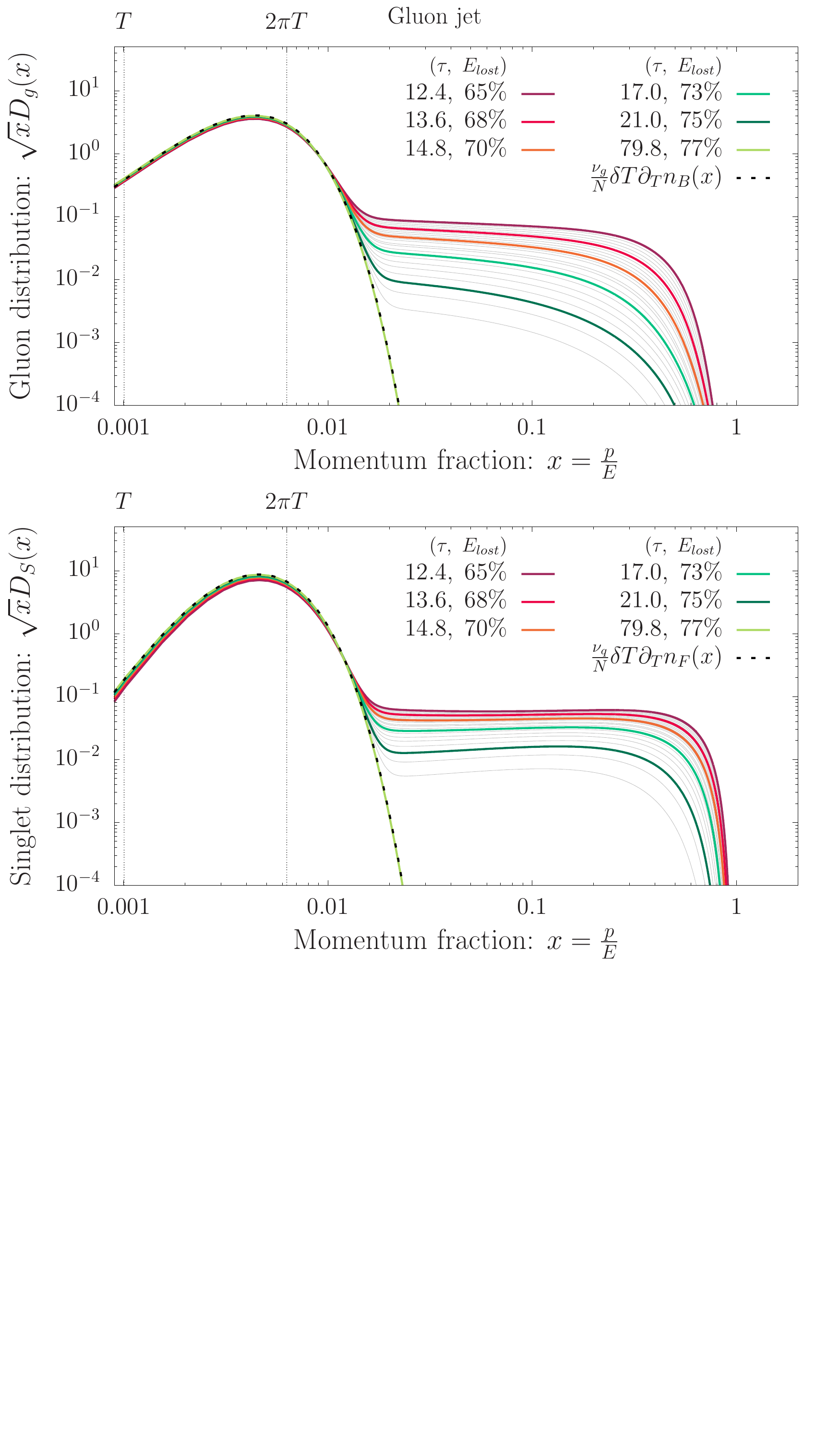}
    \caption{Evolution of the energy distributions at late times for quark (left) and gluon (right) jets. Dashed lines in each panel represent the asymptotic equilibrium distributions in Eq.~(\ref{eq:DEquilibrium}-\ref{eq:DEquilibriumLst}).}
    \label{fig:EquilibriumApproach}
\end{figure}

\subsection{Evolution towards equilibrium}
Eventually, the hard fragments of the jet have had sufficient time to undergo multiple successive quasi-democratic branchings to deposit a significant amount of their initial energy into the thermal medium. During this last stage of the evolution depicted in Fig.~\ref{fig:EquilibriumApproach}, the few remaining hard fragments continue to lose energy and valence charge thereby heating up the thermal bath and doping it with the valence charge. We find that in this regime, the in-medium jet evolution follows the characteristic pattern of ``bottom-up" thermalization \cite{Baier:2000sb,1405.6318v1,Schlichting:2019abc,Kurkela:2018vqr}, where the low energy part of the distribution ($x\sim T/E$) is well described by the (linearized) equilibrium distributions
\begin{align}
\label{eq:DEquilibrium}
    D_g^{\rm eq}(x) &=  \nu_g\delta T   ~\partial_T n_B(xE), \\
    D_S^{\rm eq}(x) &=  2N_f\nu_q\delta T   ~\partial_T n_F(xE),\\
    D_{V_{f}}^{\rm eq}(x) &= \nu_q\delta \mu_{f} ~\partial_\mu n_F(xE)|_{\mu=0}.
\label{eq:DEquilibriumLst}
\end{align}
with increasing temperature $\delta T$ and chemical potential $\delta \mu_{f}$ as a function of time, which eventually approach their final equilibrium values, indicated by the dashed lines in Fig.~\ref{fig:EquilibriumApproach}. While the soft sector is already thermalized, the evolution of the hard components of the distribution continues to be well described by the Kolmogorov-Zhakarov spectra in Eq.~(\ref{eq:Kolmogorov}) up to the highest available momentum fractions at each instant of time; as the hard components continue to lose their energy to the thermal bath this cascade proceeds towards lower and lower energies, in a fashion that is characteristic of decaying turbulence \cite{nazarenko_2011,zakharov2012kolmogorov,Mehtar-Tani:2018zba,Blaizot:2013hx}.

\begin{figure}[]
    \centering
    \hfill\includegraphics[width=0.43\textwidth]{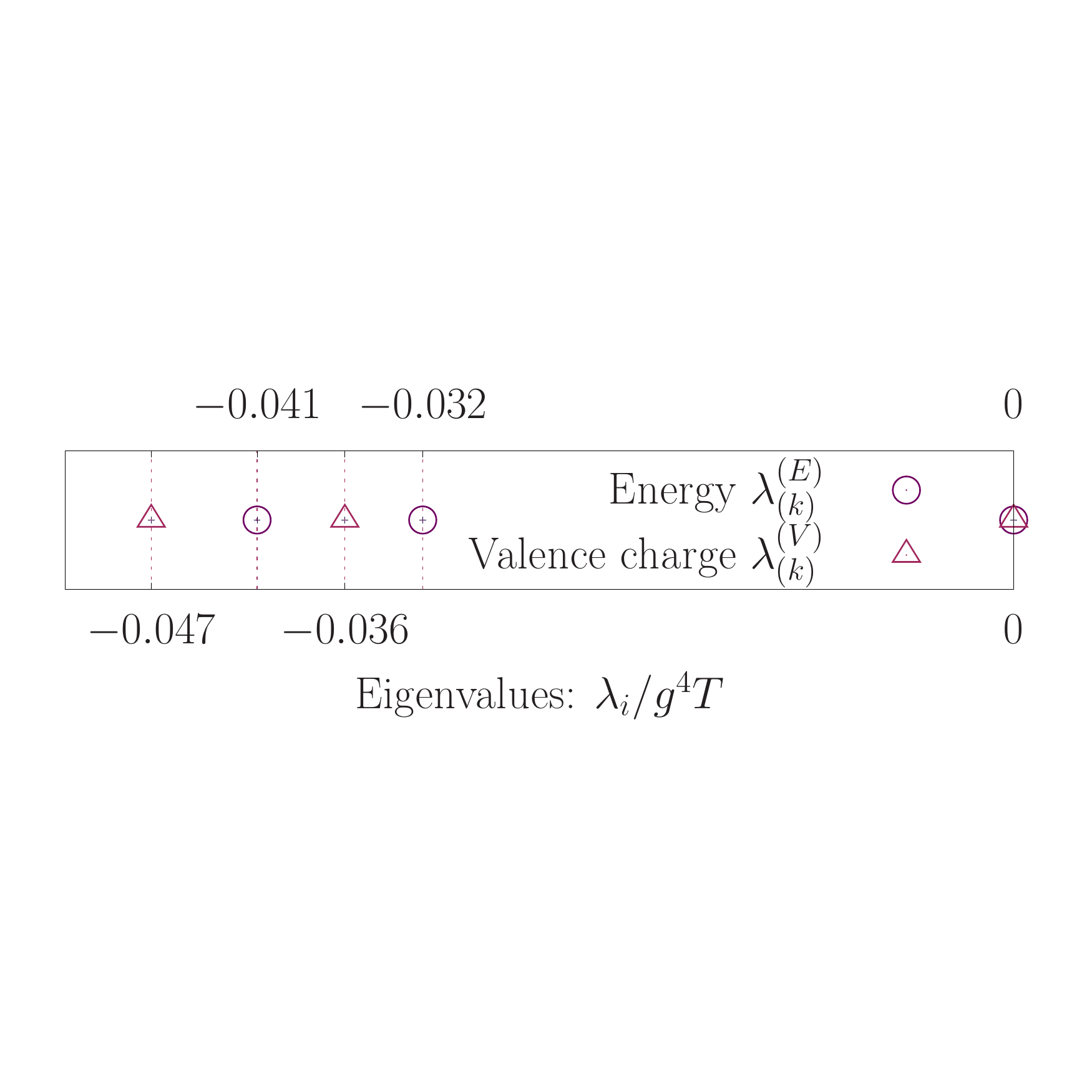}  
    \includegraphics[width=0.5\textwidth]{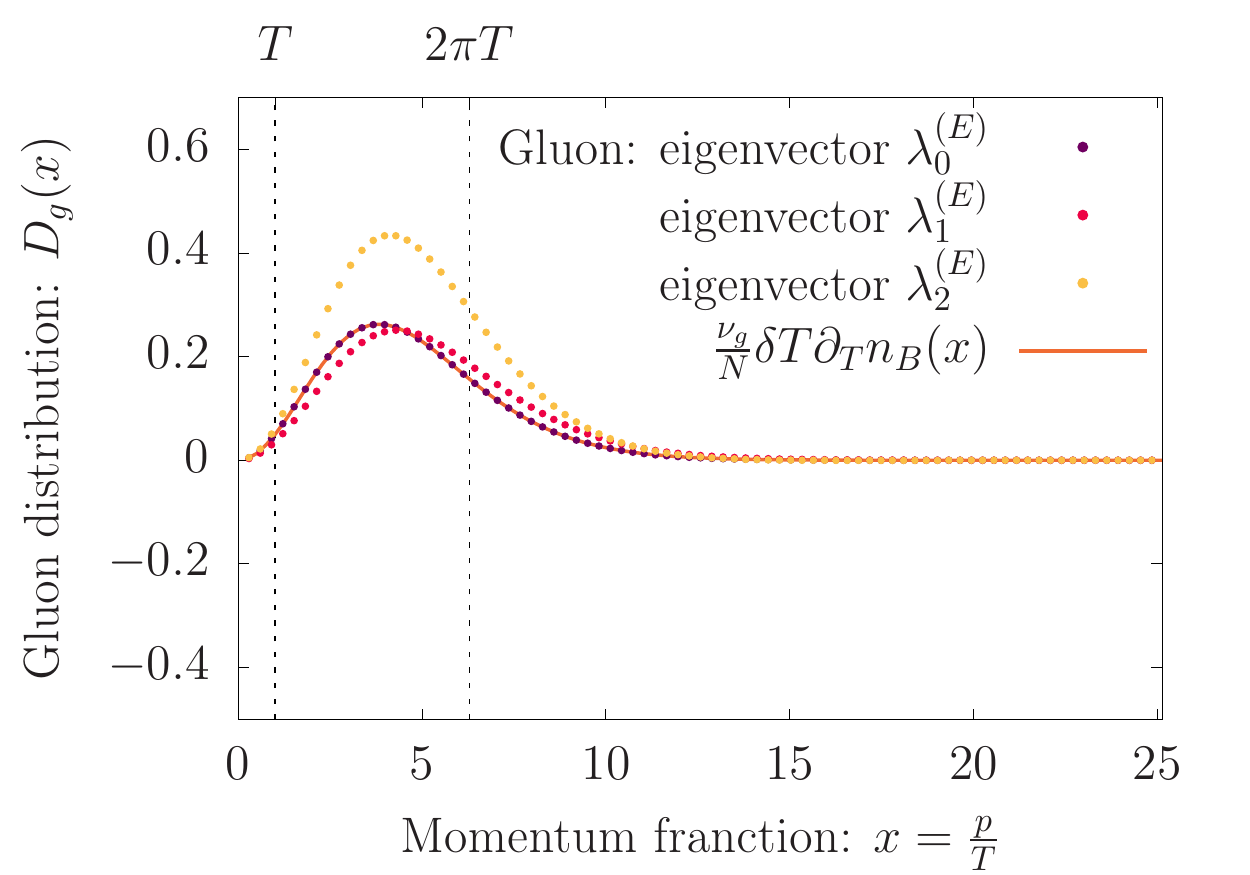}
    \includegraphics[width=0.5\textwidth]{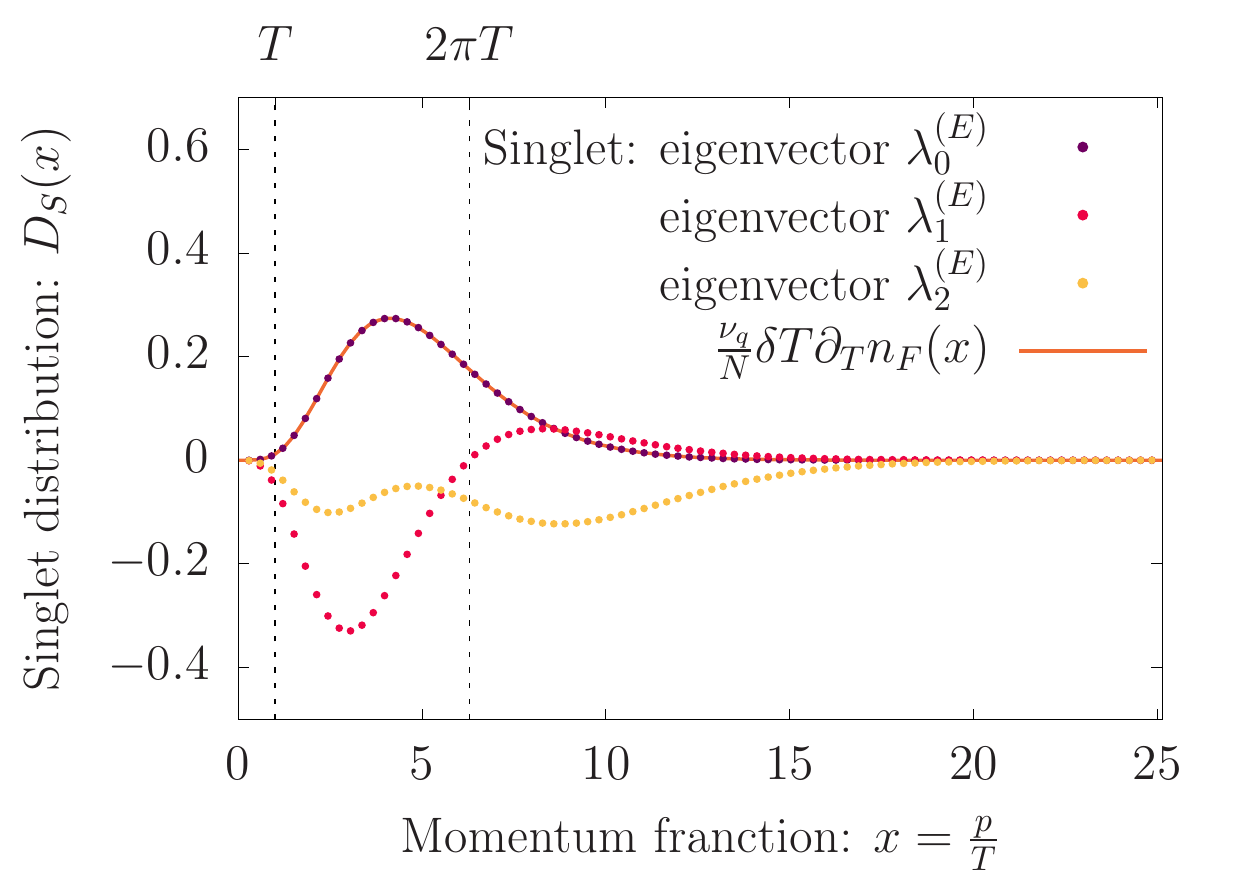}\includegraphics[width=0.5\textwidth]{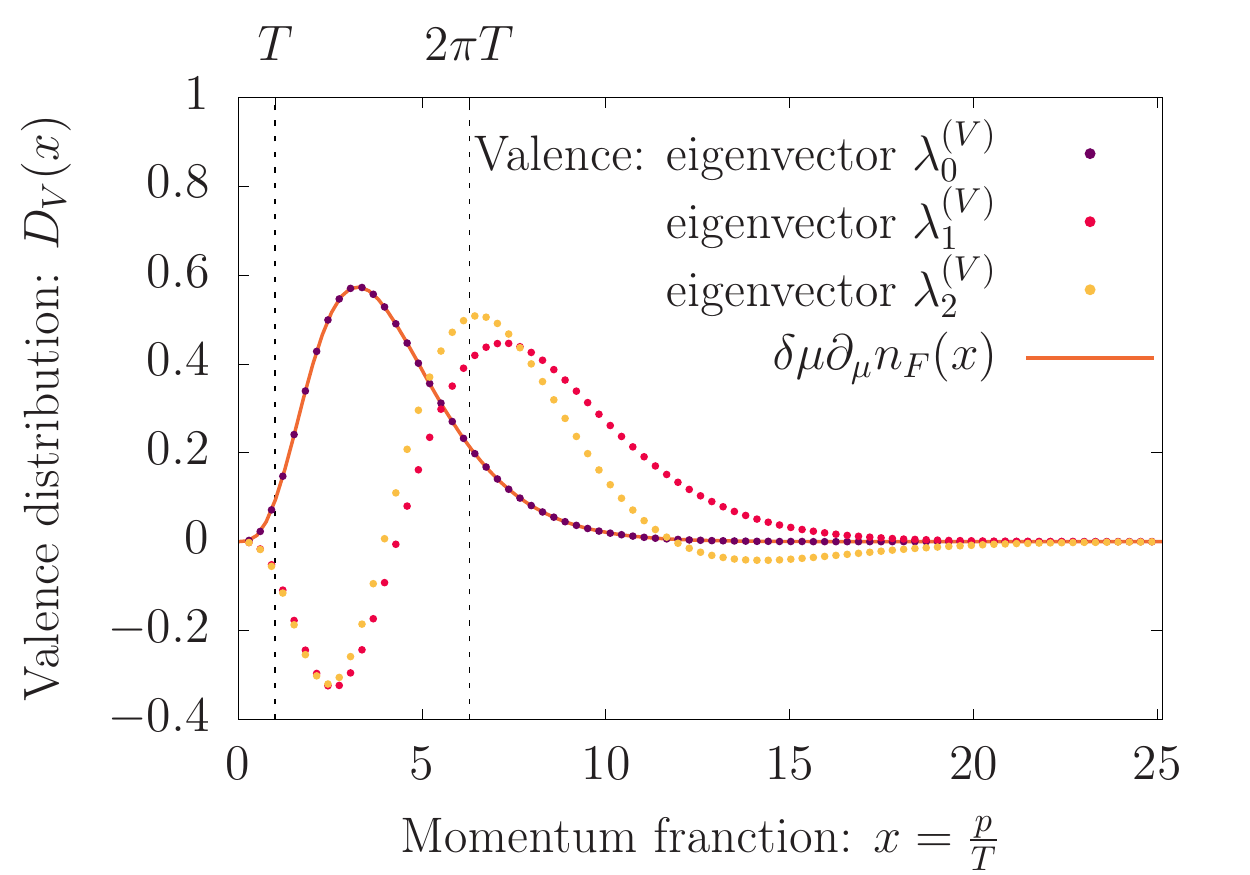}
    \caption{Spectrum of the linearized collision operator. Different panels show the low-lying eigenvalues (top left) as well as the associated eigenfunctions in the gluon (top right), singlet (bottom left) and valence charge (bottom right) channels. Eigenfunctions have been normalized according to $\int dx ~ D_{g}(x)^2+D_{S}(x)^2 =1$ in the energy sector and  $\int dx ~ D_{V}(x)^2 = 1$ in the valence charge sector.}
    \label{fig:EigenVectors}
\end{figure}

Clearly, the final stages of kinetic and chemical equilibration of jets closely resemble the thermalization patterns previously observed in the context of thermalization of the QGP at early times in heavy-ion collisions \cite{Schlichting:2019abc,Kurkela:2018vqr,Kurkela:2018oqw}. We will now further investigate to what extent the kinetic and chemical equilibration of jets is similar to the typical excitations of the medium, encoded e.g. in the transport properties of the QGP \cite{Ghiglieri:2018dgf}. Based on our effective kinetic description of in-medium jet evolution, the evolution of small perturbations around equilibrium can be compactly expressed as
\begin{eqnarray}
\partial_t D_{a}(x,t)= \int~dz~\delta C_{ab}(x,z) D_{b}(z,t)
\end{eqnarray}
indicating that the long time behavior of the distributions $D_{a}(x,t)$ is determined by the low-lying spectrum of the linearized collision operator $\delta C_{ab}(x,z)$, as quantified by the following eigenvalue equation
\begin{eqnarray}
 \int~dz~\delta C_{ab}(x,z) D_{b}(z,t) = \lambda_{(k)} D_{b}^{(k)}(x,t)
\end{eqnarray}
We provide a compact summary of our findings in Figs.~\ref{fig:EigenVectors} where we show the spectrum of the low-lying eigenvalues $\lambda_{(k)}$ along with the corresponding eigenfunctions $D_{a}^{(k)}(x,t)$, determined by numerical diagonalization of the discretized collision operator\footnote{We have checked explicitly that varying the discretization does not significantly alter the results.}.

Since the effective kinetic description in Sec.\ref{sec:Kinetic-Equation}, exactly conserves the energy $E$ and valence charges $Q_{f}$, there is a total of $N_f+1$ zero modes $\lambda_{(k)}=0$ of the collision operator, whose eigenfunctions correspond to the equilibrium solutions in Eq.~(\ref{eq:DEquilibrium}), and are correctly reproduced by our numerical analysis in  Figs.~\ref{fig:EigenVectors}. Due to the fact that energy and valence charge evolution decouple from each other in the linearized kinetic description, the linearized collision operator is block diagonal and one can further distinguish between the spectrum of modes $\lambda_{(k)}^{(E)}$ in the energy sector, spanned by the distributions $(D_{g},D_{S})$, and the $N_{f}$-fold degenerate spectrum of modes $\lambda_{(k)}^{(V)}$ in the valence charge sector $(D_{V_{f}})$. Based on our analysis, we find that both energy and charge sector feature a discrete low-lying spectrum with low-lying eigenvalues $\lambda_{1,2}^{(V)}$ and $\lambda_{1,2}^{(E)}$ of similar magnitude, which in accordance with our discussion determine the relaxation rates for energy and charge equilibration close to equilibrium. We also find that the corresponding eigenfunctions are localized at low energies $p/T$, in the sense that they decay exponentially \cite{Dusling:2009df} at large energies as can be inferred from Fig.~\ref{fig:EigenVectors}.

\begin{figure}[]
    \centering
    \includegraphics[width=\textwidth]{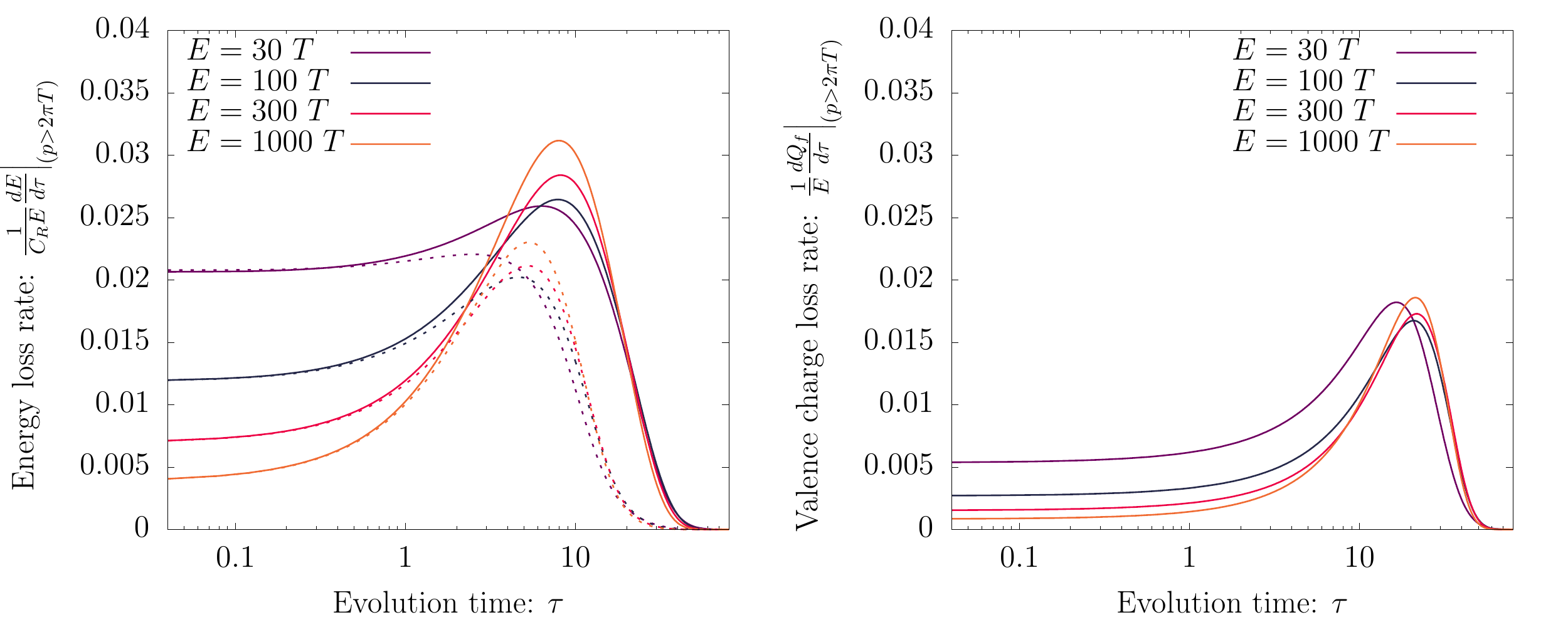}
    \includegraphics[width=\textwidth]{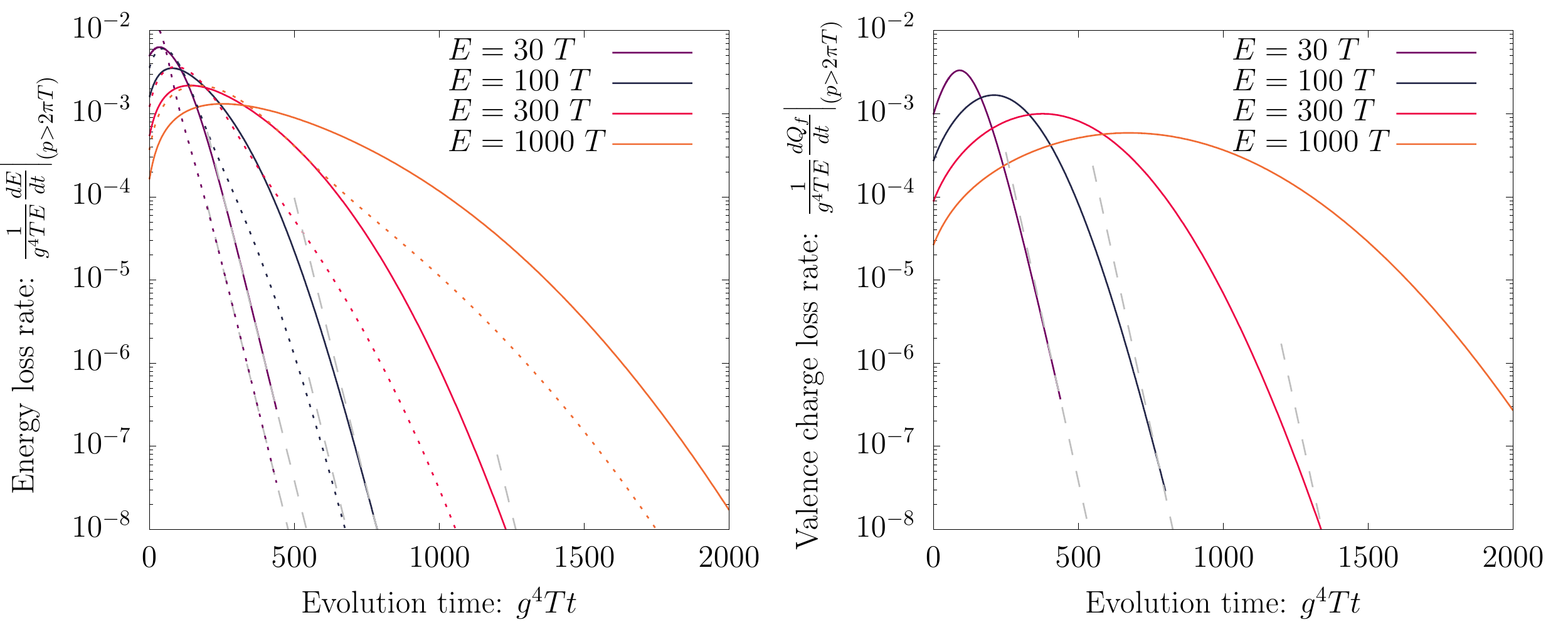}
    \caption{Comparison of energy and valence charge loss rates for quark (full lines) and a gluon (dashed lines) jets, with different initial jet energies $E=30,100,300,1000$. Dashed lines in the lower panel represent fits to an exponential decay using the first nonzero eigenvalues as the decay constant.}
    \label{fig:ExpDecay}
\end{figure}

Now that we have determined the near-equilibrium relaxation rates for energy and charge equilibrium, it is insightful to revisit the evolution of the jet energy ($E$) and valence charge ($Q_{f}$) loss rates. Numerical results for the time evolution of the energy and valence charge loss rates are presented in Fig.~\ref{fig:ExpDecay}, which compactly summarized our results for quark (dashed lines) and gluon jets (solid lines), with initial energies $E=30,100,300,1000T$. By comparing the results for different energies in the top panels, which shows the rates $\frac{1}{C_{R}} dE/d\tau$ and $dQ_{f}/d\tau$ as a function of the natural timescale $\tau=t/t_{\rm split}(E)$ for jet evolution, one finds that the leading jet energy dependence is indeed determined by the timescale for hard splittings $t_{\rm split}(E)$ and correctly captured by the scaling variable. Nevertheless, with decreasing jet energy one observes a gradual change in the energy loss pattern, where the constant energy and valence charge loss due to soft radiation and recoil starts to become increasingly important compared to the turbulent jet energy loss mechanism. Bottom panels of Fig.~\ref{fig:ExpDecay} show the same data for energy and valence charge loss, but now in units of the natural medium timescale $1/g^4T$ of the thermal medium. Based on our above discussion, one ultimately expects that at asymptotically late times, the in-medium evolution of the jet will be governed by the near-equilibrium relaxation rates, corresponding to the lowest eigenvalues $\lambda_{1}^{(E)}$ and $\lambda_{1}^{(V)}$ of the linearized collision operator. While for low energy jets, $E=30T$, such an exponential decay is clearly visible at late times, as indicated by the gray dashed lines in Fig.~\ref{fig:ExpDecay} which represent fits of the form $\frac{dE}{dt} \propto e^{\lambda_{1}^{(E)}t}$ and $\frac{dQ}{dt} \propto e^{\lambda_{1}^{(V)}t}$, it is important to note that the jets have already lost nearly all of their energy by the time that this near-equilibrium linear response treatment becomes applicable. We therefore conclude that the in-medium evolution of high-energy jets should be considered as a genuine far-from-equilibrium probe of the QGP, whose space-time dynamics can not be directly related to that of near equilibrium excitations and generally requires a detailed microscopic description.

\section{Conclusions \& Outlook}
Based on an effective kinetic description of high energetic partons in a thermal medium, we established a comprehensive picture of in-medium hard parton evolution, from the earliest stages of elastic and radiative energy loss all the way towards kinetic and chemical equilibration of hard partons inside the medium. By including the leading order small angle elastic and inelastic processes, that ensure energy and charge conservation and allow us to follow the evolution of the parton shower all the way towards equilibrium, we confirm earlier findings \cite{Mehtar-Tani:2018zba,Blaizot:2013hx} that the energy loss of highly-energetic partons is dominated by a turbulent cascade due to successive radiative branchings. By investigating the energy flux along the cascade, we explicitly demonstrated that the turbulent cascade transports energy all the way from the energy scale of the jet $x\sim 1$ to the temperature of the thermal medium $x \sim T/E$, where all leading order kinetic processes become of comparable importance. Due to multiple successive branchings, the in-medium energy distributions become insensitive to the hard structure of the jet, and display universal turbulent features in an inertial range of energy fractions $T/E \ll x \ll 1$, while the soft fragments of the distribution with $x \lesssim T/E$ rapidly thermalize inside the medium. 

Even though the in-medium evolution of high energy partons closely resembles the thermalization patterns observed in previous studies of the thermalization of the QGP at early times~\cite{Kurkela:2018vqr,Kurkela:2018oqw,Schlichting:2019abc}, it turns out that the dominant mechanism underlying jet quenching is quite different from the typical relaxation of near-equilibrium modes, indicating that highly energetic partons or jets $\gtrsim 30 T$ should really be considered as genuine non-equilibrium probes of the QGP. Conversely, the in-medium evolution of less energetic partons or jets $\lesssim 30 T$, is more sensitive to the physics at the scale of the QGP medium, and may bear the possibility to explore aspects of the physics of QGP thermalization by means of jet quenching studies in present and future experiments. Notably, the interactions of jets with the medium constituents lead to interesting changes of the chemical composition of medium modified jets (see also \cite{Mehtar-Tani:2018zba}), which should have experimentally observable consequences e.g. in ratios of identified particles ($K/\pi$, $\Lambda/\pi$, $D_0/\pi$, $\Lambda_c/\pi$, ...) inside/around heavy-ion jets. Of course, to provide detailed predictions for any such observables, one also has to include the effects of vacuum like emissions and hadronization, and it will be interesting to further explore these aspects within suitable Monte-Carlo implementations of jet evolution in heavy-ion collisions \cite{Caucal:2018ofz,Chen:2017zte,Putschke:2019yrg,Schenke:2009gb}, which can also account for vacuum like emissions and potentially important effects due to fluctuations of the jet shower and the medium.

So far our analysis has focused on energy deposition of the hard partons and the re-distribution of energy as a function of the momentum fraction $x$ on large time scales, where the hard particles can lose a significant amount of energy to the thermal medium. While in this limit it is well justified to use the radiative emission rates for an infinite medium, it would be important to explicitly include the path length $L$ dependence of the medium induced emission rates, to extend our analysis towards shorter time scales. By including large angle elastic scatterings into our framework, it would also be interesting to explore the effects of the various processes on the angular $(\theta)$ structure of the medium-induced shower, and  -- including also vacuum like emission -- explore the in-medium evolution of jet shapes. Eventually, the goal of these studies should be to develop a full picture of the out-of-cone energy loss of jets and its deposition into the thermal QGP medium (``medium response”), which would provide an important step towards a unified description of soft and hard probes in heavy-ion collisions.

\label{sec:Conclusion}
\section*{Acknowledgment}
We thank X.~Du, C.~Greiner, A.~Mazeliauskas, Y.~Mehtar-Tani, G.~D. Moore, B.~Schenke, N.~Schlusser and D.~Teaney for insightful discussions throughout this project.
This work is supported by the Deutsche Forschungsgemeinschaft (DFG, German Research Foundation) through the CRC-TR 211 'Strong-interaction matter under extreme conditions'– project number 315477589 – TRR 211.
The authors also gratefully acknowledge computing time provided by the Paderborn Center for Parallel Computing (PC$^2$).
\pagebreak
\appendix
\section{Numerical implementation}\label{ap:NumericalImplementation}
Below we provide details on the numerical implementation of the effective kinetic theory.
\subsection{Basic formalism}
Following the discrete-momentum method introduced in \cite{York:2014wja}, we discretize the distribution using “wedges” functions basis
\begin{eqnarray}
    N^{a}_i(\tau)=\int dx ~ w_i(x)\frac{D_a(x,\tau)}{x}\;,
\end{eqnarray}
where $N^{a}_i$ is the wedge coefficient for the number of particle moment for the species $a=\{g,q_f,\bar{q}_f\}$ and $w_i(x)$ is the wedge function defined as
\begin{eqnarray}
    w_i(x) = 
    \begin{cases}
        \frac{x-x_{i-1}}{x_{i}-x_{i-1}}\;, & x_{i-1} < x < x_{i} \\
        \frac{x_{i+1}-x}{x_{i+1}-x_{i}}\;, & x_{i} \leq x < x_{i+1} \\
        0\;, & x > x_{i+1} \text{~or~} x< x_{i-1}
    \end{cases}
\end{eqnarray}
with $\{x_i\}$ the discrete node points spanning the region of interest ($\sim [0,2]$). We note that the wedge functions display the following summation properties
\begin{eqnarray}
    \sum_i w_i(x) = \Theta(x_{\rm max}-x)\Theta(x-x_{\rm min})\;, \label{eq:Completenes1}\\
    \sum_i x_i w_i(x) = x\Theta(x_{\rm max}-x)\Theta(x-x_{\rm min})\;.\label{eq:Completenes2}
\end{eqnarray}
By use of these properties one finds simple relations for the number of particles and energy 
\begin{eqnarray}
    n_a(\tau) = \sum_i N^a_i(\tau) \;,\quad
    E_a(\tau) = \sum_i x_i N^a_i(\tau)\;,
\end{eqnarray}
allowing us to keep track of these quantities up to machine precision.

The collision integral is expanded in the same basis
\begin{eqnarray}
    C^{a}_i(\tau) = \int dx ~ w_i(x) \frac{C_a(x,\tau)}{x}\;. \label{eq:CollisionExpansion}
\end{eqnarray}
Because the collision integral $C_a(x,\tau)$ is linear in terms of the distribution of each species one can write $C^{a}_i(\tau)$ as a matrix vector product
\begin{eqnarray}
    C^a_i(\tau) = \sum_{j} \delta C^{ab}_{ij} N^b_j(\tau)
    &=& \delta \bar{C}^{ab}\vec{N}^b(\tau)\;,
\end{eqnarray}
by constructing the vector $\vec{N}$ and matrix $\bar{C}$ from the coefficients and matrices of the different species 
\begin{eqnarray}
    \vec{N}(\tau) &\equiv& 
    \begin{pmatrix}
        \vec{N}^g(\tau)\\
        \vec{N}^{q_f}(\tau)\\
        \vec{N}^{\bar{q}_f}(\tau)
    \end{pmatrix}\;, 
    \quad
    \bar{C}(\tau) \equiv \frac{\delta C^a_{i}(\tau)}{\delta N^b_{j}(\tau)} =
    \begin{pmatrix}
        \bar{C}_{gg}(\tau) & \bar{C}_{gq_{f}}(\tau) & \bar{C}_{g\bar{q}_f}(\tau)\\
        \bar{C}_{q_f g}(\tau)& \bar{C}_{q_f q_{f}}(\tau) & \bar{C}_{q_f \bar{q}_f}(\tau)\\
        \bar{C}_{\bar{q}_f g}(\tau)& \bar{C}_{\bar{q}_fq_{f}}(\tau) & \bar{C}_{\bar{q}_f\bar{q}_f}(\tau)
    \end{pmatrix}
    \;.
\end{eqnarray}
where $\bar{C}_{ab}(\tau)$ are matrices that characterize the contribution of the distribution of species $b$ to the collision integral of species $a$.
Although $\bar{C}(\tau)$ will not depend on $N^b_i(\tau)$ because $ C_a(\tau)$ is linear in $N^b_i(\tau)$, we will still keep track of $N^b_j(\tau)$ when we write the matrices in the following sections.

In order to recover the full distribution from the discrete values $N^a_i(\tau)$, we approximate the coefficient integral by taking $\frac{D_a(x,\tau)}{D_a^{\rm eq}(x) e^{xE/T}}$ to be constant between node points 
\begin{eqnarray}
    N^{a}_i(\tau)&=&\int dx ~ w_i(x)\frac{D_a(x,\tau)}{x~D_a^{\rm eq}(x) e^{xE/T}}D_a^{\rm eq}(x) e^{xE/T}\;,\\
    &=&\frac{D_a(x,\tau)}{D_a^{\rm eq}(x) e^{xE/T}}\mathcal{A}^{a}_{i}\;,
\end{eqnarray}
where $\mathcal{A}^{a}_{i}$ is the area defined as 
\begin{eqnarray}
    \mathcal{A}^{a}_{i} \equiv \int \frac{dx}{x} ~ w_i(x) D_a^{\rm eq}(x) e^{xE/T}\;.
\end{eqnarray}
We now can write the value of the distribution at the node points, and using a linear interpolation, one can write the full distribution as
\begin{eqnarray}
    D_a(x,\tau)&=& \sum_i w_i(x)N^{a}_{i}(\tau) \frac{D_a^{\rm eq}(x) e^{xE/T}}{\mathcal{A}^{a}_{i}}\;,\\
    &=& \sum_i x K_i(x)N^{a}_{i}(\tau)\;,
    \label{eq:DistributionInterpolation}
\end{eqnarray}
we introduced the “cardinal” function $K^{a}_i(x)\equiv  w_i(x)\frac{D_a^{\rm eq}(x) e^{xE/T}}{x\mathcal{A}^{a}_{i}}$ in the last line.

Lastly, as the basis function is constant in time, the evolution of the coefficients $\vec{N}(\tau)$ are obtained directly from the discrete collision integral as follows
\begin{eqnarray}
\label{eq:EvolutionOfVector}
    \partial_\tau \vec{N}(\tau) &=& \bar{C} \vec{N}(\tau)\;,
\end{eqnarray}
which admits the solution
\begin{eqnarray}
    \vec{N}(\tau) &=& e^{\tau \bar{C}}\vec{N}(\tau=0)\;,
\end{eqnarray}
where $e^{\tau \bar{C}}$ is a matrix exponentiation. 

The integration in Eq.~(\ref{eq:CollisionExpansion}) is done numerically using a Monte Carlo integration scheme, where at each step we update all elements of the matrix $C_{ij}$ which insures charge and energy conservation thanks to Eq.~(\ref{eq:Completenes1}) and (\ref{eq:Completenes2}). Writing the collision integral as a matrix also allows us to compute the eigenvalues and eigenfunctions discussed in Section \ref{sec:Equilibration}.
In the following sections we will provide the different matrices corresponding to each process from Section \ref{sec:Kinetic-Equation}.

\subsection{Discretization of small angle elastic collision integrals}
It is straight forward to write the hard part of the current term in Eqns.~(\ref{eq:JHardG}-\ref{eq:JHardB}) using this discretization formalism. One only needs to introduce the wedge function integration and replace the distribution by its discrete form, we obtain
\begin{eqnarray}
    -\nabla_\p\mathcal{J}^{g}_{ij}&=&N^g_j(\tau) C_A\frac{\hat q_{\rm eq}}{4T^2}\int dx~ w_i(x) \Bigg[\frac{T^2}{E^2}(\partial_x x^2\partial_x) \frac{K_j(x)}{x^2} 
     \frac{T}{E} (\partial_x x^2) \frac{K_j(x)}{x^2}(1\pm 2n_a(xE)) \Bigg]\;, \nonumber\\\\
    -\nabla_\p\mathcal{J}^{q_f}_{ij}&=&N^{q_f}_j(\tau) C_F\frac{\hat q_{\rm eq}}{4T^2}\int dx~ w_i(x) \Bigg[\frac{T^2}{E^2}(\partial_x x^2\partial_x) \frac{K_j(x)}{x^2} 
    + \frac{T}{E} (\partial_x x^2) \frac{K_j(x)}{x^2}(1\pm 2n_a(xE)) \Bigg]\;,\nonumber\\\\
    -\nabla_\p\mathcal{J}^{\bar{q}_f }_{ij}&=&N^{\bar{q}_f}_j(\tau) C_F\frac{\hat q_{\rm eq}}{4T^2}\int dx~ w_i(x) \Bigg[\frac{T^2}{E^2}(\partial_x x^2\partial_x) \frac{K_j(x)}{x^2} 
    + \frac{T}{E} (\partial_x x^2) \frac{K_j(x)}{x^2}(1\pm 2n_a(xE)) \Bigg]\;,\nonumber\\
\end{eqnarray}
for the quark/antiquark to ensure stability at the boundaries we employ an integration by parts, inspired by the “weak” form of differential equations \cite{boyd2001chebyshev}, and set the term fully integrated to zero according to the boundary conditions.

Using the same approach the recoil contribution in Eqns.~(\ref{eq:recoil1}-\ref{eq:recoil2}) are given by
\begin{eqnarray}
    -\nabla_\p\delta \mathcal J^g_{ij}
    &=&\frac{\ca \hat{\bar q}_{\rm eq}}{4 T^2}  \frac{ T\delta\bar\eta_D^{j}-\delta \hat{\bar q}^{j}}{\hat{\bar q}_{\rm eq}} \frac{\nu_g}{2\pi^2}  \frac{T}{E}
    \int \frac{dx}{2\pi^2} w_i(x) \partial_x x^2 n_a(xE)(1\pm n_a(xE))\;,\nonumber\\\\
    -\nabla_\p\delta \mathcal J^{q_f/\bar{q}_f}_{ij}
    &=&\frac{\cf \hat{\bar q}_{\rm eq}}{4 T^2}  \frac{ T\delta\bar\eta_D^{j}-\delta \hat{\bar q}^{j}}{\hat{\bar q}_{\rm eq}} \frac{\nu_g}{2\pi^2}  \frac{T}{E}
    \int \frac{dx}{2\pi^2} w_i(x) \partial_x x^2 n_a(xE)(1\pm n_a(xE))\;,\nonumber\\
\end{eqnarray}
where the recoil coefficients are now represented by vectors written as 
\begin{eqnarray}
    \delta \hat{\bar q}^{j}&=& \frac{g^4}{\pi} E^3 \int  dx~ \Bigg[ \ca  \nu_{g}^{-1} N^g_j(\tau) K_j(x)(1+2n_B(xE)) \nonumber\\
    &&+ \frac{1}{2} \sum_{f} \nu_{q}^{-1} (N^{q_f}_j(\tau) K_j(x)+N^{\bar{q}_f}_j(\tau) K_j(x))(1-2n_F(xE)) \Bigg]\;,\nonumber\\\\
    \delta\bar\eta_D^{j}&=&  \frac{g^4}{\pi} E^2 \int dx~  \frac{2}{x}\Big[ \ca  \nu_{g}^{-1} N^g_j(\tau) K_j(x) +\frac{1}{2} \sum_{f} \nu_{q}^{-1} (N^{q_f}_j(\tau) K_j(x)+N^{\bar{q}_f}_j(\tau) K_j(x)) \Big]\;.\nonumber\\
\end{eqnarray}
Similarly for the conversion term from Eqns.~(\ref{eq:ConversionGluon},\ref{eq:ConversionQuark}) we obtain
\begin{eqnarray}
    S^g_{ij} &=&\nu_{g} \frac{\mathcal{I}_{q_f}^{\rm eq}}{8T^2} \int dx~\frac{T}{xE}  
    w_i(x) \sum_{f} 
    \left\{  \nu_{q}^{-1}\Big[N^{q_f}_j(\tau) K_j(x)+N^{\bar{q}_f}_j(\tau) K_j(x)\Big](1+2n_B(xE)) \right. \nonumber\\
    && \qquad\qquad\qquad\quad\left.-  2\nu_{g}^{-1} N^g_j(\tau) K_j(x) (1-2n_F(xE))\right\}\;, 
    \\
    S^{q_f,\bar{q}_f}_{ij} &=&\nu_{g} \frac{\mathcal{I}_{q_f}^{\rm eq}}{8T^2} \int dx~\frac{T}{xE}
    w_i(x)
    \left\{  \nu_{g}^{-1}N^g_j(\tau) K_j(x) (1-2n_F(xE)) \right. \nonumber\\
    && \qquad\qquad\qquad\quad \left.-\nu_{q}^{-1} N^{q_f,\bar{q}_f}_j(\tau) K_j(x)(x)(1+2n_B(xE)) \right\}\;,
    \nonumber\\
\end{eqnarray}
and the corresponding recoil contributions are given by
\begin{eqnarray}
    \delta S^{q_f}_{ij} &=& \frac{\nu_{g}}{2\pi^2} \int dx~ w_i(x) \frac{x}{8E}\left(\delta\mathcal{I}^{q_f}_j-\delta\mathcal{I}^{\bar{q}_f}_j\right) ~n_B(xE)(1-n_F(xE)) \;,\\
    \delta S^{\bar{q}_f}_{ij} &=&\frac{\nu_{g}}{2\pi^2} \int dx~ w_i(x)\frac{x}{8E} \left(\delta\mathcal{I}_j^{\bar{q}_f}-\delta\mathcal{I}_j^{q_f}\right)~n_B(xE)(1-n_F(xE)) \;,
\end{eqnarray}
where the difference of $\delta\mathcal{I}^{q_f}_j$ and $\delta\mathcal{I}^{\bar{q}_f}_j$ is given by
\begin{eqnarray}
    \left(\delta\mathcal{I}_j^{\bar{q}_f}-\delta\mathcal{I}_j^{q_f}\right) = \frac{g^4 C_{F} \mathcal{L}}{\pi} E^2\int~dx~\frac{1}{ x}  \left(1 + 2n_B(xE)\right) \nu_q^{-1}\left(N^{q_f}_j(\tau) K_j(x)-N^{\bar{q}_f}_j(\tau) K_j(x) \right) \;.\nonumber\\
\end{eqnarray}
\subsection{Discretization of inelastic collision integrals}
Before discretizing the radiative collision integrals, we will combine both merging and splitting processes in Eqns.~(\ref{eq:Inelastic1st}-\ref{eq:Inelasticlast}) by introducing a delta function. Applying this to a general $1\leftrightarrow2$ processes, one finds
\begin{eqnarray}
    C^{1\leftrightarrow 2}_a&=&-\frac{1}{2}\int_0^1 dz \frac{d\Gamma^a_{bc}(xE,z)}{dz} \Big[  D_a(x)(1\pm n_b(zxE)\pm n_c(\bar z xE))\nonumber\\
    && \pm \frac{D_b(zx)}{z^3}( n_a(xE) \mp_b n_c(\bar z xE))\pm\frac{D_c(\bar{z}x)}{\bar{z}^3}( n_a(xE) \mp_c n_b(zxE)) \Big] \nonumber\\
    &&+\frac{\nu_b}{\nu_a}\int_0^1 dz~\frac{1}{z} \frac{d\Gamma^b_{ac}(\tfrac{xE}{z},z)}{dz} 
    \Big[ D_b(\frac{x}{z})(1\pm n_a(xE)\pm n_c(\frac{\bar z}{z}xE))
    \nonumber\\
    &&\pm \frac{D_a(x)}{z^3}( n_b(\frac{xE}{z}) \mp_a n_c(\frac{\bar z}{z} xE)) \pm \frac{D_a(x)}{\bar{z}^3}( n_b(\frac{xE}{z}) \mp_c n_a(\frac{xE}{z}))\Big]\\
    &=&\int dy~\int_0^1 dz \frac{d\Gamma^e_{kc}(yE,z)}{dz} \Big[  D_e(y)(1\pm n_k(zyE)\pm n_c(\bar z yE)) \nonumber\\
    &&\pm \frac{D_k(zy)}{z^3}( n_e(yE) \mp_k n_c(\bar z yE)) \pm\frac{D_c(\bar{z}y)}{\bar{z}^3}( n_e(yE) \mp_c n_k(zyE)) \Big]\nonumber\\
    &&
    \times\Bigg[ \frac{\nu_b}{\nu_a}
    z\delta(x-zy)\delta^{e,b}_{k,a}-\frac{1}{2}\delta(x-y)\delta^{e,a}_{k,b} \Bigg]\;, \nonumber\\
\end{eqnarray}
where we used $\mp_a$ to represent either a minus if particle $a$ is Boson or a plus if particle $a$ is a fermion. After employing the discretization scheme, one finds for the gluon collision integrals

\begin{eqnarray}
    C_{g,ij}^{g\leftrightarrow gg}&= & 
    \frac{1}{2} \int dx~ \int_0^1 dz  \frac{d\Gamma^g_{gg}(xE,z)}{dz}\Bigg[ N^g_j(\tau) K_j(x)(1+ n_B(zxE)+n_B(\bar z xE))
    \nonumber\\
    &&+\frac{N^g_j(\tau) K_j(zx)}{ z^2}(n_B(xE)-n_B(\bar zxE))+\frac{N^g_j(\tau) K_j(\bar{z}x)}{\bar z^2}(n_B(xE)-n_B(zxE)) \Bigg]\nonumber\\
    &&\times\left[ 
    w_i(\bar{z}x)+
    w_i(zx)-w_i(x) \right]\;, 
\end{eqnarray}
\begin{eqnarray}
    C_{g,ij}^{q\leftrightarrow qg}&=& \sum_{f} \int_0^1 dz  \frac{d\Gamma^{q}_{gq}\left(\frac{xE}{z}\;,z\right)}{dz}\Bigg[ N^{q_f}_j(\tau) K_j(x)\left(1+ n_B(zxE)- n_F(\bar zxE)\right)
\nonumber\\
    &&+ \frac{\nu_q }{\nu_g }\frac{N^g_j(\tau) K_j(zx)}{z^2} ( n_F(xE) - n_F(\bar{z}xE))-\frac{N^{q_f}_j(\tau) K_j(\bar{z}x)}{ \bar z^2}( n_F(xE) + n_B(z xE)) \Bigg]\nonumber\\
    &&\times w_i(zx)\;, 
\end{eqnarray}
\begin{eqnarray}
    C_{g,ij}^{\bar{q}\leftrightarrow \bar qg} &=&   \sum_{f} \int_0^1 dz  \frac{d\Gamma^{q}_{gq}\left(\frac{xE}{z}\;,z\right)}{dz}\Bigg[ N^{\bar{q}_f}_j(\tau) K_j(x)\left(1+ n_B(zxE)- n_F(\bar zxE)\right)
    \nonumber\\
        &&+ \frac{\nu_q }{\nu_g }\frac{N^g_j(\tau) K_j(zx)}{z^2} ( n_F(xE) - n_F(\bar{z}xE))-\frac{N^{\bar{q}_f}_j(\tau) K_j(\bar{z}x)}{ \bar z^2}( n_F(xE) + n_B(z xE)) \Bigg]\nonumber\\
        &&\times w_i(zx)\;, 
\end{eqnarray}
\begin{eqnarray}
    &&C_{g,ij}^{g\leftrightarrow q\bar q}= -\sum_{f}\int_0^1 dz  \frac{d\Gamma^g_{q\bar q}(xE,z)}{dz}\Bigg[  N^g_j(\tau) K_j(x)(1- n_F(zxE)- n_F(\bar z xE))
    \nonumber\\
    &&- \frac{\nu_g}{\nu_q}\frac{N^{q_f}_j(\tau) K_j(zx)}{ z^2}( n_B(xE) + n_F(\bar z xE))-\frac{\nu_g}{\nu_q}\frac{N^{\bar{q}_f}_j(\tau) K_j(\bar{z}x)}{ \bar z^2}( n_B(xE) + n_F(zxE)) \Bigg] \nonumber\\
    &&\times w_i(x)\;,
\end{eqnarray}
where for the symmetric $g\leftrightarrow gg$ and $g\leftrightarrow q\bar{q}$ processes, we symmetrized the integrand by change of variable $z\rightarrow \bar{z}$. 

Similarly, the quark collision integrals are given by 
\begin{eqnarray}
    C_{q_f,ij}^{q\leftrightarrow qg}&=&
    \sum_{f} \int_0^1 dz  \frac{d\Gamma^{q}_{gq}\left(\frac{xE}{z}\;,z\right)}{dz}\Bigg[ N^{q_f}_j(\tau) K_j(x)\left(1+ n_B(zxE)- n_F(\bar zxE)\right)
    \nonumber\\
    &&+ \frac{\nu_q }{\nu_g }\frac{N^g_j(\tau) K_j(zx)}{z^2} ( n_F(xE) - n_F(\bar{z}xE))-\frac{N^{q_f}_j(\tau) K_j(\bar{z}x)}{ \bar z^2}( n_F(xE) + n_B(z xE)) \Bigg]\nonumber\\
    &&\times \left[w_i(\bar{z}x)-w_i(x)\right]\;, 
\end{eqnarray} 
\begin{eqnarray}
    C_{q_f,ij}^{g\leftrightarrow q\bar q}= &\sum_{f}\int_0^1 dz  \frac{d\Gamma^g_{q\bar q}(xE,z)}{dz}\Bigg[  N^g_j(\tau) K_j(x)(1- n_F(zxE)- n_F(\bar z xE))
    \nonumber \qquad \qquad \qquad \qquad&
    \\
    &- \frac{\nu_g}{\nu_q}\frac{N^{q_f}_j(\tau) K_j(zx)}{ z^2}( n_B(xE) + n_F(\bar z xE))-\frac{\nu_g}{\nu_q}\frac{N^{\bar{q}_f}_j(\tau) K_j(\bar{z}x)}{ \bar z^2}( n_B(xE) + n_F(zxE)) \Bigg] \nonumber 
    &
    \\
    &\times  w_i(zx)\;.  \qquad \qquad \qquad \qquad \qquad \qquad \qquad \qquad \qquad \qquad \qquad \qquad \qquad \qquad &
\end{eqnarray}
For the antiquark channel the $\bar{q}\leftrightarrow \bar qg$ process is the same as the quark by exchange of $q_f$ with $\bar{q}_f$, while the $g\leftrightarrow q\bar q$ process is given by
\begin{eqnarray}
    C_{\bar{q}_f,ij}^{g\leftrightarrow q\bar q}=&\sum_{f}\int_0^1 dz  \frac{d\Gamma^g_{q\bar q}(xE,z)}{dz}\Bigg[  N^g_j(\tau) K_j(x)(1- n_F(zxE)- n_F(\bar z xE))
    \nonumber  \qquad \qquad \qquad \qquad &\\
    &- \frac{\nu_g}{\nu_q}\frac{N^{q_f}_j(\tau) K_j(zx)}{ z^2}( n_B(xE) + n_F(\bar z xE))-\frac{\nu_g}{\nu_q}\frac{N^{\bar{q}_f}_j(\tau) K_j(\bar{z}x)}{ \bar z^2}( n_B(xE) + n_F(zxE)) \Bigg] \nonumber &
    \\
    &\times  w_i(\bar{z}x)\;. \qquad \qquad \qquad \qquad \qquad \qquad \qquad \qquad \qquad \qquad \qquad \qquad \qquad \qquad &
\end{eqnarray}
Using the properties of the wedge function one can easily find that charge is conserved because 
\begin{eqnarray}
    \sum_i C_{q_f,ij}^{g\leftrightarrow q\bar q}- C_{\bar{q}_f,ij}^{g\leftrightarrow q\bar q}&\propto& \sum_i w_i(zx) -w_i(\bar{z}x) = 0\;, \\
    \sum_i C_{(q_f/\bar{q}_f),ij}^{(q/\bar{q}_f)\leftrightarrow (q/\bar{q}_f) g} &\propto& \sum_i w_i(\bar{z}x) - w_i(x)  = 0 \;,
\end{eqnarray}
and analogously for energy conservation, we have
\begin{eqnarray}
    \sum_i x_i [w_i(zx) + w_i(\bar{z}x)-w_i(x) ] &=& 0\;,
\end{eqnarray}
for all allowed configurations of the splitting.
\subsubsection{Inelastic effective rate}\label{ap:Inelastic}
As described in section \ref{sec:CollinearRadiation}, jet particles undergo multiple soft scattering giving rise to medium induced radiation. One also have to take into account the Landau-Pomeranchuk-Migdal (LPM) effect \cite{migdal1956bremsstrahlung} caused by interference between mean free time and the radiation formation time. Following P. Arnold \cite{Arnold:2008iy}, the infinitely many diagrams can be resummed into an effective rate written as
\begin{eqnarray}
    \label{eq:RateDefinition}
    \frac{d\Gamma^a_{bc}(p,z)}{dz} =  \frac{\alpha_s P_{ab}(z)}{[2pz(1-z)]^2} \int \frac{d^2 p_b}{(2\pi)^2} ~\text{Re}\left[ 2 \mathbf{p}_b \cdot \mathbf{g}_{(z,P)}( \mathbf{p}_b) \right]\;,
\end{eqnarray}
where $P_{ab}(z)$ are the Dokshitzer-Gribov-Lipatov-Altarelli-Parisi (DGLAP) splitting functions
\begin{eqnarray}
    P_{gg}(z) &=& 2C_A \frac{[1-z(1-z)]^2}{z(1-z)}\;,\quad
    P_{qg}(z)  = C_F \frac{1+(1-z)^2}{z}\;,\quad
    P_{gq}(z) = \frac{1}{2} \left(z^2+(1-z)^2\right) \;. \nonumber\\
\end{eqnarray}
$\mathbf{g}_{(z,P)}( p_b) $ satisfies the following integral equation
\begin{eqnarray}
    \label{eq:AMY}
    2\mathbf{p}_b &=& i \delta E(z,P,\mathbf{p}_b) \mathbf{g}_{(z,P)}(\mathbf{p}_b) + \int \frac{d^2q}{(2\pi)^2}~\frac{d\bar{\Gamma}^{\rm el}}{d^2q}~\left\{ C_{1} \left[ \mathbf{g}_{(z,P)}(\mathbf{p}_b) - \mathbf{g}_{(z,P)}(\mathbf{p}_b -\mathbf{q}) \right] +  \right. \nonumber \\
    && \left. C_{z} \left[ \mathbf{g}_{(z,P)}(\mathbf{p}_b) - \mathbf{g}_{(z,P)}(\mathbf{p}_b -z\mathbf{q}) \right]  + C_{1-z} \left[ \mathbf{g}_{(z,P)}(\mathbf{p}_b) - \mathbf{g}_{(z,P)}(\mathbf{p}_b -(1-z)\mathbf{q}) \right] \right\}\;, \nonumber\\
\end{eqnarray}
with $\frac{d\bar{\Gamma}^{\rm el}}{d^2q}$ corresponding to the elastic broadening kernel, which at leading order of perturbation is written
\begin{eqnarray}
    \frac{d\bar{\Gamma}^{\rm el}}{d^2q} = \frac{m_D^2}{q^2(q^2+m_D^2)}\;.
\end{eqnarray}
The color factors are written as
\begin{eqnarray}
    &C_1 = \tfrac{1}{2} \left( C_{z}^{R} + C_{1-z}^{R} - C_{1}^{R} \right)\;,\quad
    C_z  =  \tfrac{1}{2} \left( C_{1-z}^{R} + C_{1}^{R} - C_{z}^{R} \right)\;,& \nonumber\\
    & C_{1-z} = \tfrac{1}{2} \left( C_{1}^{R} + C_{z}^{R} - C_{1-z}^{R} \right)\;,&
\end{eqnarray}
here $C_{1,z,1-z}^{R}$ corresponds to the Casimir of the representation with momentum fraction $1,z,1-z$. 
The energy $\delta E(z,P,\mathbf{p}_b)$ is defined by
\begin{eqnarray}
    \delta E(z,P,\mathbf{p}_b) &=& \frac{\mathbf{p}_b^2}{2Pz(1-z)}+\frac{m^2_{\infty.(z)}}{2zP}+\frac{m^2_{\infty.(1-z)}}{2(1-z)P}-\frac{m^2_{\infty.(1)}}{2P}\;,
\end{eqnarray}
where the thermal masses for a plasma in equilibrium are given by 
\begin{align}
    m_{\infty,g} =& \frac{g^2}{d_A} \int \frac{d^3p}{(2\pi)^3}~ \frac{1}{p} \left[\nu_g C_A n_B(p) + 2N_f \nu_g C_F n_F(p) \right] 
    = \frac{g^2 T^2}{2} \left( \frac{N_c}{3} +\frac{N_f}{6}\right) \;,\\
    m_{\infty,q} =& m_{\infty,\bar{q}} = g^2 C_F\int \frac{d^3p}{(2\pi)^3}~ \frac{1}{p} \left[ 2n_B(p) + 2N_f n_F(p) \right] = g^2T^2 \frac{C_F}{4} \;.
\end{align}

We Fourier transform Eq.~(\ref{eq:AMY}) to impact parameter space, turning the integral equation into a differential equation, which we solve numerically following a refined version of the algorithm in~\cite{1012.3784v2}.
\subsubsection{Comparison to leading-log approximation}\label{ap:FittingHOtoTheFullRate}
\begin{figure}[t!]
    \centering
    \includegraphics[width=\textwidth]{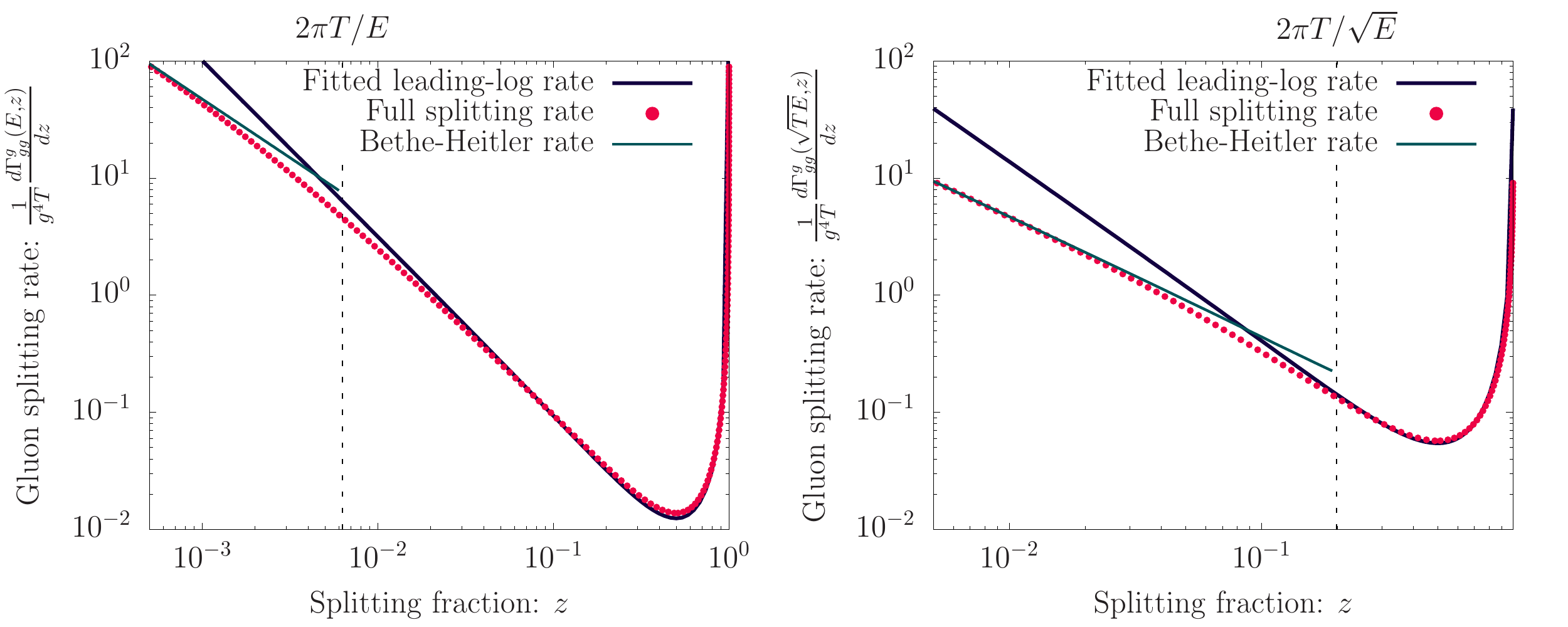}
    \caption{Comparison of the matching of the leading-log splitting rate (blue line) to the full effective rate (red dot) for $g\leftrightarrow gg$ process. We also show the Bethe-Heitler rate (green) relevant for soft radiation. On the left panel we show for a parent particle with energy $E=1000T$, and on the right panel for parent particle with energy $E=\sqrt{1000}T$.}
    \label{fig:qhatFit}
\end{figure}
For highly energetic parent particles the radiation rate in Eq.~(\ref{eq:RateDefinition}) is in the deep LPM regime which can be approximated by the Harmonic Oscillator (HO) rate \cite{Arnold:2008iy}. In order to match the HO rate, one has to choose a sensible value of $\hat{\bar{q}}$. For the early time behavior in Eqns.~(\ref{eq:SingleEmissionHOApprox}-\ref{eq:SingleEmissionHOApprox2}) we consider the parent particle to be of energy $E$ and fit $\hat{\bar{q}}(E)$ to match, as shown in the left panel of Fig.~\ref{fig:qhatFit}. In the same figure we show the rate in the Bethe-Heitler regime \cite{Schlichting:2019abc} which describes the splitting to soft fragments, one can see clearly how the full splitting rate interpolate between the leading-log rate for high energetic fragments and the BH regime for soft fragments. 
Conversely, for the successive branchings in Eqns.~(\ref{eq:TildGammag}-\ref{eq:TildGammaq}) we approximate the parent particle energy by the geometric mean between the jet energy $E$ and the temperature $T$ and fit the value of $\hat{\bar{q}}(\sqrt{TE})$ as shown in right panel of Fig.~\ref{fig:qhatFit}.

\section{Small-angle approximation}\label{ap:SmallAngle}

In this appendix we shall explain how one finds the diffusion approximation to the elastic $2\leftrightarrow 2$ QCD scatterings. We start from the following collision integral \cite{Ghiglieri:2015ala}
\begin{align}
   C_a[f]
   = & \frac 1{2|p_1|\nu_a} \sum_{bcd} \int d\Omega^{2\leftrightarrow2}
       \left|{\cal M}^{ab}_{cd}(\p_1,\p_2;\p_3,\p_4)\strut\right|^2 \mathcal{F}(\p_1,\p_2;\p_3,\p_4) \;,
\end{align}
where ${\cal M}^{ab}_{cd}(p_1,p_2;p_3,p_4)$ is the QCD matrix element and $\mathcal{F}(p_1,p_2;p_3,p_4)$ is the statistical factor given by 
\begin{align}
    \mathcal{F}(\p_1,\p_2,\p_3,\p_4) =& f_c(\p_3) \, f_d(\p_4) \, (1{\pm} f_a(\p_1)) \, (1{\pm} f_b(\p_2)) \nonumber \\
    &- f_a(\p_1) \, f_b(\p_2) \, (1{\pm} f_c(\p_3)) \, (1{\pm} f_d(\p_4))\;. 
\end{align}

\subsection{Phase-space parametrization}
We define the phase-space measure for $2\leftrightarrow2$ scatterings as follows
\begin{align}
    \int d\Omega^{2\leftrightarrow2} \equiv \int \norm{p_2}~\frac{1}{2E_2}\int\norm{p_3}~\frac{1}{2E_3}\int\norm{p_4}~\frac{1}{2E_4}(2\pi)^4 \delta^{(4)}(P_1+P_2-P_3-P_4)\;.
\end{align}
Following the parametrization in \cite{hep-ph/0302165v2}, one can use the 3 dimensional integral to apply the momentum delta function defining $q \equiv p_1-p_3 = p_4-p_2$. While the energy delta function left is cast into two delta functions, by introducing an integration over $\omega$ representing the energy exchange. The phase-space measure becomes 
\begin{align}
    \int d\Omega^{2\leftrightarrow2}=&(2\pi)\int \norm{p_2}\int \norm{q} \int d\omega ~
    \frac{1}{8p_1p_2^2q^2}\Theta(q-|\omega|)\Theta(p_1-\frac{q+\omega}{2})\Theta(p_2-\frac{q-\omega}{2})\nonumber\\
   &\delta\left(cos\theta_{1q}-\left( \frac{\omega}{q}-\frac{\omega^2-q^2}{2p_1q} \right)\right)
   \delta\left(cos\theta_{2q}-\left( \frac{\omega}{q}+\frac{\omega^2-q^2}{2p_2q} \right)\right)\;.
\end{align}
Within this parametrization the Mandelstam variables are given by
\begin{eqnarray}
    t &=& \omega^2 -q^2\;,
    \quad
    s = -2p_1p_2(1-\cos\theta_{12})\;,
    \quad
    u = -t-s\;,
\end{eqnarray}
where $\theta_{12}$ is the angle between $\p_1$ and $\p_2$. Not that here a t-channel parametrization has been used, while the u-channel can obtained by exchanging $\p_3 \leftrightarrow \p_4$ momentum and the s-channel diagrams are neglected because of the divergent nature of the t- and u-channels for small momentum exchange.

To perform the 3 dimensional $q$ integration we write the different component in the following orthonormal basis 
\begin{eqnarray}
    \begin{pmatrix}
        \vec{e}_+\\
        \vec{e}_-\\
        \vec{e}_3\\
    \end{pmatrix}
    =
    \begin{pmatrix}
        \vspace{0.15cm}
        \frac{\vec{e}_1+\vec{e}_2}{\sqrt{2+2\cos\theta_{12}}}\\
        \vspace{0.15cm}
        \frac{\vec{e}_1-\vec{e}_2}{\sqrt{2-2\cos\theta_{12}}}\\
        \frac{\vec{e}_1\times\vec{e}_2}{|\vec{e}_1\times\vec{e}_2|}\\
    \end{pmatrix}
    \;,
    \quad
    \q \equiv 
    \begin{pmatrix}
        q_+\\
        q_-\\
        q_\perp
    \end{pmatrix}
    =
    \begin{pmatrix}
        \vspace{0.15cm}
        \frac{q(\cos\theta_{1q}+\cos\theta_{2q})}{\sqrt{2+2\cos\theta_{12}}}\\
        \vspace{0.15cm}
        \frac{q(\cos\theta_{1q}-\cos\theta_{2q})}{\sqrt{2-2\cos\theta_{12}}}\\
        \pm\sqrt{q^2-q_1^2-q_2^2}\\
    \end{pmatrix}
    \;,
\end{eqnarray}
where $\theta_{1q}/\theta_{2q}$ are the angles between $\p_1/\p_2$ and $\q$. We perform a change of integration variables from $(q_+,q_-,q_\perp)$ to $(\cos\theta_{1q},\cos\theta_{2q},q)$, and combine the range of $q_\perp$ integration as follows 
\begin{eqnarray}
    \int_{-q_{\rm max}}^{q_{\rm max}} dq_\perp~ f(q_\perp)  &=&  \int_0^{q_{\rm max}} dq_\perp~ f(|q_\perp|) + f(-|q_\perp|)\;,
\end{eqnarray}
We then use the two delta functions to perform two integrations obtaining
\begin{align}
    \int d\Omega^{2\leftrightarrow2}=&2(2\pi)\int \norm{p_2}\int \frac{dq}{(2\pi)^3} \int d\omega ~
    \frac{1}{8p_1p_2^2q^2}
    \frac{q^3}{|q_\perp|\sqrt{1-\cos^2\theta_{12}}}
    \Theta \left( 1-\left|\frac{q_+}{q}\right| \right)\Theta \left( 1-\left|\frac{q_-}{q}\right| \right)\;,
\end{align}
for $q,w \ll p_1,p_2$ we can neglect earlier $\Theta$ functions restraining $\p_1$ and $\p_2$ integrations. We also have symmetrized the integrand giving rise to a factor $2$ and canceling all odd integrands of $q_\perp$.

The component of vector $\q$ in the new parametrization are written as follows
\begin{eqnarray}
    q_+ = \frac{2\omega-\tfrac{\omega^2-q^2}{2} \left( \tfrac{1}{p_1}-\tfrac{1}{p_2} \right)}{\sqrt{2+2\cos\theta_{12}}}\;, 
    \quad
    q_- = -\frac{\tfrac{\omega^2-q^2}{2} \left( \tfrac{1}{p_1}+\tfrac{1}{p_2} \right)}{\sqrt{2-2\cos\theta_{12}}}\;.
\end{eqnarray}
Since the QCD matrix element favors small angle exchange we expand the different contributions to the integrand in power of $q$ and $\omega$ in the following sections, and we use the leading order of $q$ to perform the integral. 
\subsection{Expansion of statistical terms}
Before expanding the statistical term we note that the t-channel diagrams can be written either with interaction due to a gluon exchange giving rise to the current term in the Fokker-Planck equation these require $a$ and $c$ to be the same species and likewise for $b$ and $d$, which cancels the 0-th order in $q$ of the statistical term. The matrix elements for these diagrams are proportional to $\frac{s^2}{t^2}\propto q^{-4}$ necessitating expansion of the statistical term up to second order
\begin{align}
   &\mathcal{F}^{\rm Current}(\p_1,\p_2,\q) =q_i\left\{ 
    -f_a(p_2)(1\pm f_b(\p_2))\partial^i_{p_1} f_a(p_1)+f_a(\p_1)(1\pm f_a(\p_1))\partial^i_{p_2}f_b(\p_2)\right \}\nonumber &\\
    &+\frac{q_iq_j}{2}\left\{ 
   f_b(\p_2)(1\pm f_b(\p_2))\partial^i_{p_1}\partial^j_{p_1}f_a(\p_1)+f_a(\p_1)(1\pm f_a(\p_1))\partial^i_{p_2}\partial^j_{p_2}f_b(\p_2) \right.\nonumber &\\
    &-\partial^j_{p_2}f_b(\p_2)\partial^i_{p_1}f_a(\p_1)(1\pm f_a(\p_1))
    -\partial^j_{p_1}f_a(\p_1)\partial^i_{p_2}f_b(\p_2)(1\pm f_b(\p_2))  \Bigg \}\;.&
\end{align}
Whereas the diagrams where a quark exchange takes place, give rise to the conversion processes and require $a$ and $d$ to be the same species instead, likewise for $b$ and $c$. The matrix elements for these diagrams are proportional to $\frac{s}{t}\propto q^{-2}$ which only require to take the 0-th order expansion of the statistical term 
\begin{align}
    \mathcal{F}^{\rm Conversion}(\p_1,\p_2,\q) =& f_b(p_1) \, f_a(p_2) \, (1{\pm} f_a(p_1)) \, (1{\pm} f_b(p_2)) \nonumber \\
    &- f_a(p_1) \, f_b(p_2) \, (1{\pm} f_b(p_1)) \, (1{\pm} f_a(p_2))\;.
\end{align}
\subsection{Evaluation of small angle matrix elements}
By combining the statistical terms with the matrix element one finds for the current contributions
\begin{align}
    C_a^{\rm Current } =& 2(2\pi)\int \norm{p_2}
    B_i\left\{ 
    -f_a(p_2)(1\pm f_b(\p_2))\partial^i_{p_1} f_a(p_1)+f_a(\p_1)(1\pm f_a(\p_1))\partial^i_{p_2}f_b(\p_2)\right \}\nonumber\\
    &+\frac{B_{ij}}{2}\left\{ 
    f_b(\p_2)(1\pm f_b(\p_2))\partial^i_{p_1}\partial^j_{p_1}f_a(\p_1)+f_a(\p_1)(1\pm  f_a(\p_1))\partial^i_{p_2}\partial^j_{p_2}f_b(\p_2) \right.\nonumber\\
    &\qquad\qquad
    -\partial^j_{p_2}f_b(\p_2)\partial^i_{p_1}f_a(\p_1)(1\pm f_a(\p_1))
    -\partial^j_{p_1}f_a(\p_1)\partial^i_{p_2}f_b(\p_2)(1\pm f_b(\p_2))  \Bigg \}\;,
\end{align}
and the conversion contributions can be expressed as
\begin{align}
    C_a^{\rm Conversion } =& 2(2\pi)\int \norm{p_2}~ B
    \left\{ f_b(p_1) \, f_a(p_2) \, (1{\pm} f_a(p_1)) \, (1{\pm} f_b(p_2)) \right. \nonumber\\
    &\qquad\qquad\qquad\qquad \left.
    - f_a(p_1) \, f_b(p_2) \, (1{\pm} f_b(p_1)) \, (1{\pm} f_a(p_2))\right\}\;.
\end{align}
The above equations give rise to the three following integrals  
\begin{align}
    B \equiv& \int \frac{dq}{2\pi^2} \int_{-q_+}^{q_+} d\omega~
    \frac{1}{8p_1p_2^2q^2}\frac{q^3}{|q_\perp|\sqrt{1-\cos^2\theta_{12}}} \frac{s}{t}\;,\\
    B^i \equiv& \int \frac{dq}{2\pi^2} \int_{-q_+}^{q_+} d\omega~
    \frac{1}{8p_1p_2^2q^2}\frac{q^3}{|q_\perp|\sqrt{1-\cos^2\theta_{12}}} \, q^i \, \frac{s^2}{t^2}\;,\\
    B^{ij} \equiv& \int \frac{dq}{2\pi^2} \int_{-q_+}^{q_+} d\omega~
    \frac{1}{8p_1p_2^2q^2}\frac{q^3}{|q_\perp|\sqrt{1-\cos^2\theta_{12}}} \, \frac{q^i q^j}{2} \, \frac{s^2}{t^2}\;.\\
\end{align}
Taking the integrations we find at lowest order of q 
\begin{align}
    B = & \int \frac{dq}{q}~\frac{1}{8\pi p_2}\;,
    \quad
    B^i = \int \frac{dq}{q} ~\frac{p_1}{8\pi}\left( \frac{\vec{e}_1}{p_2} -\frac{\vec{e}_2}{p_1} \right) \;,\\
    B^{ij} = & \int \frac{dq}{q} ~ \frac{p_1}{16\pi} \left(\delta^{ij}(1-\cos\theta_{12})+\tfrac{\p_1^i}{p_1}\tfrac{\p_2^j}{p_2} + \tfrac{\p_1^j}{p_1}\tfrac{\p_2^i}{p_2} 
    \right)\;.
\end{align}
\subsection{Collision integrals in small angle approximation}
After combining the integrals with the statistical term, we obtain the different collision integrals. We define the current term in the gluon channel as 
\begin{eqnarray}
    C^{\rm Current}_g[f] & = & C^{gg\overset{g}{\longleftrightarrow} gg}_g[f]+\sum_f \left( C^{gq_f\overset{g}{\longleftrightarrow} gq_f}_g[f]+ C^{g\bar{q}_f\overset{g}{\longleftrightarrow} g\bar{q}_f}_g[f]\right)\;,
\end{eqnarray}
where we only take the gluon exchange contribution of the (anti-)quark/gluon scatterings denoted by $\overset{g}{\longleftrightarrow}$. The different collision integrals are written
\begin{eqnarray}
    C^{gg \overset{g}{\longleftrightarrow} gg}_g[f] & = & \frac{g^4 C_A \mathcal{L}}{4\pi}\,\partial^i_{\p_1}\int \norm{p_2} C_A f_g(\p_2)(1+f_g(\p_2)) \partial^i_{\p_1} f_g(\p_1) \nonumber \\
    &&\qquad\qquad\qquad\qquad
    + C_A\frac{2f_g(\p_2)}{p_2}\frac{\p^i_1}{p_1} f_g(\p_1)(1+f_g(\p_1))\;,\\
    C^{gq_f \overset{g}{\longleftrightarrow} gq_f}_g[f] & = & \frac{g^4 C_A\mathcal{L}}{4\pi}\,\partial^i_{\p_1}\int \norm{p_2} 
    f_{q_f}(\p_2)(1-f_{q_f}(\p_2)) \partial^i_{\p_1} f_g(\p_1) \nonumber \\
    &&\qquad\qquad\qquad\qquad
    + \frac{2f_{q_f}(\p_2)}{p_2}\frac{\p^i_1}{p_1} f_g(\p_1)(1+f_g(\p_1)) \;,\\
    C^{g\bar{q}_f \overset{g}{\longleftrightarrow} g\bar{q}_f}_g[f] & = & \frac{g^4 C_A\mathcal{L}}{4\pi}\,\partial^i_{\p_1}\int \norm{p_2} 
    f_{\bar{q}_f}(\p_2)(1-f_{\bar{q}_f}(\p_2)) \partial^i_{\p_1} f_g(\p_1) \nonumber \\
    &&\qquad\qquad\qquad\qquad
    + \frac{2f_{\bar{q}_f}(\p_2)}{p_2}\frac{\p^i_1}{p_1} f_g(\p_1)(1+f_g(\p_1)) 
    \;,
\end{eqnarray}
where we define the logarithmic enhancement $\mathcal{L}=\int_{m_D}^\mu \frac{dq}{q}$.

Conversely, the quark exchange contribution to the scatterings lead to the conversion terms
\begin{eqnarray} 
    C^{\rm Conversion}_g[f] & = & \sum_f \left( C^{gq_f \overset{q}{\longleftrightarrow} gq_f}_g[f] +C^{g\bar{q}_f \overset{q}{\longleftrightarrow} g\bar{q}_f}_g[f]+C^{gg \overset{q}{\longleftrightarrow} q_f\bar{q}_f}_g[f] \right)\;,\\
    & =&\frac{1}{8|\p|} \sum_f
    \left[f_{q_f}(\p_1)(1+f_g(\p_1))-f_g(\p_1)(1-f_{\bar{q}_f})\right]\mathcal{I}_{q_f} \nonumber\\
    &&\qquad\quad+\left[f_{\bar{q}_f}(\p_1)(1+f_g(\p_1))-f_g(\p_1)(1-f_{q_f})\right]\mathcal{I}_{\bar{q}_f}\;, 
\end{eqnarray}
where $\mathcal{I}_{q_f}$ and $\mathcal{I}_{\bar{q}_f}$ are given by the following moments of the phase-space distribution
\begin{eqnarray}
    \mathcal{I}_{q_f}&=& \frac{g^4C_{F} \mathcal{L}}{\pi}\int \frac{d^3k}{(2\pi)^3}\frac{1}{|\ke|}\Big[f_{q_f}(\ke)(1+f_g(\ke))+f_g(\ke)(1-f_{\bar{q}_f}(\ke))\Big]\;,\\
    \mathcal{I}_{\bar{q}_f}&=&\frac{g^4C_{F} \mathcal{L}}{\pi}\int \frac{d^3k}{(2\pi)^3}\frac{1}{|\ke|}\Big[f_{\bar {q}_f}(\ke)(1+f_g(\ke))+f_g(\ke)(1-f_{q_f}(\ke))\Big]\;.
    \end{eqnarray}
Similarly, for the quark channel using only the gluon exchange part of the scatterings we write the current term  
\begin{eqnarray} 
    C^{\rm Current}_{q_f}[f] & = & C^{q_fg \overset{g}{\longleftrightarrow} q_fg}_{q_f}[f]+\sum_i \left( C^{q_f q_i\longleftrightarrow q_f q_i}_{q_f}[f]+ C^{q_f\bar{q}_i\longleftrightarrow q_f\bar{q}_i}_{q_f}[f]\right)\;,\\
    C^{q_f g \overset{g}{\longleftrightarrow} q_f g}_{q_f}[f] & = & \frac{g^4 C_F\mathcal{L}}{4\pi}\,\partial^i_{\p_1}\int \norm{p_2} 
    C_A f_{g}(\p_2)(1+f_{g}(\p_2)) \partial^i_{\p_1} f_{q_f}(\p_1) \nonumber \\
    &&\qquad\qquad\qquad\qquad
    + C_A \frac{2f_{g}(\p_2)}{p_2}\frac{\p^i_1}{p_1} f_{q_f}(\p_1)(1+f_g(\p_1)) \;,\\
    C^{q_f q_i\longleftrightarrow q_f q_i}_{q_f}[f] & = & \frac{g^4 C_F \mathcal{L}}{8\pi}\,\partial^i_{\p_1}\int \norm{p_2}  f_{q_i}(\p_2)(1-f_{q_i}(\p_2)) \partial^i_{\p_1} f_{q_f}(\p_1) \nonumber \\
    &&\qquad\qquad\qquad\qquad+ \frac{2f_{q_i}(\p_2)}{p_2}\frac{\p^i_1}{p_1} f_{q_f}(\p_1)(1-f_{q_f}(\p_1))\;,\\
    C^{q_f\bar{q}_i\longleftrightarrow q_f\bar{q}_i}_{q_f}[f] & = & \frac{g^4 C_F\mathcal{L}}{8\pi}\,\partial^i_{\p_1}\int \norm{p_2}f_{\bar{q}_i}(\p_2)(1-f_{\bar{q}_i}(\p_2)) \partial^i_{\p_1} f_{q_f}(\p_1) \nonumber \\
    &&\qquad\qquad\qquad\qquad
    +\frac{2f_{\bar{q}_i}(\p_2)}{p_2}\frac{\p^i_1}{p_1} f_{q_f}(\p_1)(1-f_{q_f}(\p_1))\;,
\end{eqnarray}
and using the quark exchange part of the scatterings we write the conversion term
\begin{eqnarray}
    C^{\rm Conversion}_{q_f}[f] & = &  C^{q_f g \overset{q}{\longleftrightarrow} q_f g}_{q_f}[f] + C^{q_f\bar{q}_f \overset{q}{\longleftrightarrow} gg}_{q_f}[f]  \;,\\
    & =&\frac{g^4 C_F^2\mathcal{L}}{4\pi} \sum_f\int \norm{p_2} 
    f_{q_f}(\p_1)(1+f_g(\p_1))\mathcal{I}_{q_f} 
    -f_g(\p_1)(1-f_{q_f})\mathcal{I}_{\bar{q}_f}\;. \nonumber\\
\end{eqnarray}
The same quark collision integrals apply to the antiquark channel after exchange of $q_f$ with $\bar{q}_f$ and vice-versa.
After summing the different contributions for each channel, one recovers the Fokker-Planck equation in section \ref{sec:Kinetic-Equation}.

\bibliography{main.bib} 
\bibliographystyle{ieeetr}
\end{document}